\documentclass[a4paper,11pt]{article}
\pdfoutput=1 

\usepackage{jheppub} 
\usepackage{amsmath}
\allowdisplaybreaks[4]        
\usepackage{amssymb}
\usepackage{euscript}     
\usepackage{color}         
\usepackage{tensor}        
\usepackage{amsthm}
\usepackage{bbm}

\usepackage{graphicx,subcaption}
 \usepackage{mathtools}

\usepackage{hyperref}
\hypersetup{colorlinks=true,allcolors=blue}

\usepackage[most]{tcolorbox}
\tcbuselibrary{skins, breakable, theorems}

\usepackage[header,title,page,titletoc]{appendix}  
\usepackage[T1]{fontenc} 
\usepackage[numbers]{natbib}

\newcommand{\eg}{{\it e.g.,}\ }
\newcommand{\ie}{{\it i.e.,}\ }

\newcommand{\mt}[1]{\textrm{\tiny #1}}

\renewcommand{\[}{\left[}

\newcommand{\CC}{\mathcal{C}}

\newcommand{\GN}{G_\mt{N}}

\newcommand{\talpha}{\tilde{\alpha}}
\newcommand{\THH}{T_\mt{H}}

\newcommand{\mC}{\mathcal{C}}

\newcommand{\mI}{\mathcal{I}}

\newcommand{\dd}{\mathrm{d}}
\newcommand{\one}{\mathbbm{1}}

\definecolor{browna}{rgb}{0.76,0.72,0.65}
\definecolor{brownb}{rgb}{0.71,0.69,0.65}
\definecolor{SpringGreen}{rgb}{0.95,0.97,0.95}
\definecolor{OliverGreen}{rgb}{0.09,0.34,0.09}
\definecolor{LeftGreen}{rgb}{0.13,0.54,0.13}

\newtcbtheorem[no counter]{theorem}{Theorem}{
	enhanced,
	sharp corners,
	attach boxed title to top left={
		yshifttext=-1mm
	},
	colback=white,
	colframe=browna,
	fonttitle=\bfseries,
	boxed title style={
		sharp corners,
		size=small,
		colback=browna,
		colframe=browna,
	} 
}{thm}

\newtcbtheorem[no counter]{Example}{Example}{
	enhanced,
	sharp corners,
	attach boxed title to top left={
		yshifttext=-1mm
	},
	colback=white,
	colframe=brownb,
	fonttitle=\bfseries,
	coltitle=white,
	boxed title style={
		sharp corners,
		size=small,
		colback=brownb,
		colframe=brownb,
	} 
}{exm}

\newtcbtheorem{myclaim}{Claim -- }
{    coltitle=OliverGreen,    
	colback=SpringGreen,    colframe=LeftGreen,    detach title,    boxrule=0pt,    leftrule=2pt,    attach title to upper,    sharp corners,    left=1mm,}
{claim}

\definecolor{browna}{rgb}{0.76,0.72,0.65}

\subheader{\today}

\title{Codimension-Zero Holographic Complexity and Spectral Three-Point Correlators}

\author{Masamichi Miyaji$^1$,~Shan-Ming Ruan$^{2,3}$,~Shono Shibuya$^4$,~Kazuyoshi Yano$^4$}

\affiliation{$^1$RIKEN Center for Interdisciplinary Theoretical and Mathematical Sciences (iTHEMS),
RIKEN, 2-1 Hirosawa, Wako, Saitama, 351-0198, Japan}
\affiliation{$^2$School of Physics, Peking University, Beijing 100871, China}
\affiliation{$^3$Center for High Energy Physics, Peking University, Beijing 100871, China}
\affiliation{$^4$Department of Physics, Nagoya University,
Nagoya, Aichi 464-8602, Japan}
\emailAdd{masamichi.miyaji@gmail.com, ruanshanming@pku.edu.cn, shibuya.shono.n8@s.mail.nagoya-u.ac.jp,
yano.kazuyoshi.h8@s.mail.nagoya-u.ac.jp}

\abstract{Holographic complexity is expected to capture the linear growth of black hole interiors, with quantum corrections ultimately leading to a late-time saturation. We develop a spectral formulation of holographic complexity measures associated with codimension-zero bulk regions in the framework of the complexity=anything proposal. The corresponding generating function of complexity is governed by three-point correlation functions of the density of states, establishing a direct connection between bulk geometric observables and nonperturbative Euclidean spacetime wormholes. We find that the codimension-zero holographic complexity reduces to the sum of two codimension-one contributions and exhibits a universal time evolution, characterized by an exponentially long regime of linear growth followed by saturation at the Heisenberg time. The late-time plateau is a direct result of the spectral level repulsion, a generic feature of quantum chaotic systems, thereby providing a universal mechanism for the saturation of holographic complexity. We also develop a complementary operator interpretation, in which Euclidean wormhole corrections render the semiclassical basis of fixed-complexity states overcomplete. Constructing a complete orthonormal basis then yields well-defined holographic complexity operators that reproduce the same universal time evolution. Our results establish a unified connection among black hole interiors, chaotic spectral statistics, and Euclidean wormholes in quantum gravity.}

\numberwithin{equation}{section}

\begin{document}

\preprint{RIKEN-iTHEMS-Report-26}

\maketitle
\flushbottom

\section{Introduction and Summary}\label{sec:introduction}

Holographic complexity has become an important probe of the black hole interior. Quantum complexity provides a quantitative measure of how difficult it is to prepare a state or implement a unitary transformation from a prescribed set of elementary operations \cite{Nielsen:2005mkt, Nielsen:2006cea}. In quantum complexity theory, the precise value of the computational complexity depends on the choice of gates, the allowed tolerance, and the cost assigned to each operation. However, its coarse features, such as the scaling with the size of the Hilbert space and the characteristic time dependence under chaotic dynamics, are expected to be much more robust. This perspective has played an important role in holography, where quantum complexity has been proposed to capture aspects of the black hole interior which are invisible to conventional entanglement measures \cite{Susskind:2014rva,Stanford:2014jda}.

A particularly sharp motivation comes from the two-sided AdS black hole, which is dual to the thermofield double state \cite{Maldacena:2001kr}. While entanglement entropy quickly reaches thermal equilibrium, the Einstein-Rosen bridge continues to grow for a much longer time \cite{Hartman:2013qma}. This observation led to the idea that the growth of the black hole interior should be dual to the growth of quantum complexity \cite{Susskind:2014rva,Stanford:2014jda}. From the bulk point of view, several concrete proposals have been extensively studied. 

\begin{figure}[t]
	\centering	\includegraphics[width=5in]{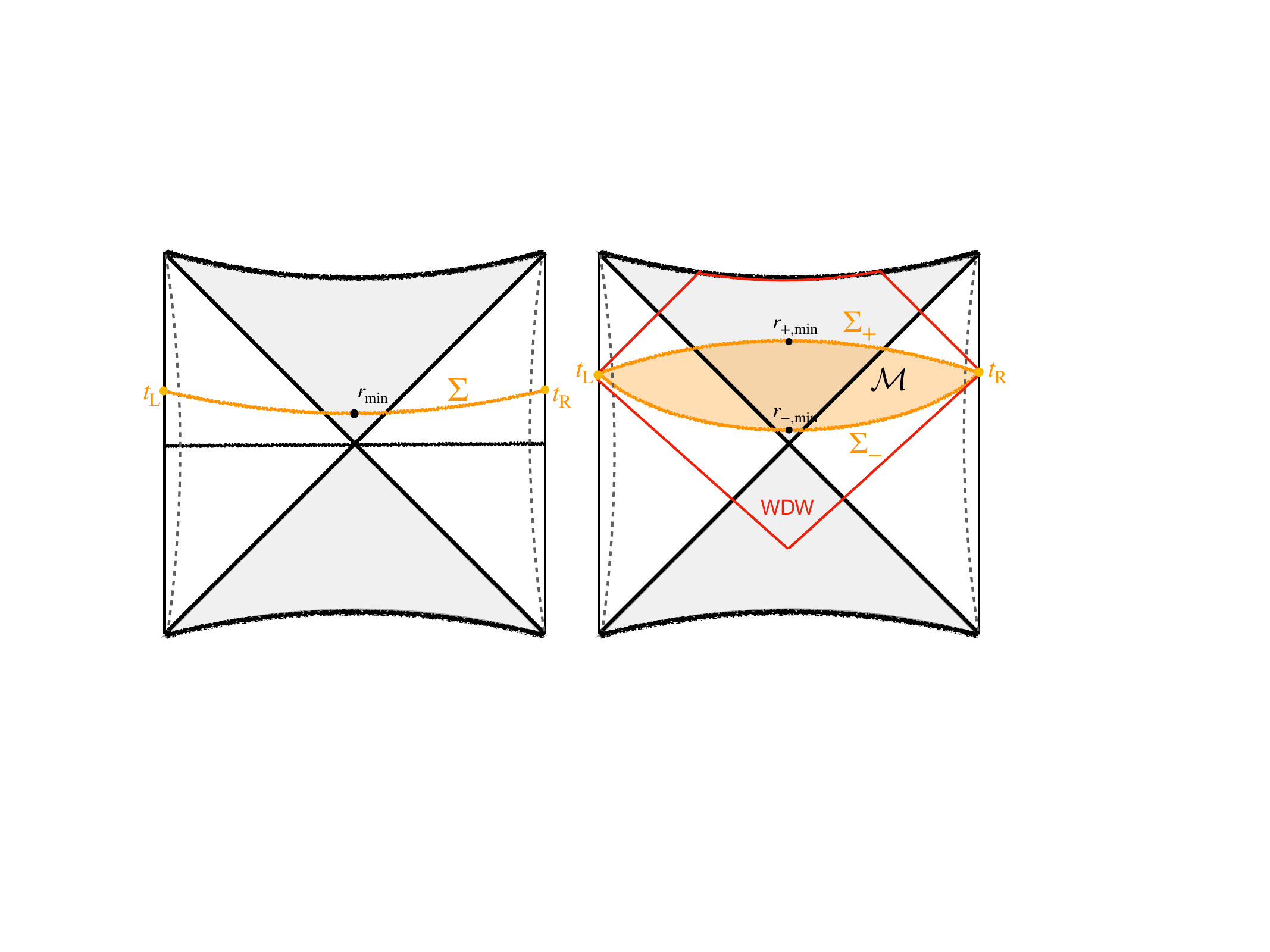}
	\caption{Schematic illustration of the Complexity=Anything proposal. The left panel represents the codimension-one observable evaluated on an extremal hypersurface $\Sigma$ anchored at the boundary times $t_L$ and $t_R$.  The right panel defines codimension-zero observables associated with a spacetime region, whose future and past boundaries are the extremal hypersurfaces $\Sigma_+$ and $\Sigma_-$.  WDW-patch observables arise as the special limit of this more general construction.}\label{fig:CAnything}
\end{figure}

This observation led to several proposals for holographic complexity, including complexity=volume (CV)~\cite{Susskind:2014rva,Stanford:2014jda}, complexity=action (CA)~\cite{Brown:2015bva,Brown:2015lvg}, and complexity=spacetime volume (CV2.0)~\cite{Couch:2016exn,Carmi:2016wjl}.  In a common normalization, these are written as
\begin{equation}
 \CC_{\rm V}(\Sigma_{\rm CFT})
  = \max_{\Sigma_{\rm CFT} = \partial \Sigma} \frac{V(\Sigma)}{\GN L}\,,
 \qquad
 \CC_{\rm A}(\Sigma_{\rm CFT})
  = \frac{I_{\rm WDW}}{\pi\hbar}\,,
 \qquad
 \CC_{\rm SV}(\Sigma_{\rm CFT})
  = \frac{V_{\rm WDW}}{\GN L^2}\,.
\label{eq:introCVCACVtwo}
\end{equation}
Here $\Sigma$ denotes a bulk hypersurface anchored on the boundary time slice $\Sigma_{\rm CFT}$, while $I_{\rm WDW}$ and $V_{\rm WDW}$ denote, respectively, the gravitational action and spacetime volume of the Wheeler-DeWitt patch. These proposals all exhibit the expected late-time linear growth for two-sided AdS black hole spacetimes. At the same time, they also contain conventional ambiguities. For example, the CV and CV2.0 proposals require the introduction of an additional length scale, while the CA proposal depends on choices associated with null boundary terms \cite{Lehner:2016vdi}. Rather than viewing these ambiguities as a defect, one may interpret them as a gravitational reflection of the ambiguities inherent in quantum complexity.

This viewpoint was developed systematically in the complexity=anything proposal \cite{Belin:2021bga,Belin:2022xmt,Jorstad:2023kmq}. In its codimension-one version, one considers a broad class of diffeomorphism-invariant observables constructed on bulk hypersurfaces. The construction involves two scalar functionals $F_1$ and $F_2$. First, one selects a special hypersurface $\Sigma_{F_2}$ by extremizing
\begin{equation}
\delta_X
\left[
\int_{\Sigma} d^d \sigma \sqrt{h}\,
F_2(g_{\mu\nu};X^\mu)
\right]
=0 \,.
\end{equation}
Then one evaluates another functional on this extremal hypersurface,
\begin{equation}
C_{F_1,\Sigma_{F_2}}= \frac{1}{\GN L}
\int_{\Sigma_{F_2}} d^d \sigma \sqrt{h}\,
F_1(g_{\mu\nu};X^\mu) \,.
\end{equation}
For $F_1=F_2=1$, this reduces to the usual extremal volume. More generally, however, a large class of such observables displays the same universal features expected of holographic complexity, including late-time linear growth and the switchback effect \cite{Shenker:2013pqa,Stanford:2014jda}. This suggests that the precise gravitational functional is not unique. Instead, holographic complexity should be characterized by universal properties shared by a large family of observables.

\begin{figure}[t]
	\centering	\includegraphics[width=3in]{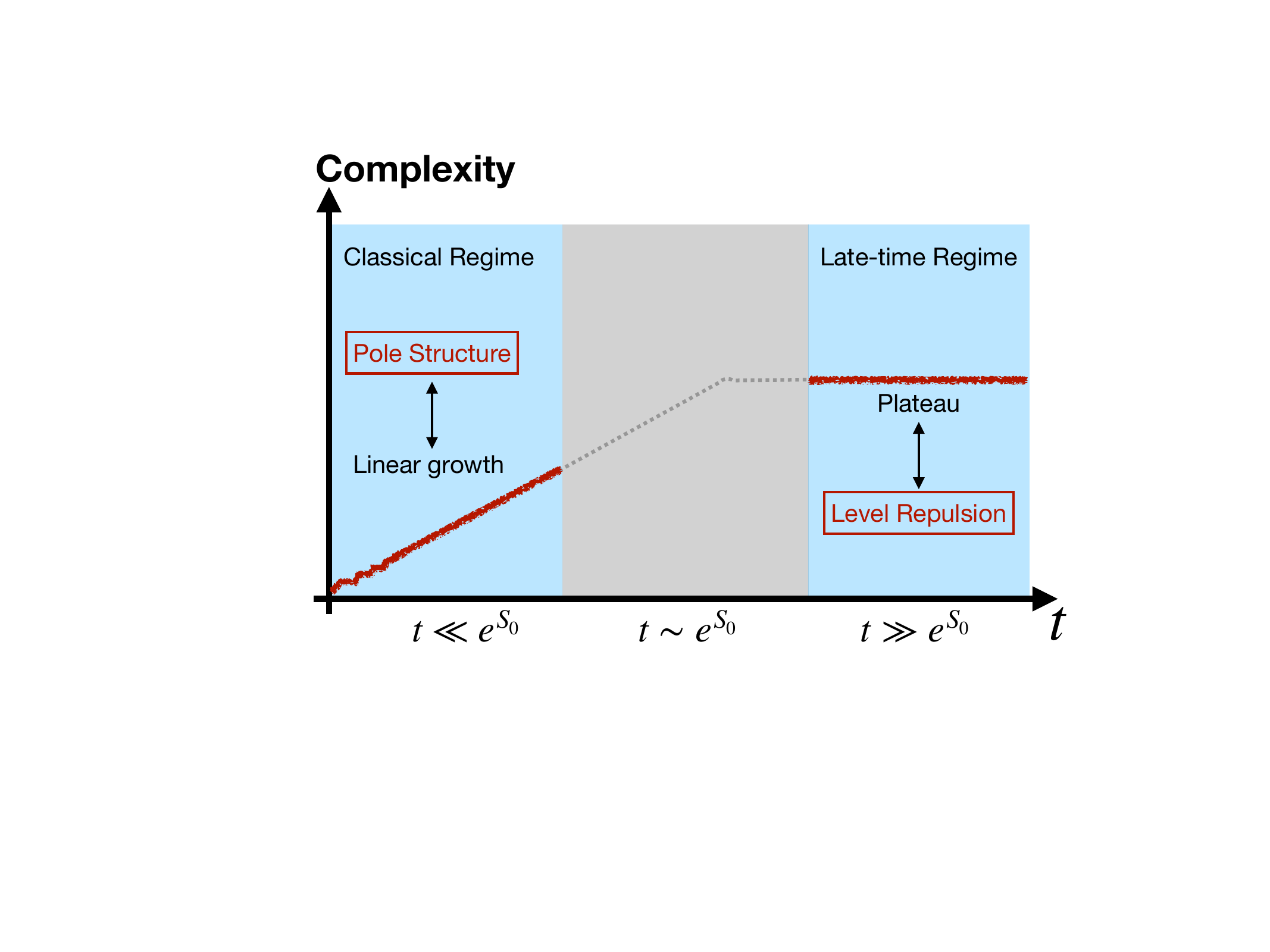}
\caption{Universal time evolution of holographic complexity. The early-time linear regime is controlled by a characteristic pole structure in the matrix elements of the holographic complexity operator. The late-time plateau is controlled by level repulsion in the discrete spectrum of chaotic systems. The crossover occurs at $t\sim e^{S_0}$, equivalently at the Heisenberg time $\THH$.}\label{fig:Universality}
\end{figure} 

The same philosophy extends naturally to codimension-zero observables \cite{Belin:2022xmt}. Instead of assigning complexity to a codimension-one hypersurface, one considers a bulk spacetime region $\mathcal{M}$ bounded by future and past codimension-one hypersurfaces $\Sigma_\pm$, both anchored on the same boundary time slice. This construction includes WDW-patch-type observables and, in appropriate limits, reproduces the standard CA and CV2.0 proposals. Moreover, these codimension-zero observables again exhibit the expected late-time linear growth in the classical black hole geometry. The crucial structural feature of the codimension-zero construction is that, after using Stokes' theorem, the observable can be represented as a sum of two boundary contributions, namely 
\begin{equation}\label{eq:Cpm}
\CC_{\rm Any} =\CC_+(\Sigma_+)+\CC_-(\Sigma_-) \,,
\end{equation}
where $\CC_\pm$ are codimension-one observables associated with the future and past boundaries of $\mathcal{M}$. Classically, the two terms in \eqref{eq:Cpm} produce the same type of linear growth as ordinary codimension-one holographic
complexities: 
\begin{equation}\label{eq:intro_classical_growth}
\CC_{\rm Any}^{\rm cl}(t)
=
\text{const}
+(M_+ + M_-)t+\cdots \,.
\end{equation}
Here the growth rates $M_\pm$ are the late-time momenta conjugate to the boundary times of $\Sigma_\pm$. For ordinary AdS black holes, these quantities are proportional to the ADM mass. Guided by this holographic identification, it was later confirmed that this linear growth behavior of the computational complexity is indeed a universal feature of chaotic systems \cite{Brandao:2019sgy, Haferkamp:2021uxo, Oszmaniec:2022srs}.

\begin{figure}[t]
	\centering	\includegraphics[width=6in]{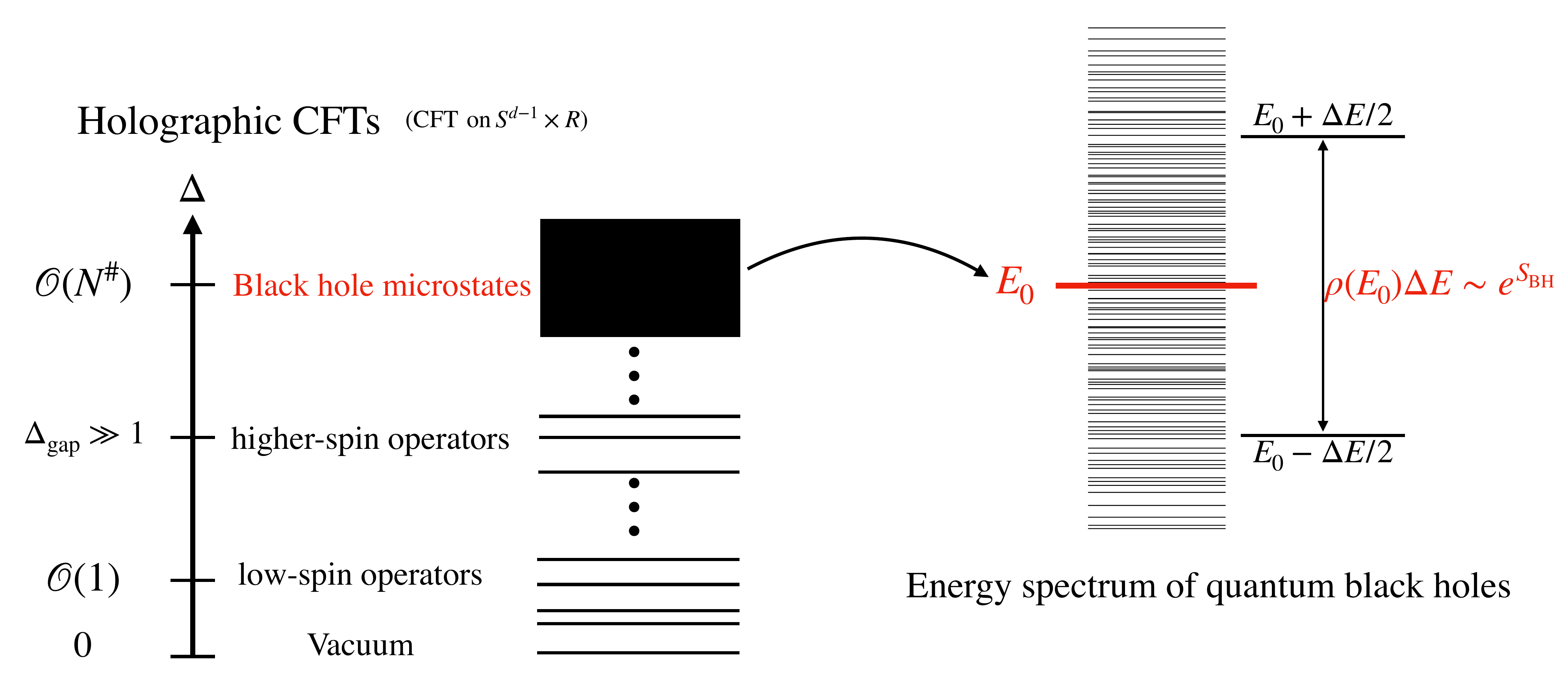}
	\caption{Discrete spectrum of holographic CFT and black hole microstates. The left part depicts the large-$N$ hierarchy: low-spin operators lie near $\Delta=\mathcal{O}(1)$, higher-spin operators are separated by a large gap, and black hole microstates appear at energies of order $N^\#$. The right part zooms into a microcanonical window $E_0-\Delta E/2<E<E_0+\Delta E/2$, containing $\rho(E_0)\Delta E\sim e^{S_{\rm BH}}$ states. The discreteness of the spectrum is expected to be visible at the Heisenberg time $T_{\mathrm{H}}$.}\label{fig:discrete}
\end{figure}

Of course, the classical prediction cannot be the complete answer for a quantum system.  A finite-dimensional quantum system cannot support indefinite linear growth.  In a chaotic system, complexity is expected to grow for an exponentially long time in the entropy and then saturate near its maximal value, as shown in figure \ref{fig:Universality}. Even when a classical black hole admits a smooth semiclassical geometry, the dual CFT has a discrete spectrum which consists of black hole microstates. Figure~\ref{fig:discrete} illustrates this from a spectral perspective. If one focuses on a microcanonical window centered at $E_0$, the relevant time scale is the Heisenberg time
\begin{equation}
 \THH := \frac{2\pi\hbar}{\hbox{average energy spacing}} \,.
\label{eq:introHeisenbergTime}
\end{equation}
This is a fundamental timescale in finite-dimensional quantum systems because it characterizes the time at which the discreteness of the system comes into play.

\paragraph{Generating functions and spectral universality.}
The spectral mechanism underlying the saturation of complexity is closely analogous to the slope-ramp-plateau structure of the spectral form factor. Semiclassical gravity captures the early growth, while finite-entropy effects are governed by correlations between nearby energy levels \cite{Bohigas:1983er,Cotler:2016fpe,Saad:2019lba}. In gravitational language, these correlations are naturally interpreted as non-perturbative Euclidean wormhole corrections. The key in our formalism to this question is not to define the holographic complexity operator directly, but instead to consider its generating function. This strategy was developed in JT gravity for fixed-length states and related observables \cite{Iliesiu:2021ari,Miyaji:2025yvm}, and was then extended to an infinite class of holographic complexity measures in \cite{Miyaji:2025jxy}. For codimension-one holographic complexity measures, the basic object is the expectation value defined as 
\begin{equation}
\left\langle e^{-\alpha \CC}\right\rangle
=
\left\langle {\rm TFD}(t) \right|
\widehat{e^{-\alpha \CC}}
\left|{\rm TFD}(t)\right\rangle \,.
\end{equation}
Its spectral representation takes the form \footnote{Here we consider only Hermitian operators $\widehat{e^{-\alpha\mathcal C}}$ with real matrix elements in the energy eigenbasis. This ensures that the generating function satisfies $G(t)=G(-t)=G^\ast(t)$. Together with the symmetric spectral kernel, the whole spectral representation is thus invariant under the exchange $E_i\leftrightarrow E_j$.}
\begin{equation}\label{eq:intro_codim_one_GF}
\left\langle e^{-\alpha \CC}\right\rangle
=
\frac{e^{2S_0}}{Z}
\int dE_i dE_j \,
e^{\pm i(E_i-E_j)t}
\overline{\left\langle D(E_i)D(E_j)\right\rangle}
\left\langle E_i\right|\widehat{e^{-\alpha \CC}}\left|E_j\right\rangle \,.
\end{equation}
This expression separates the problem into two distinct ingredients. The first is the two-point correlation function of the spectral density, whose expression at small energy separation is governed by random matrix universality in chaotic systems \cite{Bohigas:1983er,Cotler:2016fpe,Eynard:2015aea}. The second is the matrix element associated with the holographic complexity operator in the energy basis.

One of the main results of \cite{Miyaji:2025jxy} is that the classical linear growth of holographic complexity imposes a very strong constraint on the matrix element associated with holographic complexity. In the regime $\alpha\sim E_{ij}\ll 1$, the necessary and sufficient condition for the classical linear growth is given by the particular pole structure of the matrix element, \ie 
\begin{equation}
\boxed{\left\langle E_i\right|\widehat{e^{-\alpha \CC}}\left|E_j\right\rangle \equiv 
M_\alpha (E_i, E_j)  \sim \frac{1}{\pi e^{S_0}\bar D(\bar E)} \frac{\talpha}{(\talpha+iE_{ij})(\talpha-iE_{ij})} +\cdots \,,}
\label{eq:intro_pole_structure}
\end{equation}
with
\begin{equation}
\bar{E}:=\frac{E_i+E_j}{2}\,, \quad E_{ij}:= E_i -E_j \,, \quad \talpha=M\alpha+O(\alpha^2) \,. 
\end{equation}
The first-order pole is located on the imaginary $E_{ij}$ plane at $E_{ij}=\pm i\talpha$, and the coefficient is fixed by requiring that $\widehat{e^{-\alpha\CC}}$ reduces to the identity operator in the limit $\alpha\to0$. From the definition in \eqref{eq:intro_codim_one_GF}, the pole in \eqref{eq:intro_pole_structure} gives the classical contribution
\begin{equation}
\left\langle e^{-\alpha\CC}\right\rangle_{\rm cl} \simeq e^{-\talpha t} \,,
\end{equation}
and hence the corresponding holographic complexity grows linearly after taking the regulated $\alpha\to0$ limit. The late-time plateau has a different origin. At finite $S_0$, the energy spectrum is discrete, and chaotic systems exhibit level repulsion. For example, in the GUE universality class, the connected part of the two-point spectral correlator is governed by the sine kernel. In the local spectral regime,
\begin{equation}
\left\langle D(E_i)D(E_j)\right\rangle
\simeq
\bar D(E_i)\bar D(E_j) +e^{-S_0}\bar D(E_i)\delta(E_{ij}) - \frac{\sin^2\left(e^{S_0}\pi \bar D(\bar E)E_{ij}\right)}{(e^{S_0}\pi E_{ij})^2} \,.
\label{eq:intro_two_point_sine}
\end{equation}
The off-diagonal part vanishes as $E_{ij}\to0$, which is the spectral manifestation of level repulsion. Combining \eqref{eq:intro_pole_structure} with \eqref{eq:intro_two_point_sine}, we can obtain a slope-ramp-plateau structure for the generating function. After the regulated $\alpha\to0$ limit, the same mechanism gives a transition from linear growth to a plateau at the Heisenberg time,
\begin{equation}
\langle \hat\CC \rangle \approx  \text{const} +
\begin{cases}
M t\left(1-\dfrac{t}{\THH}+\dfrac{t^2}{3\THH^2}\right) \,,
& t<\THH \,,\\
\dfrac{1}{3}M \THH \,,
& t\geq \THH \,.
\end{cases}
\label{eq:intro_codim_one_result}
\end{equation}
Therefore, the linear growth and late-time saturation of complexity arise from two complementary ingredients: the pole structure of the matrix element and the level repulsion of the spectrum.

\paragraph{Codimension-zero holographic complexity and spectral three-point correlators.}

The purpose of the present work is to apply the same mechanism to the most general codimension-zero holographic complexity measures in detail. The central new point is that codimension-zero holographic observables probe richer spectral data than codimension-one observables. The starting point is to formulate the codimension-zero problem in terms of generating functions,
extending the codimension-one analysis of \cite{Miyaji:2025jxy}. This has two advantages. First, the generating function provides a regulated object whose classical limit can be matched directly to geometric observables.  Second, it cleanly separates two universal ingredients: the matrix elements of the complexity operator and the spectral correlation functions. For codimension-one observables the relevant spectral data are two-point functions of the density of states.  For codimension-zero observables, the corresponding generating function is controlled instead by a spectral three-point correlator. Since a codimension-zero complexity measure can be decomposed as in \eqref{eq:Cpm}, it is natural to define the generating functions of codimension-zero holographic complexity $\CC_{\mathrm{Any}}$ in terms of 
\begin{equation}\label{eq:intro_codim_zero_spectral}
\begin{split}
&\left\langle e^{-\alpha_+\CC_+-\alpha_-\CC_-}\right\rangle :=
\left\langle {\rm TFD}(t) \right|
\widehat{ e^{-\alpha_+\CC_+} } \cdot
\widehat{ e^{-\alpha_-\CC_-} }
\left|{\rm TFD}(t)\right\rangle \\
&\qquad=\frac{e^{3S_0}}{Z}
\int dE_1dE_2dE_3\,
e^{-iE_{12}t}
\overline{\left\langle D(E_1)D(E_2)D(E_3)\right\rangle} \mathbf{M}_{\talpha_+}(E_1,E_3)\mathbf{M}_{\talpha_-}(E_3,E_2) \\
\end{split}
\end{equation}
This is the main new feature of the present paper. The three-point spectral
correlator decomposes into several disconnected and connected components.  These components have a direct gravitational interpretation: the fully disconnected piece gives the classical growth, the pairwise connected pieces describe wormholes joining two of the bra, operator, and ket sectors, and the fully connected piece describes a wormhole joining all three sectors.  In this way, the spectral three-point correlator provides a precise organization of the wormhole corrections to codimension-zero holographic complexity. Applying  the same universal pole structure of two matrix elements in \eqref{eq:intro_codim_zero_spectral} as the codimension-one case, we show that the disconnected contribution then reproduces the classical growth in \eqref{eq:intro_classical_growth}. The connected parts of the three-point correlator are non-perturbative in $e^{-S_0}$ and become important near the Heisenberg time. More explicitly, after the appropriate regularization of the generating function and the $\alpha_\pm\to 0 $ limit, the expectation value of the codimension-zero complexity operator takes the universal form
\begin{equation}
\langle \hat\CC_{\rm Any}\rangle = \langle\hat\CC_+ \rangle + \langle\hat\CC_-\rangle
\approx \text{const} +
\begin{cases}
(M_+ + M_-) t \left( 1+  \mathcal{O}(\dfrac{t}{\THH}) \right) \,,
& t<\THH \,,\\
\sim (M_+ + M_-)\THH \,,
& t\geq \THH \,.
\end{cases}
\label{eq:intro_main_result}
\end{equation}
This result shows explicitly that quantum corrections preserve the classical linear growth for a long time, but prevent the growth from continuing indefinitely. The transition occurs at the same Heisenberg time that controls the ramp-to-plateau transition in the spectral form factor. Let us emphasize that the plateau in \eqref{eq:intro_main_result} is a statement
about the complexity itself, not merely about the generating function. The
plateau arises from a cancellation between the semiclassical contribution and the wormhole corrections organized by the spectral three-point correlator.  In this sense, the late-time saturation of holographic complexity measures is a direct consequence of spectral level repulsion in the finite black hole spectrum. 

There is also an operator interpretation of the same result.  The continuous
basis of fixed-complexity states is over-complete once non-perturbative wormhole corrections are included. Passing to a complete orthonormal basis removes this redundancy and gives a finite answer, in close analogy with the fixed-length basis in JT gravity and with finite-$N$ constructions of bulk Hilbert spaces \cite{Miyaji:2024ity,Miyaji:2025ucp}. This perspective clearly clarifies why the late-time plateau should be regarded as a consequence of the finite number of discrete black hole microstates.

The remainder of the paper is organized as follows.  In section \ref{sec:gen}, we introduce the microcanonical thermofield double state and define the generating functions for codimension-one and codimension-zero holographic complexity measures. We then derive the spectral representation of the codimension-zero generating function and explain how the spectral three-point correlator decomposes into classical and wormhole sectors.  In subsection \ref{sec:reg}, we discuss the regularization of the generating function and extract the complexity expectation value by taking the $\alpha_\pm\to0$ limit.  In section \ref{sec:prob}, we present the complementary operator interpretation in terms of probability distributions and construct the holographic complexity operator by using complete orthonormal bases. Finally, we conclude with open questions and possible extensions. Technical details of the residue calculations and several long formulae are collected in the appendices.

\section{Holographic Complexity Measures and Generating Functions}\label{sec:gen}
In this section, we develop the spectral representation of codimension-zero holographic complexity and derive the universal time evolution of holographic complexity measures. We begin by specifying our microcanonical conventions and reviewing the formalism and results for codimension-one holographic complexity. We then introduce the generating function appropriate for codimension-zero holographic observables and show that the corresponding quantum corrections are organized by the spectral three-point function. Finally, we take the $\alpha_\pm\to0$ limit from the generating function to derive the universal time evolution of holographic complexity: an exponentially long period of linear growth followed by a smooth transition to the late-time plateau.

\subsection{The microcanonical black hole sector}
Throughout this paper, we work with the microcanonical TFD state, as our focus is the two-sided AdS black hole spacetime. In particular, we consider typical black hole states whose energies lie within a window of width $\Delta E$ centered at $E_0$. The corresponding microcanonical ensemble is defined by
\begin{equation}\label{eq:MCwindow}
 E_i \in \mathcal{I}
 =\left[E_0-\frac{\Delta E}{2},E_0+\frac{\Delta E}{2}\right] \,.
\end{equation}
Let
\begin{equation}
 \rho(E)=\sum_{i\in\mathcal{H}_{\rm BH}}\delta(E-E_i)
       =e^{S_0}D(E)
\label{eq:microscopic-density}
\end{equation}
denote the microscopic density of black hole states. The total number of states in the microcanonical ensemble is therefore
\begin{equation}
 Z:=\int_{E_0-\Delta E/2}^{E_0+\Delta E/2}dE\,\rho(E)
 \simeq e^{S_0}\overline{D}(E_0)\Delta E \,.
\label{eq:microcanonical-Z}
\end{equation}
We use an overbar to denote the smooth, ensemble-averaged density, so that $\overline{\rho}(E)=e^{S_0}\overline{D}(E)$. The parameter $S_0$ represents the coarse-grained entropy, while $D(E)$ remains of order one in the large-$S_0$ limit. The microcanonical TFD state can be written in terms of energy eigenstates as
\begin{equation}\label{eq:TFD}
\begin{split}
  |\text{TFD}(t)\rangle
    &=\frac{1}{\sqrt{Z}}
      \sum_{E_n\in\mI}e^{-iE_nt}|E_n\rangle \\
    &=\frac{e^{S_0}}{\sqrt{Z}}
      \int_{E_0-\Delta E/2}^{E_0+\Delta E/2}
      dE\,D(E)\,e^{-iEt}|E\rangle \,.
\end{split}
\end{equation}
Here, $|E_n\rangle$ is shorthand for the paired state
$|E_n\rangle_{\mt{L}}\otimes|E_n\rangle_{\mt{R}}$. We fix the normalization of the TFD state by imposing
\begin{equation}
 \langle\text{TFD}(t)|\text{TFD}(t)\rangle=1 \,.
\end{equation}
The second line of \eqref{eq:TFD} is the continuum representation of the discrete sum. In the continuum approximation, the resolution of the identity and the normalization of the generalized energy states are
\begin{equation}
 \hat{\one}=e^{S_0}\int_{\mI}\dd E\,D(E)|E\rangle\langle E| \,,
 \qquad
 \langle E_i|E_j\rangle
 =\frac{\delta(E_i-E_j)}
        {e^{S_0}\overline{D}(E_i)} \,.
\label{eq:energy-normalization}
\end{equation}
Products involving the Dirac delta distribution in \eqref{eq:energy-normalization} must ultimately be interpreted as the continuum limits of the corresponding products of discrete Kronecker deltas. This distinction is immaterial for a fixed nonzero regulator but becomes essential when the regulator is removed. The corresponding Heisenberg time is then given by
\begin{equation}
 \THH=2\pi\overline{\rho}(E_0)
     =2\pi e^{S_0}\overline{D}(E_0) \,,
\label{eq:TH}
\end{equation}
with the local mean level spacing given by $\Delta_{\text{level}} = \frac{1}{\bar{\rho}(E_0)}$. We will show that, on this timescale, holographic complexity transitions from early-time linear growth to a late-time plateau.

Throughout the remainder of the paper, we generally work in the regime
\begin{equation}
 \frac{\Delta E}{E_0}\ll1 \,,
 \qquad
 t \, \Delta E\gg1 \,.
\label{eq:working-regime}
\end{equation}
The first condition allows smooth functions of energy to be approximated by their values at $E_0$, while the second suppresses endpoint effects from the microcanonical window. Another important  characteristic timescale marking the onset of the linear ramp is the so-called Thouless time $T_{\rm Th}$ \cite{Sonner:2017hxc,Gharibyan:2018jrp,Altland:2020ccq,Altland:2022xqx,Orman:2024mpw}. We thus further assume that the energy differences (like $E_{ij}$) relevant at late times lie within the universal random-matrix regime below the Thouless energy $E_{\rm Th} \sim T_{\rm Th}^{-1}$.

\subsection{Codimension-one holographic complexity: a warm-up}\label{sec:codimensionone}
Before we move to codimension-zero holographic complexity measures \eqref{eq:exppm}, it is worthwhile to review the main results of codimension-one holographic complexity measures. Although most results in this subsection have been presented in the Appendix of \cite{Miyaji:2025jxy}, this part will serve as the elementary building block for calculations in the codimension-zero case.

Given a particular measure for codimension-one holographic complexity $\CC$, we can formally define the holographic complexity operator $\hat{\CC}$ in terms of
\begin{equation}
 \hat{\CC} :=  \sum_{\CC-\text{basis}} \CC \, | \CC \rangle \langle \CC |= \int \CC | \CC \rangle \langle \CC | \, d\CC \,,
\end{equation}
where $| \mC \rangle$ denotes the quantum state with a fixed generalized volume $\mC$. A particular example is the geodesic-length operator in two-dimensional gravity \cite{Miyaji:2025yvm}. However, after including quantum corrections from Euclidean wormholes in the gravitational path integral, two distinct states $| \mC \rangle, | \mC' \rangle $ are not orthogonal to each other any more, \ie the overlap $| \langle \mC | \mC' \rangle|^2$ becomes nonzero. As a result, the holographic complexity operator $\hat{\mC}$ is not well-defined due to the over-completeness of fixed-$\mC$ basis. 

Rather than attempting to define holographic complexity operator $\CC$ directly, a complementary approach was introduced in \cite{Iliesiu:2021ari,Miyaji:2025yvm,Miyaji:2025jxy} by defining the regularized operator $\widehat{e^{-\alpha \mC}}$. The corresponding expectation value of $\widehat{e^{-\alpha \mC}}$ is called the generating function of complexity since it could reduce to $\mC$ in the $\alpha \to 0$ limit. The key point of \cite{Miyaji:2025yvm} is that the spectral decomposition of the generating functions of complexity measures takes the universal form: 
\begin{equation}\label{eq:G1}
 \langle e^{-\alpha\CC}\rangle := \frac{e^{2S_0}}{Z}\int_{\mathcal{I}} dE_idE_j\,\underbrace{e^{-i\left(E_i-E_j\right)t}}_{\text{time evolution}} \, \times \underbrace{\overline{\langle  D(E_i)D(E_j)\rangle}}_{\text{spectral correlators}} \, \times \underbrace{\langle E_i | \widehat{e^{-\alpha\CC}}  | E_j \rangle}_{\text{matrix elements}}\,.
\end{equation}
From the above decomposition, it is obvious that the full information is decoded into the spectral two-point correlator $\overline{\langle  D(E_i)D(E_j)\rangle}$ and the matrix element $\mathbf{M}_{\talpha}(E_i,E_j)$. Furthermore, the time dependence only comes from the Fourier factor $e^{-iE_{ij}t}$ which originates from the boundary condition of TFD states $|\mathrm{TFD}(t) \rangle$.   

The full structure of the matrix elements is nothing but the representation of the operator $\widehat{e^{-\alpha \mC}}$ in the eigenenergy basis. Obviously, such a representation depends on the fine definition of the operator and is usually not available, except for simple examples \footnote{See Appendix \ref{app:examples} for more explicit examples.}. However, due to the simple time dependence $e^{-i E_{ij}t}$, we found that the time evolution of the expectation value requires much less information about the details of holographic complexity operators. In particular, the dominant contributions of time evolution are determined by the pole structure of the matrix elements on the complex $E_{ij}$-plane after analytic continuation. More importantly, although we do not specify the gravitational observable for holographic complexity, the holographic input from the bulk side is that holographic complexity should present the expected linear growth 
\begin{equation}\label{eq:linear}
\left\langle\mC \right\rangle_{\rm classical}
\simeq  M t \,, 
\end{equation}
in the classical limit with $\GN \to 0$. The growth rate $M$ typically is related to the ADM energy in the case of AdS black holes. We will leave it as the input data from the holographic bulk side. This is the core idea of the complexity=anything proposal \cite{Belin:2021bga,Belin:2022xmt,Jorstad:2023kmq} which shows that there are infinitely many gravitational observables that probe the black hole interior and present the linear growth as well. With this holographic input \eqref{eq:linear}, it was shown in \cite{Miyaji:2025jxy} that the necessary and sufficient condition for the early-time linear growth of codimension-one holographic complexity is given by a particular first-order pole structure of the matrix elements, \ie 
\begin{equation} \label{eq:matrixelement}
  \mathbf{M}_{\talpha} (E_i,E_j) \equiv \langle E_i | \widehat{e^{-\alpha\CC}}  | E_j \rangle \sim \frac{\tilde\alpha}{\pi e^{S_0}\bar D(\bar E)}\frac{1}{(\tilde{\alpha} + i E_{ij} )(\tilde{\alpha} - i E_{ij})} \,,
\end{equation}
Here, we only focus on the small $\alpha$ regime with $\tilde\alpha=M\alpha+\mathcal{O}(\alpha^2)$. We also have fixed the normalization by using  
\begin{equation}\label{eq:completematrixelement}
  \lim_{\alpha \to 0}   \langle E_i | \widehat{e^{-\alpha \CC }}  | E_j \rangle =   \langle E_i  | E_j \rangle =\frac{\delta(E_{ij})}{e^{S_0} \bar{D}(E_i)} \,, 
\end{equation}
and the fact that the Dirac delta function can be represented by the following limit: 
\begin{equation}
\pi  \,  \delta (E) =  \lim_{\alpha \to 0} \frac{\alpha}{\alpha^2 + E^2} \,.
\end{equation}
The conclusion in \cite{Miyaji:2025jxy} thus indicates that, if the matrix elements of a given measure do not contain such a simple pole, it will not give rise to an exponentially long period of linear growth. In other words, the corresponding measure fails to be a candidate of holographic complexity. In the same sense, we can find that pole structure \eqref{eq:matrixelement} selects out the holographic complexity measures among all infinite gravitational observables.

On the other hand, except for the definition of holographic complexity measures, the information about the holographic system itself is indeed reflected in the spectral correlator $\overline{\langle D(E_1)D(E_2)\rangle}$. From the dual boundary perspective, we interpret this as the ensemble-averaged spectral correlator. From the viewpoint of gravity side, it corresponds to the spectral density calculated by the Euclidean gravitational path integral including high-genus saddles, \ie spacetime wormholes. Nevertheless, we will not distinguish these two perspectives. Importantly, the ensemble-averaged spectral correlator $\overline{\langle D(E_1)D(E_2)\rangle}$ indeed encodes the quantum corrections beyond the leading classical spacetime.

Although the knowledge about the full spectral correlators is absent for quantum gravity, the important lesson about holographic spacetime is that black holes are dual to quantum chaotic systems.  
Thanks to the quantum chaos nature, we thus can obtain universal features about spectral correlation function because quantum chaotic systems typically exhibit universal spectral statistics described by random matrix theory (RMT). When the energy difference of two spectra is less than a characteristic energy scale called the Thouless energy $E_{\rm Th}$, \ie $|E_i -E_j| \ll E_{\mt{Th}}$, the ensemble-averaged spectral correlator is then given by that of the appropriate Wigner–Dyson ensemble. 

\begin{figure}[t]
	\centering	\includegraphics[width=3in]{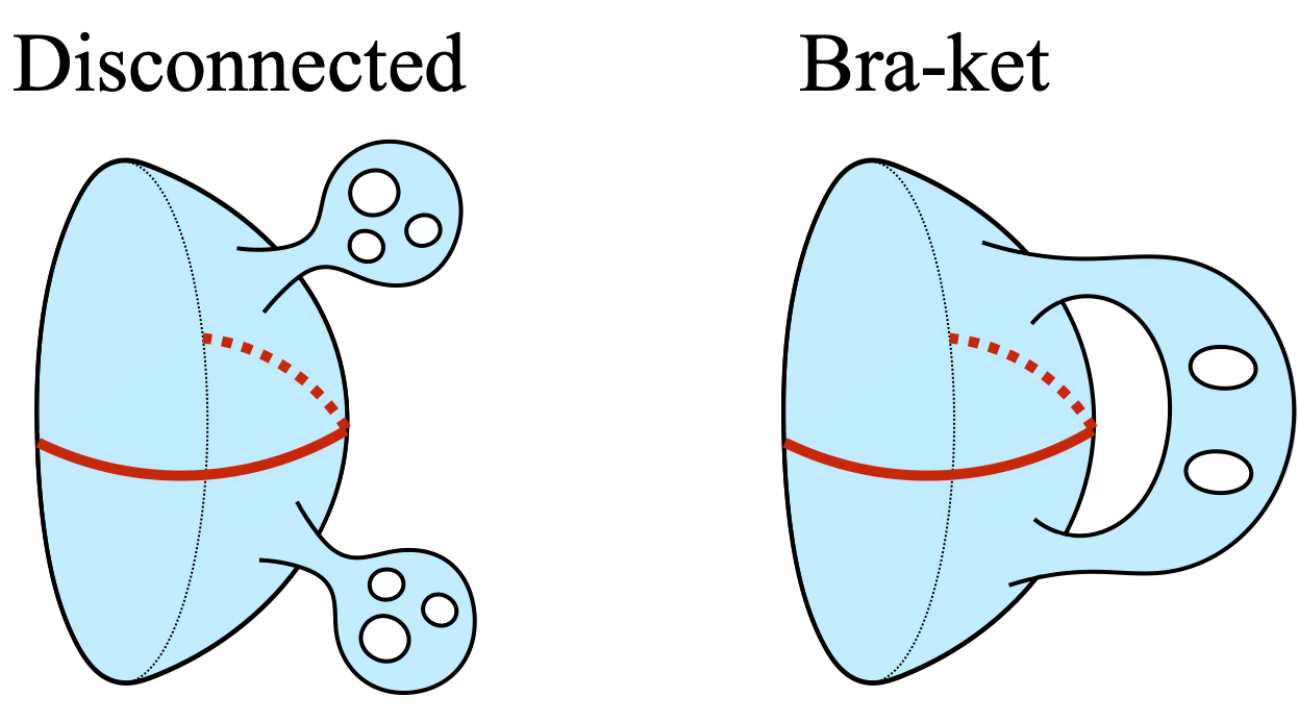}
	\caption{Schematic illustration of Euclidean wormhole saddles suggested by the decomposition of the two-point correlation function of the spectral density in eq.~\eqref{eq:DD}. Each diagram represents a distinct topology of connectivity between bra and ket.}\label{fig:wormhole}
\end{figure}

For definiteness, we specialize to the Gaussian unitary ensemble (GUE), or more generally to the $\beta=2$ unitary class, for which the eigenvalue correlations admit a determinantal representation. Correspondingly, the universal two-point correlation function of the spectral density is 
\begin{equation}\label{eq:DD}
    \overline{\langle D(E_i)D(E_j)\rangle} \simeq     \overline{D}(E_i)\overline{D}(E_j) + e^{-S_0}\overline{D}(E_i) \delta(E_i -E_j) - e^{-2S_0}K(E_i,E_j)^2\,,
\end{equation}
where $K(E_i,E_j)$ is called Christoffel-Darboux (CD) kernel. More generally, the spectral two-point correlator can be written for all
three Dyson classes as
\begin{equation}
    \overline{\langle D(E_i)D(E_j)\rangle}  \simeq   \bar D(E_i)\bar D(E_j)
 + e^{-S_0}\bar D(\bar E)\delta(E_i-E_j) - e^{-2S_0} \mathcal{Y}_2^{(\beta)}(E_i,E_j)\,.
    \label{eq:DD_general_beta}
\end{equation}
In the main text, we will only focus on the unitary class with a simpler determinantal representation and defer the discussion of the other Dyson classes to Appendix \ref{app:Dyson}. More importantly, we can find the gravitational dual interpretation to each term in the ensemble-averaged spectral correlator: the first factorized term can be viewed as a contribution from disconnected geometries in Euclidean gravitational path integral, whereas the rest can be viewed as connected spacetime wormholes (see figure \ref{fig:wormhole}).

After including the quantum corrections up to the order of $e^{-2S_0}$, one can expect that the contributions from high-genus Euclidean wormholes are not ignorable at late-times. Noting that the expectation value of holographic complexity is derived by taking $\alpha \to 0$ limit, \ie 
\begin{equation}
       \langle{e^{-\alpha \mC}} \rangle \approx 1- \alpha \, \mC + \mathcal{O}(\alpha^2)  \,,
\end{equation}
or equivalently 
\begin{equation}\label{eq:definederivative}
  {\mC}  := - \lim_{\alpha \to 0}  \partial_\alpha \left(\langle{e^{-\alpha \mC}} \rangle_{\text{reg}}  \right)\,.
\end{equation}
This limit is performed by subtracting the unphysical divergence while imposing that the generating operator reduces to the identity in the limit $\alpha \to 0$, \ie
\begin{equation}\label{eq:complete}
  \lim_{\alpha \to 0}  \widehat{e^{-\alpha \mC }} =    \hat{\mathbbm{1}} =  \sum_{\text{complete basis}}  | \mC \rangle \langle \mC |   \,. 
\end{equation}
Interestingly and crucially, the particular pole located $E_{ij}=-i \talpha$ therefore reduces to 
\begin{equation}
 \text{Pole:} \qquad E_i \to E_j \,, \qquad \text{with} \qquad \alpha \to 0 \,. 
\end{equation}
See section \ref{sec:reg} for more details of the regularization procedure. Due to this reduction, our previous paper \cite{Miyaji:2025yvm} finds that the late-time plateau of holographic complexity is derived by the level repulsion, which is the hallmark of quantum chaotic systems. More precisely, the origin of the late-time plateau of codimension-one holographic complexity is equivalent to the property of spectral correlation function:  
 \begin{equation}\label{eq:level2repulsion}
  \boxed{\overline{\langle D(E_i) D(E_j)\rangle}_{\text{off-diag}}
  = \bar{D}(E_i)\bar{D}(E_j)
  -e^{-2S_0}K(E_i,E_j)^2\underset{E_{ij}\to0}{\to}0\,.}
\end{equation}
Namely, the off-diagonal part of the spectral correlator \eqref{eq:DD}  vanishes exactly in the limit $E_{ij}\to 0$. Level repulsion is one of the key universal features of quantum chaotic systems that nearby energy levels avoid one another, in contrast to the behavior of generic integrable systems. One can easily prove this condition \eqref{eq:level2repulsion} by directly applying the simple property of the CD kernel, \ie $ K(E_i,E_i ) = \langle\rho (E_i) \rangle = e^{S_0} \bar{D}(E_i)$.

\subsubsection{Explicit results for GUE}
To obtain analytical expressions for the time evolution of both the generating functions and holographic complexity measures, 
we could further consider the explicit form of the CD kernel. Although the exact CD kernel depends on the matrix potential and on the finite matrix size, its local bulk behavior becomes universal in the large-$N$ limit. More precisely, after unfolding the spectrum by the smooth local density of states and considering energy separations of the order of the mean level spacing, the CD kernel of the GUE approaches the sine kernel, 
\begin{equation}\label{eq:sinekernelK}
    K(E_i,E_j) \simeq
    \frac{ \sin\!\left[    \pi \bar{\rho}(\bar E)(E_i-E_j)   \right] }{  \pi(E_i-E_j) }  \simeq   K(E_j,E_i)
 \,,    
\end{equation}
where $\bar{\rho}(E)$ denotes the smooth ensemble-averaged density of
states. Equivalently, introducing the unfolded separation
\begin{equation}
    s=\bar{\rho}(\bar E)(E_i-E_j)\,, 
\end{equation}
the normalized kernel satisfies \footnote{For definiteness, we work with the GUE. However, the sine-kernel description is not specific to the Gaussian potential. For a broad class of unitary invariant ensembles, the finite-$N$ Christoffel--Darboux kernel depends on the matrix potential, whereas its unfolded local bulk limit is universally given by the sine kernel. But note that the sine kernel does not apply near spectral edges, where different universal kernels arise.},
\begin{equation}
    \frac{K(E_i,E_j)}
    {\bar{\rho}(\bar E)}
    \longrightarrow
    \frac{\sin(\pi s)}{\pi s} \,. 
\end{equation}
For a generic chaotic quantum system, this random-matrix description
is expected to apply within the universal spectral window
\begin{equation}
    |E_i-E_j|    \lesssim  E_{\rm Th} \,, 
\end{equation}
where $E_{\rm Th}$ is the Thouless energy. Equivalently, the
corresponding random-matrix behavior in the time domain emerges for
times larger than the Thouless time,
\begin{equation}
    t\gtrsim t_{\rm Th}, \qquad  E_{\rm Th}\sim\frac{\hbar}{t_{\rm Th}} \,.
\end{equation}
In what follows, we restrict our analysis to this time regime and do not attempt to describe the very early-time dynamics, in particular the transient behavior before complexity enters its linear-growth regime.

For later use, we now collect the codimension-one results that will serve as the basic building blocks for the codimension-zero analysis. The generating function $\langle \widehat{e^{-\alpha\CC}}\rangle$ is naturally organized according to the decomposition of the two-point correlation function of the spectral density in eq.~\eqref{eq:DD}. We write
\begin{equation} \label{eq:G1decomposition}
   \langle e^{-\alpha\CC}\rangle 
    \equiv G^{(1)}_{\rm classical}+G^{(1)}_{\rm quantum} =G^{(1)}_{\rm classical}
    +G^{(1)}_{\rm delta}
    +G^{(1)}_{\rm sine} \,.
\end{equation}
The three terms arise, respectively, from the disconnected contribution, the contact term, and the sine-kernel contribution to the spectral correlator in eq.~\eqref{eq:DD}. The latter two encode correlations associated with the discreteness and finiteness of the spectrum and will therefore be grouped as quantum corrections: 
\begin{equation}\label{eq:Gquantum}
G_{\rm quantum}^{(1)} (\alpha, t) := G_{\rm delta}^{(1)}(\alpha) +G_{\rm sine}^{(1)}(\alpha, t) \,, 
\end{equation}
At fixed $\alpha$, these contributions vanish in the classical limit $e^{S_0}\to\infty$. Their role is nevertheless essential at times of order the Heisenberg time, when the discreteness and level correlations of the spectrum can no longer be neglected.

\begin{figure}[t]
\centering\includegraphics[width=5.5in]{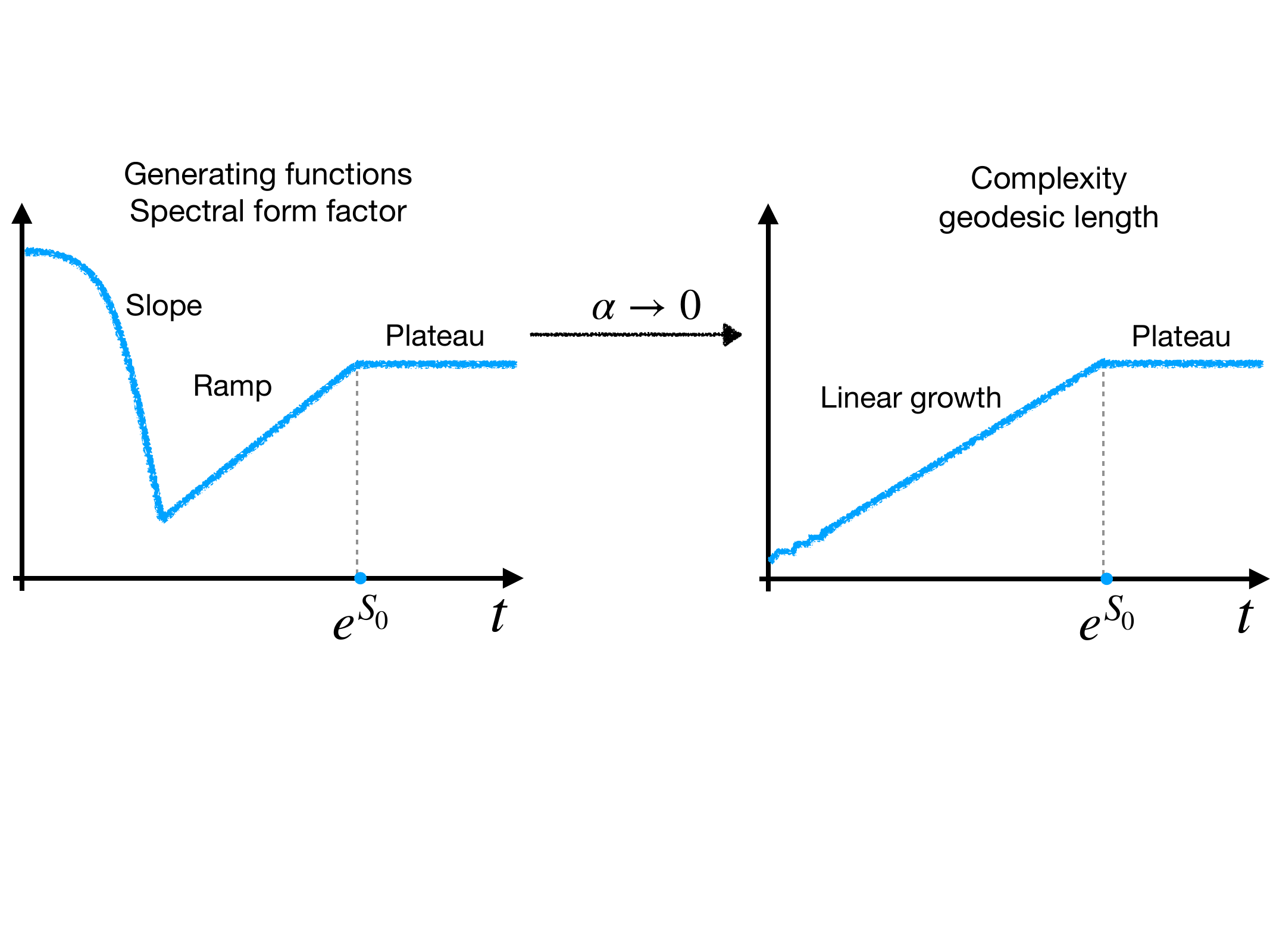}
	  \caption{	Left: Schematic time evolution of the generating function
		$\langle\widehat{e^{-\alpha\CC}}\rangle$, displaying the characteristic slope-ramp-plateau structure.
		Right: The corresponding expectation value of the complexity measure is extracted from the generating function in the limit $\alpha\to 0$.}\label{fig:GeneratingFunctions}
\end{figure}

We begin with the classical contribution that arises from the disconnected part in the spectral correlation function \eqref{eq:DD}, namely 
\begin{equation}
G_{\rm classical}^{(1)}(\alpha, t) \equiv  \frac{e^{2S_0}}{Z} \int_{E_0-\frac{\Delta E}{2}}^{E_0+\frac{\Delta E}{2}} dE_idE_j \, e^{-iE_{ij}t}  \bar{D}(E_i) \bar{D}(E_j) M_\alpha (E_i, E_j)  \,.
\end{equation}
Introducing the new variables $\bar{E}, E_{ij}$, we can find that the smooth density of states $\bar{D}(\bar{E})$ may be approximated by its value at $E_0$ due to the narrow microcanonical window. The remaining integral over $E_{ij}$ can be evaluated by contour integration. The poles of the matrix element in eq.~\eqref{eq:matrixelement} then give
\begin{equation} \label{eq:Gcl}
    G_{\rm classical}^{(1)}(\alpha, t)\approx e^{-\talpha t}\,,
\end{equation}
Here and below, the symbol $\approx$ indicates that we retain only the leading contribution and neglect corrections of order $\mathcal{O}(\Delta E/E_0)$ and $\mathcal{O}(1/(t\Delta E))$. Equation~\eqref{eq:Gcl} describes the exponentially decaying slope of the generating function as shown in figure~\ref{fig:GeneratingFunctions}. In terms of the complexity itself, this exponential dependence is the generating-function counterpart of the classical linear growth.

To incorporate quantum corrections, we consider the ensemble-averaged spectral two-point correlation function in GUE, which is explicitly given by
\begin{equation}\label{eq:sinekernel}
\begin{split}
 \overline{\langle  D(E_i) D(E_j)  \rangle }&=    \bar{D}(E_i) \bar{D}(E_j) + \langle D(E_i) D(E_j) \rangle_c \\
 &\approx    \bar{D}(E_i) \bar{D}(E_j)     +
    e^{-S_0}\delta(E_{ij}) \bar{D}(E_i)-\left(\frac{\sin (\pi e^{S_0} \bar{D}(\bar{E})E_{ij})}{e^{S_0}\pi E_{ij} }  \right)^2\,. 
    \end{split}
\end{equation} 
We next consider the contact term. Since it is supported at $E_i=E_j$, it probes only diagonal matrix elements and is therefore independent of time. Substituting the contact contribution in eq.~\eqref{eq:DD}, we obtain
\begin{equation}\label{eq:Gdelta}
\begin{split}
G_{\rm delta}^{(1)}(\alpha)\equiv \text{Constant} &\approx \frac{2}{T_{\mt{H}}}\int  \, d E_{ij}  \left( \frac{\tilde{\alpha}}{(\tilde{\alpha} + i E_{ij} )(\tilde{\alpha} - i E_{ij})}   \right) e^{-i E_{ij}t} \times  \delta(E_i - E_j)\\
&= \frac{2}{\tilde{\alpha} T_{\mt{H}}} \,.
\end{split}
\end{equation}
This time-independent contribution determines the late-time plateau of the generating function, as illustrated in figure~\ref{fig:GeneratingFunctions}.

The remaining contribution originates from the off-diagonal sine-kernel term. Its time dependence is controlled by the Fourier transform
\begin{equation}
 \int^\infty_{-\infty} dE_{ij}   \frac{\sin^2 \left( \frac{1}{2} T_{\mt{H}}E_{ij}\right)}{\left( \frac{1}{2} T_{\mt{H}}E_{ij}\right)^2} e^{- i E_{ij} t} \,. 
\end{equation}
Combining this transform with the pole structure of the matrix element $\mathbf{M}_{\talpha}(E_i,E_j)$ and evaluating the resulting contour integral gives
\begin{equation}\label{eq:Gsine}
\begin{split}
G_{\rm sine}^{(1)}(\alpha, t) 
\approx \begin{cases}
    -\dfrac{2 \talpha (\THH - t)-2 e^{-\talpha t}+e^{\talpha  (t-\THH)}+e^{-\talpha  (t+\THH)}}{\talpha ^2  T_{\mt{H}}^2} \,, \quad \,\, t< T_{\mt{H}}   \\[1.2em]
    - \dfrac{4\sinh^2 \left( \frac{1}{2} T_{\mt{H}} \talpha \right) }{\talpha^2 T_{\mt{H}}^2 } e^{- \talpha t} \,,\qquad \qquad  \qquad \quad \qquad t>T_{\mt{H}} \,, \\
    \end{cases} \,. 
\end{split}     
\end{equation}
The piecewise form arises because of the transition of the integral contour at $t=T_{\mt{H}}$. Before the Heisenberg time $t<T_{\mt{H}}$, the sine-kernel contribution contains a term linear in time and thereby generates the so-called ramp regime. Notably, this linear behavior is governed by the same Fourier-space mechanism that underlies the ramp of the spectral form factor: the second-order pole \footnote{After the decomposition $2 \sin^2(\frac{1}{2} \THH E_{ij}) = 1- \cos(\THH E_{ij})$, it origins from the constant part of the sine kernel \ie} at $E_{ij}=0$ gives rise to a term proportional to $|t|$,
\begin{equation}
    -\frac{1}{2\pi} \int_{-\infty}^{+\infty} e^{- i E t }\left( \frac{1}{(E+i\epsilon)^2} +\frac{1}{(E - i \epsilon)^2} \right) \, dE \overset{\epsilon \to 0}{=} |t| \,.
\end{equation}
For $t>T_{\mt{H}}$, the linear contribution is removed by the change in the contour prescription, and $G_{\rm sine}^{(1)}(\alpha,t)$ becomes exponentially suppressed. The time-independent contact term therefore dominates the late-time behavior and determines the plateau.

Finally, the complete time dependence follows from combining the disconnected, contact, and sine-kernel contributions. In the corresponding parametric regimes, one finds that the time evolution of generating function $\langle\widehat{e^{-\alpha \CC}} \rangle$ is described by 
\begin{equation}\label{eq:sloperampplateau}
  \langle\widehat{e^{-\alpha \CC}} \rangle \approx 
\begin{cases}
e^{-\tilde{\alpha} t} + \mathcal{O}(\frac{\tilde\alpha }{t \Delta E})\,, \quad \text{slope} \quad \tilde{\alpha} t \sim 1  \,,\\[1em]
\dfrac{2t}{\tilde{\alpha} \, T_{\mt{H}}^2} + \mathcal{O} (e^{-\tilde\alpha t}) \,,\quad \text{ramp} \quad  t <   T_{\mt{H}}  \,,\\[1em]
 \dfrac{2}{\tilde{\alpha} \, T_{\mt{H}}} + \mathcal{O}(e^{-\tilde\alpha (t-T_{\mt{H}})})\,,\quad \text{plateau} \quad  t >   T_{\mt{H}}  \,. \\
\end{cases}
\end{equation}
The generating function therefore exhibits the characteristic slope-ramp-plateau structure similar to the well-known spectral form factor, reflecting the chaotic statistics of the underlying spectrum. An analogous behavior has been obtained for geodesic lengths and time shifts in JT gravity~\cite{Miyaji:2025yvm}. As the control parameter $\alpha$ is decreased, the ramp regime becomes progressively shorter and eventually disappears in the limit $\alpha\to0$, as illustrated in figure~\ref{fig:GeneratingFunctions}.

The expectation value of the holographic complexity is obtained by differentiating the regularized generating function according to eq.~\eqref{eq:definederivative}. Taking the limit $\alpha\to0$ then yields
\begin{equation}\label{eq:Vgenvalues}
  \langle\hat\CC\rangle  \approx \text{const} + 
\begin{cases}
      M\,t\left(1-\frac{t}{T_{\mt{H}}}+\frac{t^2}{3 T_{\mt{H}}^2}\right) \,, \quad (t < T_{\mt{H}})  \\[1em]
     \frac{1}{3} MT_{\mt{H}} \,, \quad (t \ge T_{\mt{H}})
 \end{cases} \,. 
\end{equation}
At times parametrically smaller than $T_{\mt{H}}$, quantum corrections are suppressed and the leading behavior reduces to the classical result $\left\langle\mC \right\rangle_{\rm cl}
\simeq  M t $ \eqref{eq:linear}. As the evolution approaches the Heisenberg time, spectral correlations gradually suppress the growth and eventually cancel the classical linear growth. The complexity then saturates at a plateau of order $T_{\mt{H}}$. Since $T_{\mt{H}}\sim e^{S_0}$, the linear-growth regime persists for a time exponentially long in the entropy, in agreement with the expected behavior of complexity in a chaotic finite-entropy quantum system.

Motivated by the codimension-one analysis, we adopt the pole structure in eq.~\eqref{eq:matrixelement} for each of the two matrix elements appearing in the codimension-zero generating function \eqref{eq:exppm}.  In the following sections, we show how the three-point spectral correlator organizes the corresponding disconnected and wormhole contributions and leads to the universal time evolution of codimension-zero holographic complexity.

\subsection{Codimension-zero holographic complexity and spectral representation}
\label{sec:codimensionzero}

We now extend the previous analysis about codimension-one holographic complexity measures to the broader class of codimension-zero gravitational observables $\mC_{\rm Any}$ introduced in the complexity=anything proposal. The central objective of this subsection is to formulate the corresponding generating function in spectral decomposition and to identify the ingredients that control its universal time dependence. As in the codimension-one case, the semiclassical growth is encoded in the pole structure of the relevant matrix elements, whereas the quantum corrections are governed by universal  correlation function of the spectral density in the chaotic energy spectrum. The essential new feature is that codimension-zero observables involve the three-point spectral correlator rather than the two-point correlator.

As illustrated in figure \ref{fig:CAnything}, let us consider a codimension-zero bulk region $\mathcal{M}$ bounded in the future and past by two codimension-one hypersurfaces, denoted by $\Sigma_+$ and $\Sigma_-$, respectively. In the construction of refs.~\cite{Belin:2021bga,Belin:2022xmt}, Stokes' theorem allows a broad class of diffeomorphism-invariant gravitational observables defined on $\mathcal{M}$ to be expressed in terms of contributions localized on its two boundary hypersurfaces $\Sigma_\pm$. The corresponding codimension-zero holographic complexity therefore admits the decomposition
\begin{equation}
    \CC_{\rm Any} (\mathcal{M}) = \CC_+(\Sigma_+) + \CC_-(\Sigma_-)\,.
    \label{eq:Cdecomposition}
\end{equation}
Here, $\mC_+$ and $\mC_-$ are codimension-one gravitational observables associated with the future and past boundaries of $\mathcal{M}$. At the semiclassical level, each term exhibits the same characteristic linear growth as the codimension-one complexities discussed in the previous subsection. The decomposition \eqref{eq:Cdecomposition} naturally motivates the introduction of the generating function for codimension-zero holographic complexity,
\begin{equation}\label{eq:G0definition}
    G^{(0)}(\alpha_\pm;t)  :=
   \langle e^{-\alpha_+ \CC_+ -\alpha_- \CC_-} \rangle
    =
    \langle \mathrm{TFD}(t)| \widehat{e^{-\alpha_+\CC_+}} \cdot \widehat{e^{-\alpha_-\CC_-}}  |\mathrm{TFD}(t)\rangle\,.
\end{equation}
The two control parameters $\alpha_+$ and $\alpha_-$ effectively act as regulators which suppress the contributions from $\mC_\pm \to \infty$. This separate dependence will be useful both for isolating the different geometric contributions and for extracting the expectation value of $\CC_{\rm Any}$ in the limit $\alpha_\pm\to0$.

To obtain a spectral representation, we expand the TFD using eq.~\eqref{eq:TFD} and insert a complete set of energy eigenstates between the two operators in eq.~\eqref{eq:G0definition}. This gives
\begin{equation}\label{eq:exppm}
   G^{(0)}(\alpha_\pm;t)= \frac{e^{3S_0}}{Z}\int_{\mathcal{I}}dE_1dE_2 dE_3 \, e^{-iE_{12}t} \overline{\langle  D(E_1)D(E_2)D(E_3) \rangle} 
 \mathbf{M}_{\talpha_+} (E_1,E_3) \mathbf{M}_{\talpha_-} (E_3,E_2) \,. 
\end{equation}
where
\begin{equation}
    \mathbf{M}_{\talpha_\pm}(E_i,E_j)  :=  \langle E_i| \widehat{e^{-\alpha_\pm\CC_\pm}} |E_j\rangle
\end{equation}
and $\mathcal{I}$ denotes the fixed microcanonical energy window \eqref{eq:MCwindow}. 

The appearance of three energy integrations has a simple origin. The energies $E_1$ and $E_2$ are associated with the bra and ket components of the thermofield double state, while $E_3$ labels the intermediate state inserted between the two codimension-one operators. Consequently, the codimension-zero generating function probes the three-point correlation function of the spectral density, namely 
\begin{equation}
    \overline{\langle D(E_1)D(E_2)D(E_3) \rangle}\,,
\end{equation}
rather than the two-point correlator that governs the codimension-one holographic complexity measures. This is the principal structural distinction between the two cases. The evaluation of eq.~\eqref{eq:exppm} therefore requires two inputs. The first is the three-point spectral correlator, whose disconnected and connected components encode the classical and nonperturbative quantum contributions.

The second input consists of the two matrix elements $\mathbf{M}_{\talpha_\pm}(E_i,E_j)$.
The decomposition \eqref{eq:Cdecomposition} identifies $\mC_+$ and $\mC_-$ as two codimension-one gravitational observables associated with codimension-one hypersurfaces $\Sigma_\pm$, respectively. From the construction of complexity=anything proposal,  each of them exhibits linear growth at the semiclassical level, namely 
\begin{equation}
	\mC_\pm^{\rm cl}(\Sigma_\pm)= \mathrm{const} + M_\pm t +\cdots \,. 
\end{equation}
As a result, the codimension-one analysis of the previous subsection applies to the two operators $\widehat{e^{-\alpha_\pm \mC_\pm}}$ separately. In particular, the linear growth of $\CC_\pm$ requires the corresponding matrix element of $\widehat{e^{-\alpha_\pm\CC_\pm}}$ to contain the same characteristic pair of simple poles,
\begin{equation}
	\mathbf{M}_{\talpha_\pm}(E_i,E_j) \simeq
	\frac{\tilde{\alpha}_\pm}{\pi e^{S_0}\bar D(\bar E_{ij})} \frac{1}{(\tilde{\alpha}_\pm+iE_{ij})(\tilde{\alpha}_\pm-iE_{ij})},
	\qquad
	\tilde{\alpha}_\pm 	=
	M_\pm\alpha_\pm +	\mathcal{O}(\alpha_\pm^2) \,. 
	\label{eq:matrixelement_pm}
\end{equation}
Thus, the two matrix elements have the same analytic structure, but their pole locations are controlled by the generally different growth rates $M_+$ and $M_-$.

This identification also makes the origin of the classical codimension-zero
growth transparent. When the fully disconnected component of the three-point
spectral correlator is substituted into eq.~\eqref{eq:exppm}, the integrations
associated with the two matrix elements factorize. The two pairs of poles then
give
\begin{equation}
	G_{\rm classical}^{(0)}(\alpha_\pm;t)\simeq		e^{-(\widetilde{\alpha}_+ +\widetilde{\alpha}_-)t} \,. 
\end{equation}
Taking derivatives with respect to $\alpha_+$ and $\alpha_-$ therefore reproduces
\begin{equation}
	\mC_{\rm Any}^{\rm cl}(t)= \mathrm{const}+ (M_+  +M_-)t +\cdots,
\end{equation}
in agreement with the geometric decomposition \eqref{eq:Cdecomposition} for codimension-zero holographic complexity measures in the complexity=anything proposal.

The universal time evolution of the codimension-zero holographic complexity measures is therefore governed by the same two ingredients as in the codimension-one case. The pole structures of the two matrix elements in the corresponding generating function reproduce the independent semiclassical linear growth of $\CC_+$ and $\CC_-$, while the connected components of the three-point spectral correlator organize the nonperturbative quantum corrections that become important near the Heisenberg time and eventually offset the classical linear growth resulting in a late-time plateau.

\subsubsection{Spectral three-point correlation functions and spacetime wormholes} \label{sec:randmat}
We now review the spectral correlation functions that enter the codimension-zero generating function. In a quantum chaotic system, spectral correlations at sufficiently small energy separations are governed by random matrix theory, with the relevant ensemble determined by the symmetries of the Hamiltonian. In the time domain, this universal regime is reached after the Thouless time. Since our main purpose is to evaluate the time evolution explicitly by using the three-point spectral correlator, we specialize below to the Gaussian unitary ensemble (GUE). See \eg refs.~\cite{mehta2004random,Haake:2010fgh,Eynard:2015aea,livan2018introduction,Potters_Bouchaud_2020} for more detailed reviews of random matrix theory.
\footnote{More generally, the determinantal construction below applies to unitary invariant ensembles with Dyson index $\beta=2$. For the orthogonal and symplectic classes, with $\beta=1$ and $\beta=4$, respectively, the spectral correlators instead admit a Pfaffian representation governed by a matrix-valued kernel~\cite{Tracy_1998}. We discuss the similar results and differences among all three Dyson classes in Appendix \ref{app:Dyson}.}

Consider an $N\times N$ Hermitian random matrix with eigenvalue measure
\begin{equation}
    \mathcal{Z}_N
    =
    \int_{\mathbb{R}^N}
    \prod_{a=1}^{N}\left[dE_a\,e^{-V(E_a)}\right]
    \Delta(E_1,\ldots,E_N)^\beta,
    \label{eq:RMTpartition}
\end{equation}
where
\begin{equation}
    \Delta(E_1,\ldots,E_N):=\prod_{a<b}(E_a-E_b)
\end{equation}
is the Vandermonde determinant and $\beta$ is referred to as the Dyson index with $\beta=2$ for GUE. The n-point eigenvalue correlation function (factorial correlation function) is obtained by integrating over the remaining $N-n$ eigenvalues:
\begin{equation}
    R_n(E_1,\ldots,E_n) :=
   \frac{1}{\mathcal{Z}_N} \frac{N!}{(N-n)!}  \prod_{a=1}^{n}e^{-V(E_a)}
    \int \prod_{a=n+1}^{N} \left[dE_a\,e^{-V(E_a)}\right] \Delta(E_1,\ldots,E_N)^2.
    \label{eq:matrixR}
\end{equation}
With this convention, $R_n$ is normalized according to
\begin{equation}
    \int dE_1\cdots dE_n\,R_n(E_1,\ldots,E_n)   =    \frac{N!}{(N-n)!},
\end{equation}
and is therefore an $n$-level correlation function rather than a probability density normalized to unity.\footnote{A normalized joint probability density for $n$ eigenvalues may instead be defined by
\begin{equation}
    p_n(E_1,\ldots,E_n)  :=  \frac{(N-n)!}{N!}  R_n(E_1,\ldots,E_n),
\end{equation}
which satisfies
\begin{equation}
    \int dE_1\cdots dE_n\,p_n(E_1,\ldots,E_n) =  1 \,. 
\end{equation}}

For the GUE, the eigenvalues form a determinantal point process, and the complete hierarchy of correlation functions can be expressed in terms of the Christoffel--Darboux kernel:
\begin{equation}
  \boxed{  R_n(E_1,\ldots,E_n)  =  \det_{1\leq a,b\leq n}\!\left[K(E_a,E_b)\right] \,.} 
    \label{eq:RnDeterminant}
\end{equation}
The determinantal representation makes the permutation symmetry of the
eigenvalue correlation functions immediate. For any permutation
$\sigma\in S_n$, relabeling the arguments as $E_a\mapsto E_{\sigma(a)}$ induces the same permutation of the rows and columns of the kernel matrix. Since the two permutation factors cancel in the determinant, one finds
\begin{equation}
    R_n(E_{\sigma(1)},\ldots,E_{\sigma(n)}) = R_n(E_1,\ldots,E_n),
    \qquad
    \sigma\in S_n.
\label{eq:Rn-permutation-symmetry}
\end{equation}
Hence, $R_n$ is completely symmetric under permutations of its $n$ arguments. Furthermore, the kernel can be explicitly constructed from the orthonormal functions. Let $p_m(E)$ denote the orthogonal polynomials associated with the
weight $e^{-V(E)}$, normalized according to
\begin{equation}
    \int dE\,e^{-V(E)}p_m(E)p_n(E)   =  h_m\delta_{mn} \,. 
\end{equation}
Defining the corresponding weighted orthonormal functions by
\begin{equation}
    \psi_m(E)  := \frac{e^{-V(E)/2}}{\sqrt{h_m}}p_m(E) \,, 
\end{equation}
the Christoffel--Darboux kernel can be written as
\begin{equation} \label{eq:CDkernel}
   K(E_i,E_j)= \sum_{m=0}^{N-1}  \psi_m(E_i)\psi_m(E_j) =e^{-\frac{1}{2}[V(E_i)+V(E_j)]} \sum_{m=0}^{N-1}  \frac{p_m(E_i)p_m(E_j)}{h_m} \,.
\end{equation}
For a real potential on the real axis, the functions $\psi_m$ can be chosen to be real, resulting in $ K(E_i,E_j)=K(E_j,E_i)$. 

Importantly, the diagonal kernel is nothing but the ensemble-averaged density of states, \ie 
\begin{equation} \label{eq:R1}
\begin{split}
 R_1(E_i) &=K(E_i,E_i) =\bar\rho(E_i) = e^{S_0}\bar D(E_i) \,, 
\end{split}
\end{equation}
which is normalized to the total number of eigenvalues $ \int dE\,\bar\rho(E)=N$.\footnote{When the integral is restricted to the microcanonical window $\mathcal I$, it instead gives the expected number of levels in that window, $\int_{\mathcal I}dE\,\bar\rho(E)= Z$.} Furthermore, the 2-point and 3-point eigenvalue correlation functions are explicitly derived as 
\begin{equation} \label{eq:R2explicit}
    R_2(E_i,E_j)  =  K_{ii}K_{jj}-K_{ij}K_{ji}  =  \rho_i \rho_j-(K_{ij})^2 \,,
\end{equation}
and
\begin{equation}\label{eq:R3explicit}
\begin{split}
R_3\left(E_i, E_j, E_k\right) &=\operatorname{det}\left(\begin{array}{lll}
K\left(E_i, E_i\right) & K\left(E_i, E_j\right) & K\left(E_i, E_k\right) \\
K\left(E_j, E_i\right) & K\left(E_j, E_j\right) & K\left(E_j, E_k\right) \\
K\left(E_k, E_i\right) & K\left(E_k, E_j\right) & K\left(E_k, E_k\right)
\end{array}\right) \\
&= \rho_i \rho_j \rho_k + 2 K_{ij}K_{jk}K_{ki} - \rho_i (K_{jk})^2 -\rho_j (K_{ik})^2  -\rho_k (K_{ij})^2 \,,  
\end{split}
\end{equation}
respectively. Here, we have introduced the following abbreviations 
\begin{equation}
K_{ij}= K(E_i, E_j) \,, \qquad \rho_i =K_{ii} \,, 
\end{equation} 
to simplify the above expressions. 

An immediate consequence of the determinantal representation is that
$R_n$ vanishes whenever two of its arguments coincide:
\begin{equation}\label{eq:levelrepulsion}
  \boxed{  R_n(E_1,\ldots,E_a,\ldots,E_a,\ldots,E_n)=0\,.}
\end{equation}
Indeed, when two energy arguments coincide, the determinant in eq.~\eqref{eq:RnDeterminant} contains two identical rows and columns. This vanishing is the local manifestation of level repulsion: nearby energy levels are statistically suppressed from approaching one another. More quantitatively, when a single pair of energies approaches coincidence while the remaining arguments are held fixed and mutually distinct, the $n$-level correlation function behaves as
\begin{equation}
    R_n^{(\beta)}  (E_1,\ldots,E_a,\ldots,E_b,\ldots,E_n)
    \propto   |E_a-E_b|^\beta,  \qquad  E_a\to E_b \,, 
    \label{eq:Rn_level_repulsion}
\end{equation}
The proportionality coefficient depends on the local density of states and on the remaining energy arguments and the exponent $\beta$ is referred to as the Dyson index which originates directly from the pairwise Vandermonde factor in the $\beta$-ensemble measure. In particular, the two-level correlation function satisfies
\begin{equation}
    R_2^{(\beta)}(s) \propto |s|^\beta \,,  \qquad  s\to0 \,, 
    \label{eq:R2_level_repulsion}
\end{equation}
where $s$ denotes the energy separation in units of the local mean level
spacing.\footnote{The nearest-neighbor spacing distribution obeys the same short-distance scaling. After unfolding the spectrum and restricting to a fixed symmetry sector, the nearest-neighbor spacing distribution behaves as
\begin{equation}
    P_\beta(s)\propto s^\beta \,,   \qquad    s\to 0 \,, 
\end{equation}
so that the probability of finding two nearly degenerate levels vanishes at small separation. However, it should not be confused with $R_2^{(\beta)}(s)$: $P_\beta(s)$ concerns consecutive levels with no intervening eigenvalue, whereas $R_2^{(\beta)}(s)$ counts arbitrary pairs at separation $s$.} This behavior contrasts with an uncorrelated Poisson spectrum, for which the unfolded two-level correlation function is constant,
\begin{equation}
    R_2^{\rm Poisson}(s)=1\,,   \qquad  s\neq0 \,, 
\end{equation}
and hence exhibits no suppression as two distinct levels approach one another. Level repulsion is therefore a characteristic diagnostic of quantum chaos once all exact symmetries have been resolved, although it is not by itself sufficient to establish fully developed chaotic dynamics.

The functions $R_n$ describe correlations among distinct eigenvalues. By contrast, correlation functions of the \textit{exact density of states},
\begin{equation}
    \rho(E):=\sum_{a=1}^{N}\delta(E-E_a),
\end{equation}
also contain contact terms arising when two or more eigenvalue labels coincide. Correspondingly, the complete two-point spectral correlator is thus defined as 
\begin{equation}
    \overline{ \langle \rho(E_i)\rho(E_j) \rangle }  =  R_2(E_i,E_j)+\delta(E_{ij})\bar\rho(E_i) \,, \qquad \text{with} \quad  \overline{ \langle \rho(E) \rangle } \equiv \bar\rho(E) \,.
    \label{eq:rho2full}
\end{equation}
Motivated by gravitational interpretations, we can equivalently decompose it into disconnected and connected parts:
\begin{equation}
    \overline{ \langle \rho(E_i)\rho(E_j) \rangle } 
    =
    \bar\rho(E_i)\bar\rho(E_j)
    +
    \langle \rho(E_i)\rho(E_j) \rangle_{\rm c},
    \label{eq:rho1rho2}
\end{equation}
with
\begin{equation}
     \langle \rho(E_i)\rho(E_j) \rangle_{\rm c} := \bar\rho(E_i)\delta(E_{ij}) -  K(E_i,E_j)K(E_j,E_i) \,. 
    \label{eq:rho2connected}
\end{equation}
Here, the subscript $c$ denotes connected part corresponding to contributions from spacetime wormholes in Euclidean gravitational path integral. The full three-point spectral correlator is obtained by separating the contributions according to the possible coincidences among the three eigenvalue labels:
\begin{equation}
\begin{aligned}
    \overline{ \langle \rho(E_1)\rho(E_2) \rho(E_3) \rangle } 
    ={}&
    R_3(E_1,E_2,E_3)
    +\delta(E_{12})R_2(E_3,E_1)
    +\delta(E_{23})R_2(E_1,E_2)
    \\
    &+\delta(E_{31})R_2(E_2,E_3)
    +\delta(E_{12})\delta(E_{13})R_1(E_1).
\end{aligned}
\label{eq:rho3}
\end{equation}
The first term describes three distinct eigenvalues. The next three terms arise when precisely two labels coincide, while the final term corresponds to all three labels referring to the same eigenvalue.

\begin{figure}[t]
    \centering
    \includegraphics[width=5.5in]{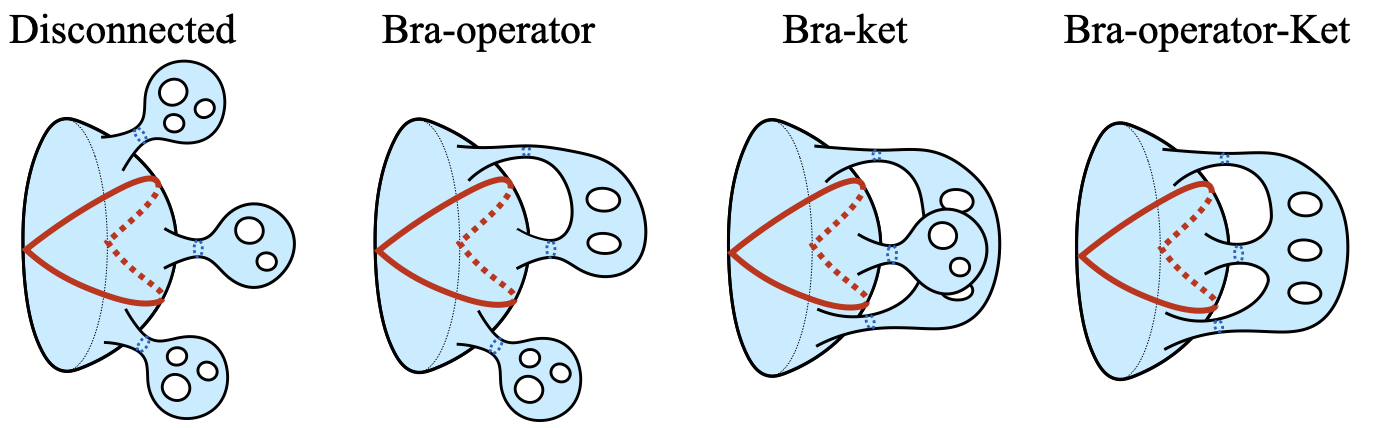}
    \caption{Schematic illustration of the Euclidean wormhole sectors suggested by the decomposition of the three-point spectral correlator in eq.~\eqref{eq:DDD}. The diagrams represent the different connectivity patterns among the bra, operator insertion, and ket sectors. Schematically, the operator-ket sector can be obtained by inverting the bra-operator sector. }
    \label{fig:wormholes}
\end{figure}

In order to make the dependence of the order of $e^{S_0}$ more explicit, we expand the codimension-zero generating function in terms of the rescaled density
\begin{equation}
    D(E):=e^{-S_0}\rho(E) \,.
\end{equation}
It is thus more convenient to express the three-point correlator in cumulant form:
\begin{equation}
\begin{aligned}
    \overline{\langle  D_1D_2D_3 \rangle }
    ={}&
    \bar D_1\bar D_2\bar D_3
    +{\langle D_1D_2D_3 \rangle }_{\rm c} + \bar D_1\,{\langle D_2D_3 \rangle}_{\rm c}+\bar D_2\, {\langle  D_3D_1 \rangle}_{\rm c} +\bar D_3\,{\langle D_1D_2 \rangle}_{\rm c} \,,
\end{aligned}
\label{eq:DDD}
\end{equation}
where $D_a:=D(E_a)$ and $\bar D_a:=\bar D(E_a)$. The connected two-point function is
\begin{equation}
    {\langle D_iD_j \rangle}_{\rm c} =
    e^{-S_0}\bar D_i \delta(E_{ij}) - e^{-2S_0}K_{ij}K_{ji} \,.
    \label{eq:DDconnected}
\end{equation}
For the GUE, the fully connected three-point function takes the compact form
\begin{equation}
\begin{aligned}
   \langle D_1D_2D_3 \rangle_{\rm c}  ={}&
    e^{-2S_0}\bar D(E_1)\delta(E_{12})\delta(E_{13})
    +2e^{-3S_0}K_{12}K_{23}K_{31}
    \\
    &-e^{-3S_0}\!\left[
        \delta(E_{23})K_{12}^2
        +\delta(E_{31})K_{23}^2
        +\delta(E_{12})K_{31}^2
    \right].
\end{aligned}
\label{eq:DDDconnected}
\end{equation}
The decomposition \eqref{eq:DDD} has a natural interpretation in a gravitational description in the Euclidean path integral. The fully disconnected contribution $\bar D_1\bar D_2\bar D_3$ corresponds to the leading contributions from the classical geometry. The three terms containing connected two-point correlators describe pairwise connections among the bra, operator, and ket sectors. In the energy ordering used in eq.~\eqref{eq:exppm},
\begin{equation}
\begin{aligned}
    \bar D_1\,\langle D_2D_3 \rangle_{\rm c}
    &:\,\, \text{operator--ket},
    &
    \bar D_2\,\langle D_3D_1 \rangle_{\rm c}
    &:\,\, \text{bra--operator},
    &
    \bar D_3\,\langle D_1D_2 \rangle_{\rm c}
    &:\,\, \text{bra--ket}.
\end{aligned}
\end{equation}
These terms are naturally associated with Euclidean wormholes connecting the corresponding pairs of sectors. The fully connected cumulant ${\langle D_1D_2D_3 \rangle}_{\rm c}$ is similarly associated with a geometry connecting all three sectors, which we refer to as the bra-operator-ket contribution. These connectivity patterns are illustrated in figure~\ref{fig:wormholes}. Note that only the classical contribution, \ie disconnected geometries in figure~\ref{fig:wormholes} survives in the limit $S_0\to\infty$.

Although the finite-$N$ CD kernel depends on the
matrix potential, its unfolded microscopic limit is universal at
large $N$ away from the spectral edges. Accordingly, to evaluate the
three-point spectral correlator explicitly, we adopt the same GUE
sine-kernel approximation introduced in section~\ref{sec:codimensionone}. Each term in eqs.~\eqref{eq:DDconnected} and \eqref{eq:DDDconnected} is constructed from pairwise CD kernels, so the sine-kernel approximation \eqref{eq:sinekernelK} applies separately to each pair of energies:
\begin{equation}
    K(E_i,E_j)  \simeq     \frac{\sin\!\left[\pi\bar\rho(\bar E)(E_i-E_j)\right]}{\pi(E_i-E_j)}\approx \frac{\sin\!\left(\frac{1}{2}T_{\mt{H}}E_{ij}\right)}{\pi E_{ij}} \,.
    \label{eq:sinekernel01}
\end{equation}
The second approximation with $T_{\mt{H}}=2\pi\bar\rho(E_0)$ is often used in later calculations since we focus on the narrow microcanonical window where the smooth density may be approximated by its value at $E_0$. 

To characterize the semiclassical limit, it is useful to interpret the
sine kernel as a distribution. When integrated against test functions
that vary smoothly on energy scales much larger than the mean level
spacing, one has
\begin{equation}
    \frac{\sin(\Lambda x)}{\pi x}
    \mathop{\rightharpoonup}_{\Lambda\to\infty}
    \delta(x),
    \qquad
    \frac{\sin^2(\Lambda x)}{\pi\Lambda x^2}
    \mathop{\rightharpoonup}_{\Lambda\to\infty}
    \delta(x),
    \label{eq:sineDistributionalLimits}
\end{equation}
where $\rightharpoonup$ denotes convergence in the distributional
sense. Applying these identities to the sine kernel, with $\Lambda=\pi\bar\rho(\bar E_{ij})$, gives
\begin{equation} \label{eq:kernelDistributionalLimits}
\lim_{e^{S_0} \to \infty}   K(E_i, E_j)   \rightharpoonup \delta (E_{ij}) \,, \qquad 
  \lim_{e^{S_0} \to \infty}   K(E_i, E_j)^2   \rightharpoonup \bar{\rho}(\bar{E})\delta (E_{ij}) \,. \\
\end{equation}
These distributional relations make the cancellations within the
connected correlators manifest. In the two-point function, the contact
term is canceled at leading order by the squared sine kernel:
\begin{equation}
    {\langle \rho(E_i)\rho(E_j) \rangle }_{\rm c}
    =
    \bar\rho(E_i)\delta(E_{ij}) -  K(E_i,E_j)^2 \mathop{\rightharpoonup}_{e^{S_0}\to\infty}  0 \,.
    \label{eq:rho2ConnectedClassical}
\end{equation}
Here we ignore $O(1)$ term in the right hand side. Similarly, in the fully connected three-point function, the double-contact, contact kernel, and triple-kernel terms cancel at leading order, so that
\begin{equation}
 \lim_{e^{S_0} \to \infty}  { \langle \rho(E_i)\rho(E_j)\rho(E_k) \rangle }_{\rm c} \rightharpoonup   0.
\label{eq:rho3ConnectedClassical}
\end{equation}
Here, the limits are understood after smearing over energy scales much larger than the local mean level spacing. Thus, the $n$-point correlation function of the spectral density factorizes in the classical limit:
\begin{equation}
 \begin{split}
     \lim_{e^{S_0} \to \infty} \overline{\left\langle\rho(E_i) \rho(E_j)\right\rangle} & \rightharpoonup  \bar{\rho}(E_i) \bar{\rho}(E_j) \,,  \\
 \lim_{e^{S_0} \to \infty} \overline{\left\langle\rho(E_i) \rho(E_j) \rho(E_k)\right\rangle} &\rightharpoonup \bar{\rho}(E_i) \bar{\rho}(E_j) \bar{\rho}(E_k)  \,,
 \end{split}
\end{equation}
This is the precise sense in which the disconnected sector gives the leading classical contribution, while the connected sectors encode finite-entropy quantum corrections.

At finite entropy, however, the connected correlators remain nontrivial when the energy differences are of the order of the local mean level spacing,
\begin{equation}
    E_{ij}  \sim  \bar\rho(E_0)^{-1}  \sim  e^{-S_0}\,. 
\end{equation}
These microscopic correlations survive in unfolded variables and become important in Fourier transforms at times of order the Heisenberg time. They provide the nonperturbative spectral corrections that gradually cancel the semiclassical linear growth of holographic complexity and produce its late-time plateau.

\subsection{Evaluation of the codimension-zero generating function}
\label{sec:G0evaluation}

\begin{figure}[t]
    \centering
    \includegraphics[width=6in]{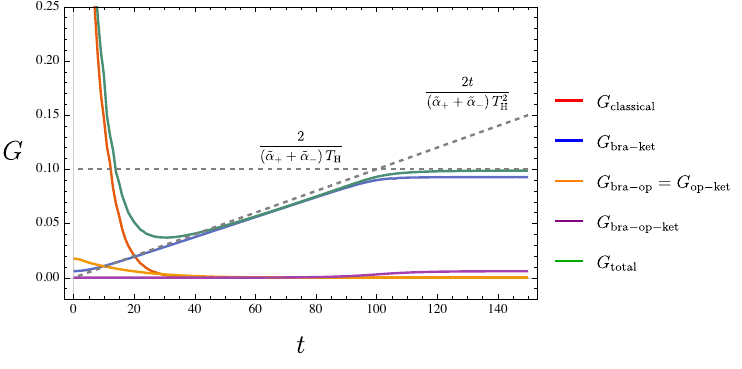}
    \caption{Contributions to the codimension-zero generating function $G^{(0)}(\alpha_\pm;t)$
    \eqref{eq:Gzerototal} for  $\Delta E=1$, $T_{\mt{H}}=100$ and  $\widetilde{\alpha}_\pm=0.1$. Green: the total generating function; red: $G_{\rm classical}$; blue: $G_{\rm bra-ket}$; orange: $G_{\rm bra-op}$ and $G_{\rm op-ket}$ contributions; purple: $G_{\rm bra-op-ket}$. For these finite control parameters $\alpha_\pm$, $G_{\rm classical}$ and $G_{\rm bra-ket}$ provide the dominant contributions to the slope-ramp-plateau profile of the codimension-zero generating function $G^{(0)}(\alpha_\pm;t)$. The dashed lines indicate the leading ramp and plateau approximations in the regime $\Delta E\gg\widetilde\alpha_\pm$.}
    \label{fig:G0s}
\end{figure}

Having identified the two ingredients entering the spectral representation \eqref{eq:exppm}, namely the pole structure of the two matrix elements and the decomposition of the three-point spectral correlator, we now evaluate the codimension-zero generating function explicitly. Substituting the cumulant decomposition \eqref{eq:DDD} into \eqref{eq:exppm} separates the answer into five contributions distinguished by their connectivity among the bra, operator, and ket sectors:
\begin{equation}
   \boxed{ G^{(0)}(\alpha_\pm;t) = G_{\rm classical}  +G_{\rm bra-op} +G_{\rm op-ket}  +G_{\rm bra-ket} +G_{\rm bra-op-ket}\,.}
    \label{eq:Gzerototal}
\end{equation}
The first term arises from the fully disconnected part of the three-point correlator, the next three from its pairwise connected components, and the final term from the fully connected three-point cumulant. The resulting energy integrals can be evaluated using the pole structure \eqref{eq:matrixelement_pm}, together with the sine-kernel approximation introduced above. Figure~\ref{fig:G0s} previews the resulting time dependence. For finite source parameters $\alpha_\pm$, the fully disconnected and bra-ket sectors dominate the leading slope-ramp-plateau profile, while the remaining connected sectors are parametrically suppressed.

We begin with the fully disconnected contribution,
\begin{equation}
    G_{\rm classical} := \frac{e^{3S_0}}{Z} \int_{\mathcal I^3}  dE_1\,dE_2\,dE_3\, e^{-iE_{12}t}\,  \bar D(E_1)\bar D(E_2)\bar D(E_3)\,
    \mathbf{M}_{\talpha_+}(E_1,E_3)  \mathbf{M}_{\talpha_-}(E_3,E_2) \,, 
    \label{eq:Gclassical}
\end{equation}
which gives the classical contribution. As anticipated from the decomposition
$\CC_{\rm Any}=\CC_++\CC_-$, it factorizes at leading order into two codimension-one generating functions. To make this factorization explicit, we can change the three integration variables $E_i$ as
\begin{equation}
    E:=E_3,
    \qquad
    E_{13}:=E_1-E_3,
    \qquad
    E_{32}:=E_3-E_2\,. 
    \label{eq:GclassicalVariablesDefinition}
\end{equation}
with the Fourier phase separating as
\begin{equation}
    e^{-iE_{12}t} =  e^{-iE_{13}t}e^{-iE_{32}t} \,. 
\end{equation}
In terms of these variables, eq.~\eqref{eq:Gclassical} reduces to
\begin{equation}
\begin{split}
    G_{\rm classical} =
    \frac{e^{3S_0}}{Z}
    \int dE\,dE_{13}\,dE_{32}\,
    e^{-i(E_{13}+E_{32})t}\,
    &\bar D(E+E_{13})\bar D(E-E_{32})\bar D(E)
    \\
    &\times \mathbf{M}_{\talpha_+}(E_1,E_3)  \mathbf{M}_{\talpha_-}(E_3,E_2) \,, 
\end{split}
\label{eq:GclassicalVariables}
\end{equation}
where the exact integration domain is determined by the requirement that $E$, $E+E_{13}$, and $E-E_{32}$ all lie in the microcanonical window $\mathcal I$. Within the narrow microcanonical window, we neglect the variation of the smooth density of states and the mean-energy dependence of the matrix elements. In particular,
\begin{equation}
    \bar D(E+E_{13})
    \approx \bar D(E-E_{32}) \approx  \bar D(E)  \approx \bar D(E_0)\,. 
\end{equation}
The matrix elements may similarly be evaluated at fixed mean energy $E_0$, while retaining their dependence on the energy differences $E_{13}$ and $E_{32}$. Moreover, in the regime $t\Delta E\gg1$, endpoint effects are suppressed and the two energy-difference integrations may be extended to the real line. Using the leading microcanonical normalization
\begin{equation}
    Z \approx e^{S_0}\bar D(E_0)\Delta E \,, 
\end{equation}
and retaining only the leading terms, one can obtain
\begin{equation}
    G_{\rm classical}\approx
    \left[  e^{S_0}\bar D(E_0)I_+(\alpha_+,t) \right]
    \times \left[e^{S_0}\bar D(E_0)I_-(\alpha_-,t) \right]\,. 
\end{equation}
where
\begin{equation}
    I_\pm(\alpha_\pm,t) :=\int_{-\infty}^{\infty} dE_{ij}\, e^{-i E_{ij}t}  \mathbf{M}_{\talpha_\pm}(E_i, E_j)\,, 
\end{equation}
Each factor is precisely the disconnected contribution to a codimension-one generating function evaluated with the corresponding control parameter $\alpha_\pm$. Hence,
\begin{equation}
    G_{\rm classical} (\alpha_\pm;t)
    \approx
    G_{\rm classical}^{(1)}(\alpha_+,t)
    G_{\rm classical}^{(1)}(\alpha_-,t)\,, 
    \label{eq:GclassicalFactorization}
\end{equation}
up to corrections of order $\mathcal{O}(\Delta E/E_0)$ and $\mathcal{O}(1/(t\Delta E))$.

Finally, substituting the pole structure \eqref{eq:matrixelement_pm} and using the codimension-one result \eqref{eq:Gcl}, we find 
\begin{equation}
  \boxed{ G_{\rm classical}(\alpha_\pm;t)
    \approx e^{-\widetilde{\alpha}_+t} e^{-\widetilde{\alpha}_-t}
    = e^{-(\widetilde{\alpha}_++\widetilde{\alpha}_-)t}.}
    \label{eq:GclassicalResult}
\end{equation}
This exponentially decaying contribution produces the initial slope of the codimension-zero generating function shown in figure~\ref{fig:G0s}. After differentiation with respect to the two source parameters, it reproduces the classical growth rate $M_++M_-$ implied by the geometric decomposition \eqref{eq:Cdecomposition}.

Among the connected contributions, the bra-ket sector is particularly important because it dominates the leading ramp and plateau of the codimension-zero generating function. We will show that this contribution reduces to the quantum part of a codimension-one generating function with an effective pole parameter
\begin{equation}
    \talpha_{\rm eff}  := \talpha_+   +  \talpha_-,  \qquad \talpha_\pm
    = M_\pm\alpha_\pm + \mathcal{O}(\alpha_\pm^2).
    \label{eq:alphaeff}
\end{equation}
The bra-ket contribution is defined by
\begin{equation} \label{eq:Gbra-ket}
\begin{split}
 G_{\rm bra-ket}&:=\frac{e^{3S_0}}{Z}\int_\mathcal{I} dE_1dE_2 dE_3 \, e^{-iE_{12}t} \bar{D}(E_3) \langle D(E_1) D(E_2)\rangle_c \mathbf{M}_{\talpha_+}(E_1,E_3)\mathbf{M}_{\talpha_-}(E_3,E_2)   \,.
\end{split}
\end{equation}
To evaluate this expression, we instead use another three integration variables
\begin{equation}
    E:=E_2,  \qquad  E_{12}:=E_1-E_2,  \qquad   E_{23}:=E_2-E_3.
\end{equation}
The connected two-point correlator depends only on $E_1$ and $E_2$, whereas the dependence on the intermediate energy $E_3$ is contained in the two matrix elements. Within the narrow microcanonical window, the averaged density of states can be approximated by $\bar D(E_3)\simeq\bar D(E_0)$. We may therefore first perform the integral over $E_{23}$. Applying the same pole structure \eqref{eq:matrixelement_pm} and using the convolution identity
\begin{equation}
    \int_{-\infty}^{\infty}
    dx\,
    \frac{a}{a^2+(x+y)^2}  \frac{b}{b^2+x^2}
    =
    \frac{(a+b)\pi}{(a+b)^2+y^2}\,,
    \label{eq:LorentzianConvolution}
\end{equation}
we obtain
\begin{equation}
    \int_{-\infty}^{\infty}  dE_{23}\, \mathbf{M}_{\talpha_+}(E_1,E_3) \mathbf{M}_{\talpha_-}(E_3,E_2) 
    \approx
    \frac{4\pi}{T_{\mt{H}}^2} \frac{\talpha_{\rm eff}} {\talpha_{\rm eff}^2+E_{12}^2}
     = \frac{2\pi}{T_{\mt{H}}}  \mathbf{M}_{\talpha_{\rm eff}}(E_1,E_2)\,. 
\label{eq:GbraketEffectiveMatrix}
\end{equation}
Thus, the convolution of the two matrix elements has precisely the same pole structure as a single codimension-one matrix element, with $\talpha$ replaced by $\talpha_{\rm eff}$. Substituting eq.~\eqref{eq:GbraketEffectiveMatrix} into
eq.~\eqref{eq:Gbra-ket}, we then arrive at
\begin{equation}
\begin{split}
    G_{\rm bra-ket}
    \approx
    \frac{e^{2S_0}}{Z} \int_{\mathcal I}  d E \int_{-\infty}^{\infty} dE_{12}\,e^{-iE_{12}t}\,
    {\langle D(E_1)D(E_2) \rangle}_{\rm c}\,  \mathbf{M}_{\talpha_{\rm eff}}(E_1,E_2).
\end{split}
\label{eq:GbraketReduced}
\end{equation}
This expression is precisely the quantum contribution to the codimension-one generating function evaluated at $\talpha=\talpha_{\rm eff}=\talpha_+ +\talpha_-$. Therefore,
\begin{equation}
    \boxed{
    G_{\rm bra-ket}(\talpha_\pm;t)
    \approx
    G_{\rm quantum}^{(1)}
    \left(\talpha_{\rm eff},t\right)
    }.
    \label{eq:GbraketResult}
\end{equation}

Combining the above result with the fully disconnected contribution, we find
\begin{equation}
\begin{split}
    G_{\rm classical} + G_{\rm bra-ket}
    &\approx
   G_{\rm classical}^{(1)}\left(\talpha_{\rm eff},t\right)+ G_{\rm quantum}^{(1)}\left(\talpha_{\rm eff},t\right) = G^{(1)}(\talpha_{\rm eff},t).
\end{split}
\label{eq:GclassicalBraketCodimOne}
\end{equation}
The sum of the classical and bra--ket sectors therefore reproduces the complete codimension-one generating function with the replacement $\talpha\to\talpha_{\rm eff}$. This reduction has a direct geometric interpretation. The bra--ket sector is described by the same type of connected Euclidean wormhole geometry that produces the quantum contribution in the codimension-one gravitational observables. In the present codimension-zero configuration, the operator region is bounded by the two codimension-one hypersurfaces $\Sigma_+$ and $\Sigma_-$, carrying the observables $\CC_+$ and $\CC_-$, respectively. Since the bra--ket wormhole connects the two external sectors across the entire operator region, it probes these two boundary contributions only through their combined insertion. Accordingly, within this wormhole sector, the pair of codimension-one observables is effectively equivalent to a single codimension-one observable characterized by $ \talpha_{\rm eff} = \talpha_+ + \talpha_-$. This explains why $G_{\rm classical}+G_{\rm bra-ket}$ in \eqref{eq:GclassicalBraketCodimOne} reproduces the codimension-one generating function after the replacement $\talpha \to \talpha_{\rm eff}$. 
More importantly, as we will show below, the remaining wormhole sectors are parametrically subleading to the fully disconnected and bra--ket contributions. In other words, the codimension-zero generating function $G^{(0)}$ \eqref{eq:Gzerototal} is dominated by 
\begin{equation}
\boxed{G^{(0)}(\alpha_\pm;t) \approx  G_{\rm classical}  +G_{\rm bra-ket}  \,.}
\end{equation}
In the corresponding parametric regimes, we thus conclude that 
\begin{tcolorbox}
\begin{equation}\label{eq:sloperampplateau_zero}
 G^{(0)}(\alpha_\pm;t) \approx  
\begin{cases}
e^{-({\tilde\alpha}_++ {\tilde\alpha}_-) t} + \mathcal{O}(\frac{{\tilde\alpha}_++ {\tilde\alpha}_-}{t \Delta E})\,, \quad \text{slope} \quad ({\tilde\alpha}_++ {\tilde\alpha}_-) t \sim 1  \,,\\[1em]
\dfrac{2t}{({\tilde\alpha}_++ {\tilde\alpha}_-) \, T_{\mt{H}}^2} + \mathcal{O} (e^{-({\tilde\alpha}_++ {\tilde\alpha}_-) t}) \,,\quad \text{ramp} \quad  t <   T_{\mt{H}}  \,,\\[1em]
 \dfrac{2}{({\tilde\alpha}_++ {\tilde\alpha}_-) \, T_{\mt{H}}} + \mathcal{O}(e^{-({\tilde\alpha}_++ {\tilde\alpha}_-) (t-T_{\mt{H}})})\,,\quad \text{plateau} \quad  t >   T_{\mt{H}}  \,. \\
\end{cases}
\end{equation}
\end{tcolorbox}
The results obtained so far are therefore already sufficient to establish the universal time evolution of the codimension-zero generating function, as illustrated in figure~\ref{fig:G0s}. In particular, it exhibits the characteristic slope-ramp-plateau structure familiar from the spectral form factor. Its slope-ramp-plateau structure is therefore inherited directly from the codimension-one analysis. More specifically, the bra--ket sector contains the same universal two-point spectral correlations $\langle D(E_i)D(E_j) \rangle_c$ that generate the ramp and plateau in the codimension-one case. The appearance of these regimes in the codimension-zero generating function thus reflects the same underlying spectral statistics and level repulsion characteristic of quantum-chaotic spectra, as illustrated in figure~\ref{fig:G0s}.
 
We next consider the pairwise connected contributions involving the operator region. Since the bra--operator sector follows from the operator--ket sector by exchanging
$\talpha_+\leftrightarrow\talpha_-$, it is sufficient to evaluate one of these two sectors, \eg 
\begin{equation} \label{eq:Gop-ket}
\begin{split}
 G_{\rm op-ket}&:=\frac{e^{3S_0}}{Z}\int_{\mathcal{I}} dE_1dE_2 dE_3 \, e^{-iE_{12}t} \bar{D}_1 \langle D(E_2) D(E_3)\rangle_c \mathbf{M}_{\talpha_+}(E_1,E_3) \mathbf{M}_{\talpha_-}(E_3,E_2)   \,.
\end{split}
\end{equation}
Using the same three integration variables as eq.~\eqref{eq:GclassicalVariablesDefinition}, the Fourier phase factorizes as $e^{-iE_{12}t}=e^{-iE_{13}t}e^{-iE_{32}t}$. Within the same narrow-window approximation used above, we evaluate the smooth mean-energy dependence at $E_0$ and extend the integrations over $E_{13}$ and $E_{32}$ to the real line. The result then separates into
\begin{equation}
\begin{split}
    G_{\rm op-ket}
    \approx
    &  
    \left[
    e^{S_0}\bar D(E_0)
    \int_{-\infty}^{\infty}dE_{13}\,
    e^{-iE_{13}t}\,
    \mathbf{M}_{\talpha_+}(E_1,E_3)
    \right]
    \\
    &\times
    \left[
    \frac{e^{2S_0}}{Z}
    \int_{\mathcal I}dE_3
    \int_{-\infty}^{\infty}dE_{32}\,
    e^{-iE_{32}t}\,
    \left\langle D(E_3)D(E_2)\right\rangle_{\rm c}\,
    \mathbf{M}_{\talpha_-}(E_3,E_2)
    \right].
\end{split}
\label{eq:GopketFactorization}
\end{equation}
The first bracket is precisely the classical contribution to the codimension-one generating function with pole parameter $\talpha_+$, while the second bracket is its quantum contribution with pole parameter $\talpha_-$. Therefore,
\begin{equation}
    G_{\rm op-ket}(\talpha_+,\talpha_-;t)
    \approx
    G_{\rm classical}^{(1)}(\talpha_+,t)\,
    G_{\rm quantum}^{(1)}(\talpha_-,t).
    \label{eq:Gop-ket_result}
\end{equation}
The bra--operator sector is obtained by exchanging the two
codimension-one contributions:
\begin{equation}
    G_{\rm bra-op}(\talpha_+,\talpha_-;t)
    \approx
    G_{\rm quantum}^{(1)}(\talpha_+,t)\,
    G_{\rm classical}^{(1)}(\talpha_-,t).
    \label{eq:Gbra-op_result}
\end{equation}
These factorizations directly reflect the connectivity of the corresponding Euclidean wormhole geometries. In the operator--ket sector, the operator and ket regions are connected by wormholes, while the bra region remains disconnected. Accordingly, the contribution associated with $\CC_-$ is governed by the quantum part of the codimension-one generating function, whereas the contribution associated with $\CC_+$ retains its classical form. The bra--operator sector has the analogous structure with the two boundary contributions exchanged.

For fixed nonzero pole parameters satisfying $\talpha_\pm T_{\mt{H}}\gg1$, both mixed sectors are parametrically subleading to $G_{\rm classical}$ and $G_{\rm bra-ket}$. During the ramp and plateau regimes, the quantum factor is accompanied by an
exponentially damped classical factor from $ G_{\rm classical}^{(1)}(\talpha_\pm,t)$. Consequently, $G_{\rm op-ket}$ and $G_{\rm bra-op}$ do not modify the leading slope-ramp-plateau profile of the generating function at finite $\talpha_\pm$. Nevertheless, these sectors must be retained when taking the regulated $\alpha_\pm\to0$ limit to extract the codimension-zero holographic complexity.

We have now evaluated all contributions arising from the pairwise connected
components of the three-point spectral correlator. It remains to analyze
the fully connected bra--operator--ket sector. Recalling
eq.~\eqref{eq:DDDconnected},
\begin{equation}
\begin{aligned}
    \left\langle D_1D_2D_3\right\rangle_{\rm c}
    ={}&
    e^{-2S_0}\bar D(E_1) \delta(E_{12})\delta(E_{13})  +2e^{-3S_0}K_{12}K_{23}K_{31}
    \\
    &\quad -e^{-3S_0} \left[ \delta(E_{23})K_{12}^{2} +\delta(E_{13})K_{23}^{2}+\delta(E_{12})K_{13}^{2}
    \right],
\end{aligned}
\label{eq:DDDconnectedRepeated}
\end{equation}
we decompose this contribution into the five terms associated with the
corresponding structures in the connected correlator:
\begin{equation}
\begin{split}
    G_{\rm bra-op-ket}(\talpha_\pm;t)
    :={}& G_{\delta_{12}\delta_{13}}(\talpha_\pm)
    +G_{\delta_{23}K_{12}^{2}}(\talpha_\pm;t)
    +G_{\delta_{13}K_{23}^{2}}(\talpha_\pm;t)
    \\
    &+
    G_{\delta_{12}K_{13}^{2}}(\talpha_\pm) +G_{K_{12}K_{23}K_{31}}(\talpha_\pm;t).
\end{split}
\label{eq:GbraopketDecomposition}
\end{equation}

The first three terms can once again be expressed directly in terms of
the codimension-one building blocks. We begin with the double-contact
contribution:
\begin{equation}
\begin{split}
    G_{\delta_{12}\delta_{13}}(\talpha_\pm)
    &:=
    \frac{e^{3S_0}}{Z}\int_{\mathcal{I}^3} dE_1dE_2dE_3\,e^{-iE_{12}t}
    \mathbf{M}_{\talpha_+}(E_1,E_3)\mathbf{M}_{\talpha_-}(E_3,E_2)
    \\
    &\hspace{3.5cm}\times
    \left[e^{-2S_0}\delta(E_{12})\delta(E_{31})\bar D(E_1)\right]
    \\
    &=
    \frac{e^{S_0}}{Z}\int_{\mathcal I}dE\,\bar D(E)\, \mathbf{M}_{\talpha_+}(E,E)
    \mathbf{M}_{\talpha_-}(E,E).
\end{split}
\end{equation}
Using the diagonal limit of the matrix elements,
\begin{equation}
    \mathbf{M}_{\talpha_\pm}(E,E)
    \approx
    \frac{2}{T_{\mt{H}}\talpha_\pm},
\end{equation}
together with the leading microcanonical normalization, we obtain
\begin{equation}
    G_{\delta_{12}\delta_{13}}(\talpha_\pm)
    \approx
    \frac{2}{T_{\mt{H}}\talpha_+} \frac{2}{T_{\mt{H}}\talpha_-}
    =G_{\rm delta}^{(1)}(\talpha_+) G_{\rm delta}^{(1)}(\talpha_-).
\label{eq:GdeltadeltaResult}
\end{equation}
This term is independent of time because the two delta functions force
all three energies to coincide.

The two contributions containing one contact term and one squared kernel factorize in an analogous way. For the $\delta(E_{23})K_{12}^{2}$ contribution, the delta function sets $E_3=E_2$, leaving a diagonal matrix element:
\begin{equation}
\begin{split}
    G_{\delta_{23}K^2_{12}}(\talpha_\pm,t)
    &:=
    \frac{e^{3S_0}}{Z} \int_{\mathcal{I}^3} dE_1dE_2dE_3\,e^{-iE_{12}t} \mathbf{M}_{\talpha_+}(E_1,E_3)\mathbf{M}_{\talpha_-}(E_3,E_2)
    \\
    &\hspace{3.5cm}\times \left[ -e^{-3S_0}\delta(E_{23})K(E_1,E_2)^2\right]
    \\
    &\approx
    \mathbf{M}_{\talpha_-}(E_0,E_0)
    \frac{e^{2S_0}}{Z} \int_{\mathcal{I}^2}dE_1dE_2\,e^{-iE_{12}t}
    \left[  -e^{-2S_0}K(E_1,E_2)^2 \right] \mathbf{M}_{\talpha_+}(E_1,E_2).
\end{split}
\end{equation}
It therefore reduces to
\begin{equation}
    G_{\delta_{23}K^2_{12}}(\talpha_\pm,t)
    \approx
    G_{\rm delta}^{(1)}(\talpha_-)\,G_{\rm sine}^{(1)}(\talpha_+,t).
\end{equation}
Similarly, the $\delta(E_{13})K_{23}^{2}$ term gives
\begin{equation}
\begin{split}
    G_{\delta_{13}K_{23}^{2}}(\talpha_\pm;t)
    &\approx G_{\rm delta}^{(1)}(\talpha_+)\,G_{\rm sine}^{(1)}(\talpha_-,t).
\end{split}
\label{eq:Gdelta13K23Result}
\end{equation}
Thus, each of these terms contains one diagonal codimension-one contribution and one sine-kernel contribution. This factorized structure is consistent with the connectivity of the corresponding wormhole sectors.

The remaining two terms are genuinely new in the sense that they cannot be reduced to products of codimension-one generating functions. We first consider the contribution proportional to $\delta(E_{12})K_{13}^{2}$. Integrating over $E_2$ yields 
\begin{equation}\label{eq:GdeltaKK}
\begin{split}
    G_{\delta_{12}K^2_{13}}(\talpha_\pm)
    &:=
    \frac{e^{3S_0}}{Z}\int_{\mathcal{I}^3}dE_1dE_2dE_3\,e^{-iE_{12}t}
    \mathbf{M}_{\talpha_+}(E_1,E_3)\mathbf{M}_{\talpha_-}(E_3,E_2)
    \\
    &\hspace{3.5cm}\times
    \left[-e^{-3S_0}\delta(E_{12})K(E_1,E_3)^2\right]
    \\
    &=  -\frac{1}{Z} \int_{\mathcal{I}^2}dE_1dE_3\,K(E_1,E_3)^2\mathbf{M}_{\talpha_+}(E_1,E_3) \mathbf{M}_{\talpha_-}(E_3,E_1).
\end{split}
\end{equation}
Because the delta function imposes $E_1=E_2$, the Fourier phase becomes trivial and the resulting contribution is independent of time. Introducing $\bar E=(E_1+E_3)/2$ and $E_{13}=E_1-E_3$, and substituting the sine kernel together with the pole structures of the two matrix elements, the mean-energy integral yields
\begin{equation}
\begin{split}
    G_{\delta_{12}K_{13}^{2}}(\talpha_\pm)
    \approx
    -\frac{8\talpha_+\talpha_-}{\pi T_{\mt{H}}^{3}}
    \int_{-\infty}^{\infty}dE_{13}\,
    \frac{    \sin^{2}\!\left(\frac{T_{\mt{H}}E_{13}}{2}\right)}{ E_{13}^{2}  \left(E_{13}^{2}+\talpha_+^{2}\right)
    \left(E_{13}^{2}+\talpha_-^{2}\right)
    }.
\end{split}
\label{eq:Gdelta12K13Integral}
\end{equation}
Evaluating the remaining integral gives
\begin{equation}
\begin{split}
    G_{\delta_{12}K_{13}^{2}}(\talpha_\pm)
    \approx
    \frac{4}{T_{\mt{H}}^{3}}\frac{
        \talpha_+^{3} \left(1-e^{-\talpha_-T_{\mt{H}}}\right)-\talpha_-^{3} \left(1-e^{-\talpha_+T_{\mt{H}}}\right)
     }{\talpha_+^2\talpha_-^2  \left(\talpha_+^{2}-\talpha_-^{2}\right)
    } - \frac{4}{\talpha_+\talpha_-\THH^2}.
\end{split}
\end{equation}
It is obvious that the full expression is manifestly symmetric under $\talpha_+\leftrightarrow\talpha_-$, which corresponds to the symmetry between the ket and the bra part. It is also straightforward to find that the equal-parameter limit $\talpha_+=\talpha_-$ is still regular since 
\begin{equation}
G_{\delta_{12}K_{13}^{2}}(\talpha,\talpha)
    \approx -\frac{2}{T_{\mt{H}}^{3}\talpha^{3}}
    \left[  2T_{\mt{H}}\talpha-3
        + \left(T_{\mt{H}}\talpha+3\right) e^{-T_{\mt{H}}\talpha}
    \right].
\label{eq:Gdelta12K13equal}
\end{equation}
However, this term still develops a regulator-dependent divergence as $\talpha_\pm\to 0$, and must therefore be treated using the regularization prescription introduced in section~\ref{sec:reg}.

We finally consider the contribution involving three Christoffel--Darboux kernels:
\begin{equation}
\begin{split}
    G_{K_{12}K_{23}K_{31}}(\talpha_\pm,t)
    &:=  \frac{e^{3S_0}}{Z}  \int_{\mathcal{I}^3} dE_1dE_2dE_3\, e^{-iE_{12}t}   \mathbf{M}_{\talpha_+}(E_1,E_3)  \mathbf{M}_{\talpha_-}(E_3,E_2)
    \\
    &\hspace{2cm}\times
    \left[ 2e^{-3S_0}  K(E_1,E_2)K(E_2,E_3)K(E_3,E_1) \right].
\end{split}
\label{eq:GKKK}
\end{equation}
This integral may be evaluated by using $E_3$, $E_{13}=E_1-E_3$, and $E_{23}=E_2-E_3$ as integration variables and applying the residue theorem to the resulting pole structure. The complete expression is lengthy and is therefore relegated to Appendix \ref{app:KKK}. More importantly, all five components in the fully connected bra--operator--ket sector are suppressed by $T_{\mt{H}}^{-2}$ for fixed nonzero parameters $\talpha_\pm$. 

For later reference, we summarize the contributions obtained in this
subsection:
\begin{equation} \label{eq:Gzero}
\begin{split}
    G_{\rm classical}(\talpha_\pm,t)&=G_{\rm classical}^{(1)}(\talpha_+,t)\,G_{\rm classical}^{(1)}(\talpha_-,t)\\
     G_{\rm bra-ket}(\talpha_\pm,t)&=G_{\rm quantum}^{(1)}({\talpha}_++ {\talpha}_-,t)\\    
    G_{\rm bra-op}(\talpha_\pm,t)&=G_{\rm quantum}^{(1)}({\talpha}_+,t)\,G_{\rm classical}^{(1)}(\talpha_-,t)\\
    G_{\rm op-ket}(\talpha_\pm,t)&=G_{\rm quantum}^{(1)}({\talpha}_-,t)\,G_{\rm classical}^{(1)}(\talpha_+,t)\\
    G_{\rm bra-op-ket}(\talpha_\pm,t)&=G_{\rm delta}^{(1)}({\talpha}_+)\,G_{\rm delta}^{(1)}({\talpha}_-)+G_{\rm delta}^{(1)}({\talpha}_+)\,G_{\rm sine}^{(1)}({\talpha}_-,t)\\
     &+G_{\rm delta}^{(1)}({\talpha}_-)\,G_{\rm sine}^{(1)}({\talpha}_+,t)+G_{\delta_{12}K_{13}^2}({\talpha}_\pm,t)+G_{K_{12}K_{23}K_{31}}({\talpha}_\pm,t) \,. 
\end{split}
\end{equation}
Crucially, the leading slope-ramp-plateau behavior of the codimension-zero generating function is dominated by $G_{\rm classical}$ and $G_{\rm bra-ket}$, while the remaining pairwise and wormhole sectors contribute subleading corrections at the order of $T_{\mt{H}}^{-2}$. As we will discuss in the next subsection, all these sectors would contribute when the regulated $\alpha_\pm\to 0$ limit is taken to extract the time evolution of holographic complexity itself. However, the time evolution of complexity is still dominated by $G_{\rm classical}$ and $G_{\rm bra-ket}$ since the remaining contributions cancel each other.

\subsection{From generating functions to holographic complexity}
\label{sec:reg}
Once the generating function has been properly defined, one can expect to extract the expectation value of the holographic complexity $\CC$ from its
expansion around $\alpha=0$:
\begin{equation}
    \left\langle e^{-\alpha\CC}\right\rangle \approx 
    1-\alpha\langle \hat{\mC} \rangle +\mathcal{O}(\alpha^2).
\label{eq:G1expansion}
\end{equation}
Equivalently,
\begin{equation}
\langle \hat{\mC} \rangle \equiv \langle  \text{TFD}(t)|\, \hat{C} \, |\text{TFD}(t) \rangle = \lim_{\alpha \to 0} \left(  - \partial_\alpha   \langle \widehat{e^{-\alpha \mathcal{C}}} \rangle  \right) \,. 
\label{eq:complexity-from-G}
\end{equation}
However, the implementation of this limit is nevertheless subtle. As we will discuss in detail in section~\ref{sec:prob}, divergences may arise from the continuum approximation to the spectrum and from the overcompleteness of the continuous $\mathcal C$-basis. At the level of spectral correlation functions, these divergences manifest themselves as ill-defined products of Dirac delta functions associated with repeated diagonal constraints. Before taking the $\alpha\to0$ limit, we therefore first establish how such diagonal products should be interpreted in the continuum description of an underlying finite discrete spectrum.

\subsubsection{Regularization of diagonal terms}

The same issue is already visible in the normalization of the TFD state.
By construction, the TFD state defined in eq.~\eqref{eq:TFD} satisfies
\begin{equation}
    \langle\text{TFD}|\text{TFD}\rangle=1.
\end{equation}
In the ensemble-averaged continuum representation, however, its norm is
written as
\begin{equation}
\begin{split}
    \overline{\langle\text{TFD}|\text{TFD}\rangle}
    =
    \frac{e^{2S_0}}{Z}
    \int_{\mathcal I^2}
    dE_i\,dE_j\,
    e^{-iE_{ij}t}
    \left\langle D(E_i)D(E_j)\right\rangle
    \frac{\delta(E_i-E_j)}
    {e^{S_0}\bar D(E_i)}.
\end{split}
\label{eq:TFDnorm}
\end{equation}
Naively, one encounters a product of Dirac delta functions since the decomposition of the spectral two-point correlation function contains the contact term:
\begin{equation}
   \overline{ \left\langle D(E_i)D(E_j)\right\rangle }
    =
    e^{-2S_0}
    \left[
        R_2(E_i,E_j)
        +\delta(E_i-E_j)\bar\rho(E_i)
    \right]\,.
\label{eq:DD-R2}
\end{equation}
Since $R_2$ describes correlations between distinct eigenvalues, level
repulsion implies 
\begin{equation}\label{eq:R2delta0}
    R_2(E_i,E_j)\delta(E_i-E_j)=0.
\end{equation}
Only the contact term therefore contributes to
eq.~\eqref{eq:TFDnorm}, leaving
\begin{equation}
    \overline{\langle\text{TFD}|\text{TFD}\rangle}
    =
    \frac{1}{Z}
    \int_{\mathcal I^2}
    dE_i\,dE_j\,
    e^{-iE_{ij}t}
    \delta(E_i-E_j)\delta(E_i-E_j).
\label{eq:TFDnorm2}
\end{equation}
The square of a Dirac delta function is not well-defined as a distribution. This apparent divergence is an artifact of replacing a finite discrete spectrum by a continuum description. In the exact discrete Hilbert space, the corresponding diagonal sum is finite.

For a discrete spectrum $\{E_a\}_{a=1}^{N}$, the exact density of states
and the energy-basis normalization are
\begin{equation}
    \rho(E)=\sum_{i=1}^{N}\delta(E-E_i),
    \qquad
    \langle E_i|E_j\rangle=\delta_{ij},
    \qquad
    \hat{\mathbbm 1} = \sum_{i=1}^{N}|E_i\rangle\langle E_i|\,,
\label{eq:discrete-spectrum}
\end{equation}
where $N$ represents the total number of states. In particular, for any smooth function $f(E)$,
\begin{equation}
    \sum_{i=1}^{N}f(E_i)= \int_{\mathcal I}dE\,\rho(E)f(E).
\end{equation}
After ensemble averaging or coarse graining, the exact density is
replaced by the smooth mean density,
\begin{equation}
   \sum_{i=1}^{N}
    \quad\longleftrightarrow \quad 
    \int_{\mathcal I}dE\,\bar\rho(E)= \int_{\mathcal I}dE\,e^{S_0}\bar D(E)\,.
\label{eq:sum-to-integral}
\end{equation}
For the microcanonical ensemble considered here,
\begin{equation}
    Z =N =
    \int_{\mathcal I}dE\,\bar\rho(E)\approx e^{S_0}\bar D(E_0)\Delta E\,. 
\label{eq:microcanonical-number}
\end{equation}
With respect to the continuum measure $\bar\rho(E)\,dE$, the Kronecker delta is thus represented by
\begin{equation}\label{eq:deltaij}
 \boxed{ \delta_{ij}  \longleftrightarrow \langle E_i | E_j \rangle = \frac{\delta\left(E_i-E_j\right)}{e^{S_0}  \bar{D}(E_j)}   \,.} 
\end{equation}
Indeed, this correspondence reproduces a single diagonal sum:
\begin{equation}
\begin{split}
    \sum_{j=1}^{N}\delta_{ij}f(E_i,E_j)
    \quad\longleftrightarrow\quad
    \int_{\mathcal I}dE_j \,
    \bar\rho(E_j) \times 
    \frac{\delta(E_i-E_j)}{\bar\rho(E_i)} \times f(E_i,E_j)  =  f(E_i,E_i)\,.
\end{split}
\label{eq:single-delta-sum}
\end{equation}
The corresponding relation for two diagonal constraints follows from the
idempotency of the Kronecker delta:
\begin{equation}
  \delta_{ij}^2 = \delta_{ij} \longrightarrow   \sum_{i,j=1}^{N}
    \delta_{ij}^2\,f(E_i,E_j) =   \sum_{i=1}^{N}f(E_i,E_i) \,. 
\label{eq:doubledeltasum}
\end{equation}

The continuum prescription must reproduce this finite discrete result, especially the idempotency of the diagonal constraint in the underlying discrete spectrum. Since $\delta_{ij}^{\,2}=\delta_{ij}$, its continuum counterpart should satisfy
\begin{equation}
    \left[
        \frac{\delta(E-E')}{\bar\rho(E)}
    \right]^2
    \overset{\rm reg}{\Longrightarrow}
    \frac{\delta(E-E')}{\bar\rho(E)},
\end{equation}
by using the identification \eqref{eq:deltaij} from the discrete-to-continuum map. The product of two Dirac delta functions is therefore regularized as
\begin{equation}
    \boxed{
    \delta(E-E')^2
    \overset{\rm reg}{\Longrightarrow}
    \bar\rho(E)\delta(E-E') = e^{S_0}\bar D(E)\delta(E-E').
    }
\label{eq:twodelta}
\end{equation}
The arrow in eq.~\eqref{eq:twodelta} denotes a finite-level-spacing regularization and should not be interpreted as an identity between distributions. The regularization prescription follows directly from the fact that the local mean level spacing is still finite, namely 
\begin{equation}
    \Delta_{\rm level}(E) =\frac{1}{\bar\rho(E)} =  \frac{1}{e^{S_0}\bar D(E)}.
\end{equation}
At this resolution, the extra Dirac delta function should be understood as a
regulated peak whose width is of order $\Delta_{\rm level}(E)$ and whose
height is therefore of order $\Delta_{\rm level}(E)^{-1}$. More
precisely, the discrete and continuum diagonal constraints are related
locally by
\begin{equation}
    \delta_{ij}
    \quad\longleftrightarrow\quad
    \Delta_{\rm level}(E_i)\,
    \delta(E_i-E_j)
    =
    \frac{\delta(E_i-E_j)}{\bar\rho(E_i)}.
\label{eq:discrete-continuum-delta}
\end{equation}
Accordingly, the regulated coincidence value of the extra Dirac delta function should be set by the inverse mean level spacing:
\begin{equation}
   \delta(E-E')\big|_{E' \to E } \overset{\rm reg}{\Longrightarrow}  \frac{1}{\Delta_{\rm level}(E)}  =\bar\rho(E)
    = e^{S_0}\bar{D}(E) \,. 
\label{eq:delta-zero-reg}
\end{equation}
More explicitly, for a smooth test function $f(E,E')$,
\begin{equation}
\begin{split}
    \int dE\,dE'\,
    f(E,E')\delta(E-E')^2
    &\overset{\rm reg}{\Longrightarrow}
    \int dE\,dE'\,
    f(E,E')\bar\rho(E)\delta(E-E') \\
    &=\int dE\,\bar\rho(E)f(E,E)\,. 
\end{split}
\label{eq:double-delta-test}
\end{equation}
This is precisely the continuum counterpart of the discrete relation
\eqref{eq:doubledeltasum}.

As a consistency check, applying eq.~\eqref{eq:twodelta} to
eq.~\eqref{eq:TFDnorm2} gives
\begin{equation}
\begin{split}
    \overline{\langle\text{TFD}|\text{TFD}\rangle}
    &\overset{\rm reg}{=}
    \frac{1}{Z}\int_{\mathcal I^2}dE_i\,dE_j\, e^{-iE_{ij}t} \bar\rho(E_i)\delta(E_i-E_j)
    \\
    &= \frac{1}{Z} \int_{\mathcal I}dE\,e^{S_0} \bar{D}(E) = 1\,, 
\end{split}
\label{eq:TFDnorm-regularized}
\end{equation}
which restores the correct normalization of the TFD state.

\subsubsection{Regularization of the codimension-one generating function}

Having established the regularization prescription for repeated diagonal
constraints, we can now examine the $\alpha\to0$ limit of the generating
function and extract the corresponding holographic complexity. We first
consider the codimension-one case, postponing the codimension-zero
analysis to the next subsection.

Recall that the codimension-one generating function decomposes into
three contributions associated with the disconnected, contact, and
sine-kernel terms in the spectral two-point function:
\begin{equation}\label{eq:G1again}
\begin{split}
    \langle e^{-\alpha\CC} \rangle&=\frac{e^{2S_0}}{Z}\int_{\mathcal{I}^2} dE_idE_j\,e^{-iE_{ij}t} \overline{\langle  D(E_i)D(E_j)\rangle}\langle E_i | \widehat{e^{-\alpha \CC }}| E_j \rangle\\
    &=G_{\rm classical}^{(1)}+G_{\rm delta}^{(1)}+G_{\rm sine}^{(1)}\,.
\end{split}
\end{equation}
The regularized operator entering this expression is formally defined by
\begin{equation}
    \widehat{e^{-\alpha\CC}}
    :=
    \int d\mC\,
    e^{-\alpha\mC}
    |\mC\rangle\langle\mC|.
\end{equation}
For a complete and properly normalized $\mC$-basis, the resolution of
the identity requires
\begin{equation}\label{eq:Cbasis-completeness}
 \boxed{  \hat{\mathbbm{1}}  =   \sum_{\text{complete basis}}  | \mC \rangle \langle \mC | =  \lim_{\alpha \to 0}  \widehat{e^{-\alpha \mC }} \,. }
\end{equation}
Consequently, the generating function with respect to any complete basis must satisfy
\begin{equation}
    \lim_{\alpha\to0} \left\langle e^{-\alpha\CC}\right\rangle = 1 \,.
\label{eq:Gdelta2}
\end{equation}

A direct substitution of the results
\eqref{eq:Gcl}, \eqref{eq:Gdelta}, and \eqref{eq:Gsine}, however, gives
a divergent contact contribution,
\begin{equation}
 \lim_{\alpha\to0}\langle e^{-\alpha\CC} \rangle =  \lim_{\alpha\to0} G_{\rm delta}^{(1)} = \lim_{\alpha\to0} \frac{2}{\tilde\alpha \THH} \to \infty 
\end{equation}
while the disconnected and sine-kernel terms cancel:

\begin{equation}\label{eq:GclGsincancel}
 \lim_{\alpha\to0} \left[  G_{\rm classical}^{(1)}(\alpha, t)+G_{\rm sine}^{(1)}(\alpha, t)  \right] =0 \,. 
\end{equation}
The latter cancellation is a direct consequence of level repulsion, which forces the noncontact part of the two-point spectral correlator to vanish at coincident energies. The remaining divergence is therefore entirely associated with the continuum representation of the diagonal contact term.

This origin becomes transparent if the $\alpha\to0$ limit is taken at the level of the spectral integral. From eq.~\eqref{eq:Cbasis-completeness}, the matrix element reduces to the continuum representation of the identity operator:
\begin{equation}\label{eq:matrixelementlimit}
   \lim_{\alpha\to0} \langle E_i | \widehat{e^{-\alpha \CC }}| E_j \rangle=\langle E_i| \hat{\mathbbm{1}} |E_j \rangle=\frac{\delta\left(E_i-E_j\right)}{e^{S_0}  \bar{D}(E_i)}\,. 
\end{equation}
The level-repulsion condition,
\begin{equation}
\lim_{E_j \to E_i} \left(  \bar{D}(E_i) \bar{D}(E_j) -\left(\frac{\sin (\pi e^{S_0} \bar{D}(\bar{E})(E_i-E_j))}{e^{S_0}\pi (E_i-E_j) }  \right)^2 \right) =0 \,. 
\end{equation}
eliminates the noncontact contribution. The generating function then
reduces formally to
\begin{equation}
\begin{split}
 \lim_{\alpha\to0} \langle e^{-\alpha\CC} \rangle  &=\frac{e^{2S_0}}{Z}\int_{\mathcal{I}^2} dE_idE_j\,e^{-iE_{ij}t} \overline{\langle  D(E_i)D(E_j)\rangle}\left( \lim_{\alpha\to0}  \langle E_i | \widehat{e^{-\alpha \CC }}| E_j \rangle \right) \\
 &= \frac{1}{Z}\int_{\mathcal{I}^2} dE_idE_j\,\delta (E_i-E_j) \times \delta (E_i-E_j)\,. \\
\end{split}
\end{equation}
The apparent divergence is precisely the double-delta problem discussed above. Applying the regularization prescription \eqref{eq:twodelta},
\begin{equation}
    \delta(E_i-E_j)^2 \overset{\rm reg}{\Longrightarrow}  e^{S_0}\bar D(E_i)\delta(E_i-E_j)\,, 
\end{equation}
we recover the desired result,  
\begin{equation}
\lim_{\alpha\to0} \langle e^{-\alpha\CC} \rangle_{\rm reg}   = \frac{1}{Z}\int_{\mathcal{I}^2} dE_idE_j\,\delta (E_i-E_j) \times e^{S_0} \bar{D}(E_i) = 1 \,. 
\end{equation}

Equivalently, in order to deal with the $\alpha \to 0$ limit, one may replace the divergent contact contribution by a regularized function satisfying
\begin{equation}
\begin{split}
     G_{\rm delta}^{(1),{\rm reg}}(\alpha)  &=G_{\rm delta}^{(1)}(\alpha)-\left(\frac{2}{\tilde\alpha \THH}-1+\mathcal{O}(\alpha)\right)\\
     &=1+\mathcal{O}(\alpha)\,,
\end{split}
\end{equation}
The normalization condition determines the zeroth-order term uniquely, but it does not yet fix the coefficient linear in $\alpha$. We therefore obtain
\begin{equation}\label{eq:alphapluszero}
 \boxed{\lim_{\alpha \to 0}\langle \text{TFD} | \widehat{e^{-\alpha \mC }} | \text{TFD} \rangle  \equiv   \lim_{\alpha\to0} \langle e^{-\alpha\CC} \rangle_{\rm reg} = \lim_{\alpha \to 0}  G_{\rm delta}^{(1),{\rm reg}}(\alpha)  = 1 \,.}  
\end{equation}


\subsubsection{Time evolution of codimension-one holographic complexity}
To extract the expectation value of complexity measures, the generating function must be determined to linear order in $\alpha$. The normalization condition alone does not fix the linear term in $G_{\rm delta}^{(1),{\rm reg}}$. However, we find that its leading $T_{\mt H}$-dependence can nevertheless be constrained by requiring consistency with the semiclassical limit. For fixed nonzero $\alpha$, the exact quantum contribution must vanish as $e^{S_0}\to\infty$, namely 
\begin{equation}\label{eq:deltareg}
 \lim_{e^{S_0} \to \infty}  G_{\rm quantum}^{(1)}(\alpha , t) = \lim_{e^{S_0} \to \infty}  \left[ G_{\rm delta}^{(1)}(\alpha) +G_{\rm sine}^{(1)}(\alpha, t) \right]=0 \,.
\end{equation}
Within the small-$\alpha$ expansion, this requirement implies that all terms growing with $T_{\mt H} \sim e^{S_0}$ must cancel between the regularized contact term and the sine-kernel contribution.

Expanding the explicit expression \eqref{eq:Gsine} around $\alpha=0$
gives
\begin{equation}
\begin{split}
G_{\rm sine}^{(1)}(\alpha, t) = -1+\frac{\talpha}{3}\left(\THH+\mathcal{O}\left(\left(\frac{1}{\THH}\right)^{n\geq 1}\right)\right) + \mathcal{O}(\alpha^2)\,.
\end{split}     
\end{equation}
Consistency with eq.~\eqref{eq:deltareg} therefore fixes the coefficient of $\talpha$ term in $G_{\rm delta}^{(\rm reg)}$ as follows:
\begin{equation}\label{eq:Gdeltaregular}
\begin{split}
      G_{\rm delta}^{(1),{\rm reg}}(\alpha) &=1-\frac{\tilde\alpha}{3} \THH +\talpha\,\mathcal{O}\left(\left(\frac{1}{\THH}\right)^{n\geq 1}\right)+\mathcal{O}(\alpha^2)\,.
\end{split}
\end{equation}
The remaining subleading contributions are not fixed by this argument. This freedom represents a regularization-scheme dependence and gives rise only to a time-independent additive constant in the complexity.

In order to perform the $\alpha \to 0$ limit, we define the regularized codimension-one generating function by
\begin{equation}
   \langle e^{-\alpha\CC} \rangle_{\rm reg} :=G_{\rm classical}^{(1)}+  G_{\rm delta}^{(1),{\rm reg}}+G_{\rm sine}^{(1)} \,.
\end{equation}
Taking the $\alpha \to 0$ limit for the regularized generating function with $t<T_{\mt H}$ thus yields 
\begin{equation}
\begin{split}
\langle e^{-\alpha\CC} \rangle_{\rm reg} &\overset{\alpha \to 0}{\approx} 
  e^{- t \tilde{\alpha} } + \left( 1 - \talpha \frac{T_{\mt{H}}}{3}  + \mathcal{O}(\talpha^2) \right)- \left( e^{- t \tilde{\alpha} } + \talpha \frac{(t- T_{\mt{H}})^3}{3 T_{\mt{H}}^2}  + \mathcal{O}(\talpha^2)\right) \,,\\
&\approx 1  - \talpha \left( t - \frac{t^2}{T_{\mt{H}}} + \frac{t^3}{3 T_{\mt{H}}^2}\right)  +  \mathcal{O}(\talpha^2)  \,.
\end{split}
\end{equation}
For $t\geq T_{\mt H}$, one instead finds
\begin{equation}
\begin{split}
\langle e^{-\alpha\CC} \rangle_{\rm reg} &\overset{\alpha \to 0}{\approx} 
  e^{- t \tilde{\alpha} } + \left( 1 - \talpha \frac{T_{\mt{H}}}{3}  + \mathcal{O}(\talpha^2) \right)- \left( e^{- t \tilde{\alpha} } + \mathcal{O}(\talpha^2)\right) \,,\\
&\approx 1  - \talpha \frac{T_{\mt{H}}}{3} +  \mathcal{O}(\talpha^2)  \,.
\end{split}
\end{equation}
The expectation value of the codimension-one complexity follows from eq.~\eqref{eq:G1expansion}:
\begin{equation}\label{eq:Cone}
\boxed{ \langle \hat{\CC} \rangle  = -\lim_{\alpha \to 0}\partial_\alpha \langle e^{-\alpha\CC} \rangle_{\rm reg} \approx \text{const.} + 
\begin{cases}
  M  \, 
t\left(1-\frac{t}{T_{\mt{H}}}+\frac{t^2}{3 T_{\mt{H}}^2}\right)\,,\,\, (t < T_{\mt{H}})  \\[1em]
   \frac{1}{3} M \,T_{\mt{H}} \,, \quad\qquad\
   \,\, (t \ge T_{\mt{H}}) 
 \end{cases}\,.}
\end{equation}
after using the location of the pole with $\talpha=M\alpha+\mathcal O(\alpha^2)$. We stress that the additive constant cannot be determined within the present regularization scheme. It receives contributions both from the
subleading term in $G_{\rm delta}^{(1),{\rm reg}}$ and from the subleading normalization of the pole parameter $\talpha$. The time-dependent part is nevertheless universal. For $t\ll T_{\mt H}$, the complexity grows linearly as $\langle\CC\rangle\simeq Mt$. The growth rate then decreases smoothly as the Heisenberg time is approached, and the complexity finally saturates at the plateau value for $t\geq T_{\mt H}$. An analogous behavior was obtained in JT gravity in \cite{Miyaji:2025yvm} by integrating the finite time derivative of the complexity, without explicitly implementing the regularization of the generating function.

It is important to distinguish the plateau of the generating function at fixed nonzero $\alpha$ from the plateau of the complexity extracted in the $\alpha\to0$ limit. For example, a generating function of the form
\begin{equation}
    e^{-\alpha t}+\text{const.}
\end{equation}
approaches a constant at late times for every fixed $\alpha>0$, but its derivative at $\alpha=0$ still contains a term growing linearly with $t$. Saturation of the complexity therefore requires a more precise cancellation
between the linear growth generated by the classical contribution and the corresponding time-dependent quantum correction.

This cancellation, \ie the late-time plateau of complexity can be traced to the level repulsion of quantum chaotic systems. The key point is that the growth rate of complexity is given by the residue at the pole
$E_{ij}=-i\talpha$. Its coefficient in $\talpha \to 0$ limit satisfies
\begin{equation}
\text{Residue at $E_{ij}= - i \tilde{\alpha}$}  \quad \longrightarrow \quad   \lim_{E_{ij}= - i \tilde{\alpha}  \to 0}  \left( \bar{D}^2   - \dfrac{\sinh^ 2(\pi e^{S_0} \bar{D} \tilde{\alpha} )}{(e^{S_0}\pi\tilde{\alpha} )^2}  \right) \times  t \, e^{-t \tilde{\alpha}}=0 \,.
\end{equation}
This condition is the analytic continuation of the property of the CD kernel, which follows from
\begin{equation}
    \lim_{E_{ij}\to0} R_2(E_i, E_j) = 
   e^{2S_0}\left[   \bar D(E_i)\bar D(E_i) - e^{-2S_0}K(E_i,E_i)^2 \right]= 0 \,. 
\label{eq:CD-level-repulsion}
\end{equation}
This is nothing but the spectral manifestation of level repulsion in a quantum chaotic system. As a summary, we find that the late-time plateau of the codimension-one complexity is therefore not a generic consequence of the plateau in the generating function alone. It originates from the universal short-distance spectral correlations, \ie the level repulsion that guarantees the precise cancellation between the classical linear growth and quantum contributions generated by Euclidean wormholes.

\subsubsection{Codimension-zero generating function and complexity}
We now apply the regularization developed above to the codimension-zero generating function. Before extracting the expectation value of holographic complexity operator itself, it is important to verify that the generating function satisfies the expected consistency condition when either of the two parameters $\alpha_\pm$ is switched off. More precisely, completeness of the $\CC_\pm$-bases requires the
codimension-zero generating function to reduce to its codimension-one counterpart in the limit $\alpha_-\to0$, or equivalently $\alpha_+\to0$. At the operator level,
\begin{equation}\label{eq:aminuszero}
 \boxed{\lim_{\alpha_- \to 0} \langle e^{-\alpha_+ \CC_+ -\alpha_- \CC_-} \rangle = \langle \text{TFD} | \widehat{e^{-\alpha_+ \CC_+ }} \cdot  \hat{\mathbbm{1}}  | \text{TFD} \rangle  =  G^{(1)}(\alpha_+;t)\,.}
\end{equation}
Although this relation is immediate in the exact discrete Hilbert space, its continuum representation in the $\alpha \to 0$ limit contains products of Dirac delta functions at coincident energies, which must be treated using the
regularization prescription introduced above.

To see this explicitly, we use the diagonal limit of the matrix element in
eq.~\eqref{eq:completematrixelement}, which is associated with a complete $\mC$-basis. Taking the $\alpha_-\to0$ limit gives
\begin{equation}\label{eq:G0reduction}
\begin{split}
 \frac{e^{3S_0}}{Z}\int dE_1dE_2 dE_3 \, e^{-iE_{12}t}  \overline{\langle  D(E_1)D(E_2)D(E_3) \rangle} \times \mathbf{M}_{\talpha_+}(E_1,E_3) \, \langle E_3 | \hat{\mathbbm{1}} | E_2 \rangle   \,.
\end{split}
\end{equation}
The nontrivial part of this reduction is therefore the contraction
\begin{equation}
\int dE_3  \, \overline{\langle  \rho(E_1)\rho(E_2) \rho(E_3)\rangle} \times \frac{\delta(E_2 -E_3)}{\bar{\rho}(E_2)}\,. 
\end{equation}
As we introduced before, the three-point spectral correlation function can be decomposed as the joint eigenvalue distributions: 
\begin{equation}
\left\langle\rho_i \rho_j \rho_k \right\rangle  = R_3(i,j,k)+\delta_{i,j} R_2(k,i)+\delta_{j,k} R_2(i,j)+\delta_{k,i} R_2(j,k)+\delta_{i,j} \delta_{j,k} \rho_i \,.
\end{equation}
Crucially, many terms multiplied by the external factor $\delta(E_{23})$ would automatically vanish due to the level repulsion, \ie 
\begin{equation}
 R_3(j,i,i)=0 \,, \qquad R_2(i,i) =0 \,. 
\end{equation}
Collecting all non-vanishing terms and regularizing the double Dirac delta functions according to the regularization scheme \eqref{eq:twodelta}, one can obtain 
\begin{equation}
\begin{split}
\int dE_3  \, \overline{\langle  \rho_1\rho_2 \rho_3\rangle } \times \frac{\delta(E_2 -E_3)}{\bar{\rho}_2} &= \int dE_3 \delta(E_{23})\frac{\delta(E_{23})}{\bar{\rho}(E_2)} \left[ R_2\left(E_1, E_2\right) + \delta\left(E_{12}\right) \bar{\rho}\left(E_1\right) \right]  \\
&\overset{\rm reg}{\Longrightarrow}  R_2\left(E_1, E_2\right)+\delta\left(E_1-E_2\right) \bar{\rho}\left(E_1\right)\\
&= \overline{\left\langle\rho(E_1)\rho(E_2)\right\rangle} \,.
\end{split}
\end{equation}
Thus, the regularized contraction of the three-point spectral correlation function with the identity kernel reproduces the full two-point spectral correlator. Substitution into eq.~\eqref{eq:G0reduction} therefore yields the codimension-one generating function and establishes eq.~\eqref{eq:aminuszero} directly in the continuum description.

It is also instructive to verify this reduction using the decomposition \eqref{eq:Gzerototal} of the codimension-zero generating function, 
\begin{equation}
G^{(0)}(\alpha_\pm;t) = G_{\rm classical}  +G_{\rm bra-op} +G_{\rm op-ket}  +G_{\rm bra-ket} +G_{\rm bra-op-ket}\,.
\end{equation}
This sector-by-sector analysis makes explicit how the different wormhole topologies reorganize themselves when one of the operators is replaced by the identity. The preceding analysis has established the regularized limits of codimension-one generating function as follows  
\begin{equation}
\begin{split}
  \lim_{\alpha \to0} G_{\rm classical}^{(1)}(\alpha,t) = 1 \,, \quad  \lim_{\alpha \to 0} G_{\rm delta}^{(1),\rm{reg}}(\talpha,t) = 1, \qquad   \lim_{\alpha \to0} G_{\rm sine}^{(1)}(\alpha,t) = -1 \,. 
\end{split}
\end{equation}
Using the factorized expressions obtained in eq.~\eqref{eq:Gzero}, it is thus easy to get 
\begin{equation}
\begin{split}
    \lim_{\alpha_-\to0}G_{\rm classical}(\talpha_\pm,t) &=G_{\rm classical}^{(1)}(\talpha_+,t) \,, \\
       \lim_{\alpha_-\to0}G_{\rm bra-op}(\talpha_\pm,t) &=G_{\rm quantum}^{(1)}({\talpha}_+,t) = \lim_{\alpha_-\to0}G_{\rm bra-ket}(\talpha_\pm,t) \,, \\
        \lim_{\alpha_-\to0}G_{\rm op-ket}(\talpha_\pm,t)&=0 \,.  
\end{split}
\end{equation}
The fully connected sector $G_{\rm bra-op-ket}$ requires some additional care. In the single limit, the second matrix element becomes the continuum
representation of the identity,
\begin{equation}
 \lim_{\alpha_-\to0} \mathbf{M}_{\alpha_-}(E_i, E_j) = \frac{\delta(E_i- E_j)} {\bar{\rho}(E_i)} \,. 
\end{equation}
Applying this relation to the definitions of $G_{\delta_{12}K_{13}^{2}}$ \eqref{eq:GdeltaKK} and $G_{K_{12}K_{23}K_{31}}$ \eqref{eq:GKKK}, together with some straightforward algebra, gives
\begin{equation}
 \lim_{\alpha_-\to0}  G_{\delta_{12}K^2_{13}}({\talpha}_\pm) =  -G_{\rm delta}^{(1)}({\talpha}_+) \,, \qquad   \lim_{\alpha_-\to0}G_{K_{12}K_{23}K_{31}}(\talpha_\pm,t)=-2G_{\rm sine}^{(1)}(\talpha_{+},t)\,,
\end{equation}
Combining these limits with other three factorized terms in $G_{\rm bra-op-ket}$ \eqref{eq:GbraopketDecomposition}, one obtains
\begin{equation}
\begin{split}
    \lim_{\alpha_-\to0} G_{\rm bra-op-ket}(\talpha_\pm,t)    &=-G_{\rm quantum}^{(1)}({\talpha}_+,t)\,. 
\end{split}
\end{equation}
The fully connected sector therefore removes one of the two identical quantum contributions arising from $G_{\rm bra-op}$ and $G_{\rm bra-ket}$. Summing all sectors, we recover 
\begin{equation}
    \lim_{\talpha_-\to0} G^{(0)}(\alpha_\pm;t) =G_{\rm classical}^{(1)}({\talpha}_+,t)+G_{\rm quantum}^{(1)}({\talpha}_+,t) =\langle e^{-\alpha_+ \CC_+} \rangle\,.
\end{equation}
This confirms the expected reduction to the codimension-one generating function. By symmetry, the same conclusion holds upon exchanging $\alpha_+$ and $\alpha_-$.

Having established this consistency condition, we move to derive the time evolution of codimension-zero holographic complexity by taking the derivative from the regularized generating function, namely 
\begin{equation}
   \boxed{ \langle\CC_{\rm Any}\rangle:=-\lim_{\alpha_\pm \to 0}(\partial_{\alpha_+}+\partial_{\alpha_-})\langle e^{-\alpha_+ \CC_+ -\alpha_- \CC_-} \rangle_{\rm reg} \,. }
\end{equation}
For example, if we focus on the classical regime, we can find that the fully disconnected contribution automatically yields 
\begin{equation}\label{eq:CAnyclassical}
\begin{split}
    -\lim_{\alpha_\pm \to 0}(\partial_{\alpha_+}+\partial_{\alpha_-})G_{\rm classical}(\talpha_+,\talpha_-,t)&=(M_++M_-)t\\
\end{split}
\end{equation}
by using the location of poles $\talpha_\pm=M_\pm\alpha_\pm+\mathcal O(\alpha_\pm^2)$. This is precisely the sum of the two classical codimension-one
holographic complexity measures associated with the future and past hypersurfaces $\Sigma_\pm$. It reproduces the linear growth of classical codimension-zero holographic complexity as the complexity=anything proposal.

As we have calculated in the previous subsection, the time evolution of the codimension-zero generating function at the regime with $\talpha_\pm T_{\mt H} \gg 1$ presents the typical slope-ramp-plateau structure. More importantly, this is dominated by $G_{\rm classical}$ and $G_{\rm bra-ket}$, 
\begin{equation}
G^{(0)}(\alpha_\pm;t) \approx  G_{\rm classical}  +G_{\rm bra-ket} \,, \qquad \text{for} \qquad \text{finite $\alpha \THH$} \,. 
\end{equation}
Applying the decomposition structure:
\begin{equation}
\boxed{G_{\rm classical}  +G_{\rm bra-ket} \approx G_{\rm classical}^{(1)}(\talpha_+,t)\,G_{\rm classical}^{(1)}(\talpha_-,t) + G_{\rm quantum}^{(1)}(\talpha_+ + \talpha_-,t) \,, }
\end{equation}
it is thus straightforward to obtain 
\begin{equation}\label{eq:Ccl+Cbraket}
    -\lim_{\alpha_\pm \to 0}(\partial_{\alpha_+}+\partial_{\alpha_-})(G_{\rm classical}(\talpha_{\pm},t)+G_{\rm bra-ket}(\talpha_{\pm},t))  = \langle \mC_+ \rangle +\langle \mC_- \rangle  \,. 
\end{equation}
where $\langle \mC_\pm \rangle $ denotes the corresponding regularized codimension-one holographic complexity \eqref{eq:Vgenvalues} with a pole parameter $M_\pm$. Equation~\eqref{eq:Ccl+Cbraket} already suggests the final answer, but the remaining wormhole sectors cannot simply be discarded. In contrast,  although these terms are subleading in the generating function at finite $\alpha_{\pm} \THH$, 
their derivatives can remain the same order as that from $G_{\rm classical}$ and $G_{\rm bra-ket}$ after taking the $\alpha_{\pm}\to 0$ limit. The expectation value of codimension-zero holographic complexity measures must therefore be extracted only after all sectors have been included. Remarkably, as we demonstrate below, the additional contributions are individually non-vanishing but cancel exactly in their
sum, leaving the result derived from $G_{\rm classical}+G_{\rm bra-ket}$ unchanged.

We begin with the pairwise connected bra--operator sector $G_{\rm bra-op}$. Using its factorized form together with the regularized limit $G_{\rm quantum}^{(1),{\rm reg}}(0,t)=0$, we obtain
\begin{equation}
    \begin{split}
       & -\lim_{\alpha_\pm \to 0}(\partial_{\alpha_+}+\partial_{\alpha_-})G_{\rm bra-op}(\talpha_\pm,t) \\
        &=-\lim_{\alpha_\pm \to 0} \left[ \partial_{\alpha_+}G_{\rm quantum}^{(\rm reg)}({\talpha}_+,t)\,G_{\rm classical}^{(1)}(\talpha_-,t) + 
       G_{\rm quantum}^{(\rm reg)}({\talpha}_+,t)\,\partial_{\alpha_-}G_{\rm classical}^{(1)}(\talpha_-,t) \right]\,\\
        &= -\lim_{\alpha_+ \to 0}\partial_{\alpha_+}G_{\rm quantum}^{(\rm reg)}({\talpha}_+,t) \\ 
        &= \langle\hat\CC_+\rangle_{\rm quantum}\,.
    \end{split}
\end{equation}
where $\langle\hat \CC_\pm \rangle_{\rm quantum}$ represents all quantum contributions (as derived in eq.~\eqref{eq:Vgenvalues}) associated with a parameter $M_\pm$. The contribution from $G_{\rm op-ket}$ thus takes a similar form, 
\begin{equation}
    \begin{split}
        -\lim_{\alpha_\pm \to 0}(\partial_{\alpha_+}+\partial_{\alpha_-})G_{\rm op-ket}(\talpha_\pm,t)&= \langle\hat\CC_-\rangle_{\rm quantum}\,.
    \end{split}
\end{equation}
The appearance of $\langle\hat\CC_\pm\rangle_{\rm quantum}$ in these two sectors has a direct geometric interpretation. In the bra--operator sector, the bra
and operator regions are connected by Euclidean wormholes, while the
ket region remains disconnected; the resulting contribution therefore
reproduces all quantum corrections associated with $\CC_+$. The operator--ket sector analogously yields the quantum correction associated with $\CC_-$. 

The cancellation of these additional quantum contributions is supplied by the fully connected bra--operator--ket sector $G_{\rm bra-op-ket}$, which can be evaluated by further considering the limit of $G_{\delta_{12}K_{13}^2}$ and $G_{K_{12}K_{23}K_{31}}$. In the limit $\alpha_\pm \to 0$, we also need to regularize $ G_{\delta_{12}K^2_{13}}$ involving products of Dirac delta functions. Similarly, we can apply the same regularization scheme and obtain 
\begin{equation}\label{eq:GKdelta}
    \begin{split}
        G_{\delta_{12}K^2_{13}}^{(\rm reg)}({\talpha}_\pm,t)&\approx-\left(1-\frac{\THH}{3}\talpha_++\mathcal{O}(\alpha_+^2)\right)\left(1-\frac{\THH}{3}\talpha_-+\mathcal{O}(\alpha_-^2)\right) \,,
    \end{split}
\end{equation}
which yields 
\begin{equation}\label{eq:C-GdeltaK}
\begin{split}
    -\lim_{\alpha_\pm \to 0}(\partial_{\alpha_+}+\partial_{\alpha_-})G_{\delta_{12}K_{13}^2}^{\rm(reg)}(\talpha_\pm)&=-(M_++M_-)\frac{\THH}{3}  = -\left(  \langle\hat\CC_+\rangle_{\rm delta} +\langle\hat\CC_-\rangle_{\rm delta} \right) \,. 
\end{split}
\end{equation}
The triple-kernel contribution follows from the small-$\alpha$ expansion of the explicit result in Appendix \ref{app:KKK}:
\begin{equation}
\begin{split}
    -\lim_{\alpha_\pm \to 0}(\partial_{\alpha_+}+\partial_{\alpha_-})G_{K_{12}K_{23}K_{31}}(\talpha_\pm,t)&=\begin{cases}
        2\left(M_++M_-\right)\left( \frac{T_{\mt{H}}}{3}+\frac{t^2}{T_{\mt{H}}}- \frac{t^3}{3T_{\mt{H}}^2}\right)\,, &t<T_{\mt{H}}\\
        2\left(M_++M_-\right)\,t\,, &t>T_{\mt{H}}\,.
    \end{cases}\\
\end{split}
\end{equation}
Equivalently, this is proportional to the sine-kernel contributions of the two codimension-one complexity measures:
\begin{equation}
\begin{split}
    -\lim_{\alpha_\pm\to0}
    \left(   \partial_{\alpha_+} +\partial_{\alpha_-} \right) G_{K_{12}K_{23}K_{31}}
    =
    -2\left( \left\langle\CC_+\right\rangle_{\rm sine}+ \left\langle\CC_-\right\rangle_{\rm sine} \right).
\end{split}
\label{eq:C-GKKK-sine}
\end{equation}
Combining the contributions from the first three factorized terms in $G_{\rm bra-op-ket}$ with eqs.~\eqref{eq:C-GdeltaK} and \eqref{eq:C-GKKK-sine}, the complete fully connected sector becomes
\begin{equation}
\begin{split}
    -\lim_{\alpha_\pm \to 0}(\partial_{\alpha_+}+\partial_{\alpha_-})G_{\rm bra-op-ket}(\talpha_\pm,t)&=- \left(  \langle\hat\CC_+\rangle_{\rm quantum}  + \langle\hat\CC_-\rangle_{\rm quantum} \right)\,.
\end{split}
\end{equation}
The fully connected bra--operator--ket contribution therefore cancels precisely the quantum corrections generated by the bra--operator and operator--ket sectors. This cancellation explains why the result \eqref{eq:Ccl+Cbraket} inferred from $G_{\rm classical}+G_{\rm bra-ket}$ remains valid even after all subleading wormhole topologies are included.

In summary, after collecting the contributions from every sector, we arrive at the
additive relation
\begin{equation}
 \boxed{ \langle\hat\CC_{\rm Any}\rangle = \langle\hat\CC_+ \rangle + \langle\hat\CC_-\rangle\,.} 
\end{equation}
Thus, the expectation value of the codimension-zero holographic complexity is given by the sum of the two associated codimension-one complexity measures. The classical additivity underlying the complexity=anything proposal (\ie eq.~\eqref{eq:Cpm}) therefore persists after the nonperturbative corrections from Euclidean wormholes are included. More explicitly, the resulting universal time dependence associated with the CD kernel in GUE is given by 
\begin{equation}
\langle\hat\CC_{\rm Any}\rangle \approx \text{const.}
    + 
 \begin{cases}
  (M_++M_-)  \, t\left(1-\frac{t}{T_{\mt{H}}}+\frac{t^2}{3 T_{\mt{H}}^2}\right)\,,\,\, (t < T_{\mt{H}})  \\[1em]
   \frac{1}{3} (M_++M_-) \,T_{\mt{H}} \,, \quad\qquad\
   \,\, (t \ge T_{\mt{H}}) 
 \end{cases} \,. \\
\end{equation}
The additive constant inherits the regularization ambiguity already present in the codimension-one result. The time-dependent part, however, is universal: the complexity initially grows linearly with rate $M_++M_-$, its growth gradually slows as the Heisenberg time is approached, and it ultimately saturates at a plateau of order $T_{\mt H}$.

As in the codimension-one analysis, the late-time saturation is ultimately enforced by level repulsion, \ie\footnote{Due to the permutation symmetry of $R_n$, level repulsion means that the joint probability density $R_n$ for $n$ eigenvalues vanishes whenever any two arguments of $R_n$ coincide.}
\begin{equation}
    R_2(E_i,E_i)=0 \,, \qquad  R_3(E_i,E_j,E_j)=0\,,
\label{eq:R2R3-level-repulsion-final}
\end{equation}
in terms of the $n$-point eigenvalue correlation function $R_n$. These coincident-energy properties ensure the precise cancellation between the classical linear growth and the quantum corrections associated with the connected Euclidean wormhole sectors. The universal transition from linear growth to late-time plateau in the codimension-zero holographic complexity is therefore a direct consequence of the short-distance spectral correlations characteristic of quantum-chaotic systems.

\section{Holographic Complexity Operators and Expectation Values} \label{sec:prob}
In section~\ref{sec:reg}, we encountered a formal divergence in the unregularized spectral representation of the generating function when extracting the expectation value of the complexity measure $\CC$ in the limit $\alpha\to0$. We resolved this divergence by introducing a finite-level-spacing regularization consistent with the finite number of independent states in the nonperturbative Hilbert space. In this section, we develop a complementary interpretation of this prescription from the perspective of constructing a proper holographic complexity operator in an orthonormal basis. 

More precisely, we show that the divergence originates from the non-normalizability of the probability distribution defined with respect to the naive continuous family of fixed-$\CC$ states. Once nonperturbative Euclidean wormhole corrections are included, these states acquire nonvanishing overlaps, and the continuous family becomes overcomplete. A complete orthonormal basis can instead be constructed through Gram--Schmidt orthogonalization, allowing us to define a well-behaved holographic complexity operator $\hat{\mC}$. We show that the expectation value of this operator reproduces the regularized result obtained in section~\ref{sec:reg}. Explicit implementations of this orthogonalization procedure have been used to study the geodesic length in JT gravity~\cite{Miyaji:2024ity} and the chord number in the double-scaled SYK model~\cite{Miyaji:2025ucp}.

\subsection{Fixed-complexity wavefunctions and the pole structure}
\label{sec:wavefunction_pole}

The analysis of section~\ref{sec:gen} characterized the universal time evolution of holographic complexity directly through the analytic structure of the matrix elements of its generating operator. In particular, semiclassical linear growth requires a particular type of poles \eqref{eq:intro_pole_structure} near the origin of the complex energy-difference plane. However, this pole structure could be derived from the perspective of states with fixed holographic complexity $\mC$. In particular, this information is encoded in the wavefunction representation of these fixed-$\mC$ states. 

Let $|\mC\rangle$ denote the formal family of states associated with a fixed value of the renormalized holographic complexity measure $\mC$. In terms of the energy eigenbasis, their spectral profiles are given by
\begin{equation}
    \phi_E(\mC) := \langle E|\mC\rangle,
\end{equation}
so that the fixed-$\mC$ states admit the expansion
\begin{equation}
    |\mC\rangle
    =
    e^{S_0}
    \int_{\mathcal I}dE\,D(E)\,
    \phi_E(\mC)|E\rangle,
\label{eq:fixedC_spectral_expansion}
\end{equation}
where $\mathcal I$ denotes the microcanonical energy window. At this stage, the family $|\mC\rangle$ should be regarded as a semiclassical continuous family rather than an exact orthonormal basis of the nonperturbative Hilbert space. This distinction will become important when we discuss the normalization of the corresponding probability distribution and the construction of the complexity operator.

Formally resolving the generating operator in terms of these states gives
\begin{equation}
\label{eq:pole02}
    \langle E_i|
    \widehat{e^{-\alpha\mC}}
    |E_j\rangle
    :=
    \int_{\mathcal D_{\mC}}d\mC\,
    \phi_{E_i}(\mC)\phi_{E_j}^{*}(\mC)
    e^{-\alpha\mC},
\end{equation}
where $\mathcal D_{\mC}$ denotes the range of the renormalized observable. If the fixed-$\mC$ states formed a complete orthonormal basis, this expression would be the ordinary spectral decomposition of the operator. Here it instead serves as a useful semiclassical representation whose nonperturbative completion will be examined below.

Since we are considering a renormalized gravitational observable like holographic complexity $\mC$, its kinematic range generally need not be bounded from below. For example, the renormalized geodesic length in JT gravity takes values on the full real line. Nevertheless, its wavefunction has very different asymptotic behaviors in the two regimes: it is oscillatory at sufficiently large positive length, whereas the contribution from large negative length is doubly exponentially suppressed \cite{Miyaji:2025yvm}.\footnote{For the renormalized geodesic length $\ell$ in JT gravity, one has $\ell\in(-\infty,\infty)$. The approximate fixed-length wavefunction is oscillatory for $Ee^\ell\gg1$, whereas contributions from $Ee^\ell<1$ are doubly exponentially suppressed. See eqs.~(2.20) and (2.21) of ref.~\cite{Miyaji:2025yvm}.} Motivated by this characteristic example, we thus expect that contributions originating from the negative holographic complexity should be ignorable. Moreover, we will find that the pole relevant for linear growth is determined entirely by the oscillatory regime. We may therefore ignore the suppressed negative-$\mC$ regime and restrict the analysis to
\begin{equation}
    \mC\in[0,\infty)
\end{equation}
without changing the universal nonanalytic structure. A finite change of the lower bound affects only terms analytic in $\alpha$ and corresponds to an additive constant ambiguity in the renormalized complexity.

As shown in section~\ref{sec:gen}, the semiclassical linear growth of holographic complexity is governed by the characteristic pole structure in eq.~\eqref{eq:intro_pole_structure}. We can also reverse the logic of section~\ref{sec:gen}. Rather than postulating the pole and subsequently evaluating the generating function, we ask what asymptotic fixed-$\mC$ wavefunction produces this pole structure which results in the linear growth of holographic complexity measures. From the perspective of fixed-$\mC$ states, the pole originates from the oscillatory behavior of the wavefunction kernel. In the near-diagonal regime  $E_i\simeq E_j$, the required asymptotic form is 
\begin{equation}
\label{eq:wavefunction}
    \phi_{E_i}(\mC)\phi_{E_j}^{*}(\mC)
    \underset{E_i\approx E_j}{\approx}
    \frac{1}
    {\pi e^{S_0}\bar D(\bar E)M}
    \cos\left(E_{ij}t_{\mC}\right),
    \qquad
    t_{\mC}:=\frac{\mC}{M},
\end{equation}
with $E_{ij}:=E_i-E_j, \bar{E}:=\frac{E_i+E_j}{2}$. The parameter $M$ determines the semiclassical growth rate of the observable. It is straightforward to find that substitution into eq.~\eqref{eq:pole02} yields
\begin{equation}
\begin{split}
    \langle E_i|
    \widehat{e^{-\alpha\mC}}
    |E_j\rangle
    &\underset{E_i\approx E_j}{\approx}
    \frac{1}
    {\pi e^{S_0}\bar D(\bar E)M}
    \int_0^\infty d\mC\,
    e^{-\alpha\mC}
    \cos\left(\frac{E_{ij}\mC}{M}\right)=
    \frac{1}
    {\pi e^{S_0}\bar D(\bar E)}
    \frac{\talpha}
    {E_{ij}^2+\talpha^2},
\end{split}
\label{eq:pole_from_fixedC_wavefunction}
\end{equation}
with $\talpha=M\alpha+\mathcal O(\alpha^2)$. The oscillatory behavior of the fixed-$\mC$ states therefore generates the two simple poles
\begin{equation}
    E_{ij}=\pm i\talpha
\end{equation}
that govern the semiclassical linear growth. This construction applies to the broad class of holographic observables possessing the universal pole structure discussed in section~\ref{sec:gen}, including the geodesic length in JT gravity~\cite{Miyaji:2024ity} and the chord-number operator in the double-scaled SYK model~\cite{Miyaji:2025ucp}. The latter also provides a realization of the Krylov spread-complexity operator.

Interestingly, the standing wavefunction in eq.~\eqref{eq:wavefunction} contains two distinct oscillatory components,
\begin{equation}
\label{eq:black_white_branches}
    2\cos\left(E_{ij}t_{\mC}\right)= \underbrace{e^{iE_{ij}t_{\mC}}}_{\text{black-hole branch}}+
    \underbrace{e^{-iE_{ij}t_{\mC}}}_{\text{white-hole branch}}.
\end{equation}
These terms represent the two possible time orientations of the semiclassical Lorentzian geometry. The first describes the future-oriented branch on which the black hole interior expands, while the second describes the time-reversed branch associated with a contracting white hole. Their geometric interpretation becomes manifest by considering the probability distribution of fixed-$\mC$ states with respect to the thermofield double state.

Following our previous study on the geodesic length in JT gravity \cite{Miyaji:2025yvm}, we consider the ensemble-averaged squared overlap, namely 
\begin{equation}
\begin{split}
    P^{(1)}_t(\mC) &:=
    \overline{\langle\mathrm{TFD}(t)|\mC\rangle\langle\mC|\mathrm{TFD}(t)\rangle }
    \\
    &=
    \frac{e^{2S_0}}{Z}\int_{\mathcal I}dE_i\,dE_j\,e^{-iE_{ij}t}\,\phi_{E_i}(\mC)\phi_{E_j}^{*}(\mC)\, \overline{\langle D(E_i)D(E_j) \rangle }\,.
\end{split}
\label{eq:fixedC_TFD_overlap}
\end{equation}
The squared overlap between (normalized) states can be naturally interpreted as the transition probability or the probability density with continuous variables. More importantly, the squared overlap $P^{(1)}_t(\mC)$ provides a direct way to calculate the generating function, \ie 
\begin{equation}
\label{eq:genC02}
    \left\langle e^{-\alpha\mC}\right\rangle = \int_{\mathcal{D}_\mC}d\mC\, P^{(1)}_t(\mC)e^{-\alpha\mC}\,.
\end{equation}
Similarly, we can formally define the expectation value of the holographic complexity operator in terms of 
\begin{equation}
\label{eq:defineC02}
    \langle\hat{\mC}\rangle  =  \int_{\mathcal{D}_\mC} d\mC\, P^{(1)}_t(\mC)\,\mC\,. 
\end{equation}
Once the nonperturbative spectral correlations are included, the continuous family $|\mC\rangle$ becomes overcomplete, and the resulting $P^{(1)}_t(\mC)$ fails to be normalizable due to the constant plateau for large-$\mC$ regime. The divergence appearing in eq.~\eqref{eq:defineC02} is therefore the state-space counterpart of the divergence at the limit $\alpha \to 0$ encountered in the generating-function analysis. The construction of a complete orthonormal basis, and the corresponding finite complexity operator, will be developed in section~\ref{sec:operator}.

The quantum corrections to the squared overlap, \ie nonperturbative Euclidean wormhole effects enter eq.~\eqref{eq:fixedC_TFD_overlap} through the spectral correlation functions of the density of states. In the notation of section~\ref{sec:gen},
\begin{equation}
    e^{2S_0}\, \overline{ \langle D(E_i)D(E_j) \rangle } = R_2(E_i,E_j) +  e^{S_0}\bar D(E_i)\delta(E_{ij}).
\label{eq:fixedC_two_point_decomposition}
\end{equation}
The factorial correlator $R_2$ and the coincident-energy contribution therefore provide the quantum corrections to the semiclassical overlap. At the leading semiclassical level, the spectral correlator factorizes into the product of the smooth densities. Combining the TFD phase with the two terms in eq.~\eqref{eq:black_white_branches} yields
\begin{equation}
    e^{-iE_{ij}t}e^{iE_{ij}t_{\mC}}= e^{-iE_{ij}(t-t_{\mC})}\,,
    \qquad
    e^{-iE_{ij}t}e^{-iE_{ij}t_{\mC}}=e^{-iE_{ij}(t+t_{\mC})}\,.
\end{equation}
The first contribution is consequently localized near $t=t_{\mC}$, whereas the second is localized near $t=-t_{\mC}$. More explicitly, we focus on a microcanonical window of width $\Delta E$. After treating the smooth density as approximately constant within that window, the classical distribution takes the form 
\begin{equation}
\begin{split}
    P^{(1)}_{t,\mathrm{cl}}(\mC)
    \approx \frac{2}{\pi M\Delta E}
    \Bigg[
    \frac{\sin^2\left[\frac{\Delta E}{2}(t-t_{\mC}) \right]}{(t-t_{\mC})^2}
    +
    \frac{\sin^2\left[   \frac{\Delta E}{2}(t+t_{\mC}) \right]}{(t+t_{\mC})^2}
    \Bigg]\,. 
\end{split}
\label{eq:fixedC_classical_two_peaks}
\end{equation}
The two sharply localized peaks provide a direct geometric interpretation of the branch decomposition. The peak at $t=t_{\mC}$ corresponds to the future-oriented black-hole saddle, while the peak at $t=-t_{\mC}$ follows the past-oriented white-hole saddle. In particular, for positive boundary time and $t_{\mC}\geq0$, the first peak lies in the physical semiclassical region, whereas the center of the second peak lies outside
the positive-$t_{\mC}$ domain. The quantum corrections to the probability distribution $P^{(1)}_{t,\mathrm{quantum}}(\mC)$ have been analyzed in greater detail for fixed-length states in Ref.~\cite{Miyaji:2025yvm}. Since the resulting structure is closely analogous to the one considered here, we do not repeat the analysis in the present work. Instead, we will focus on the generalized analysis for codimension-zero complexity measures.

Without loss of generality, we will restrict the following discussion to positive Lorentzian times. In this regime, the semiclassical evolution of interest is dominated by the future-oriented black-hole branch. To isolate the black-hole branch, it is convenient to replace the real wavefunction \eqref{eq:wavefunction} by a single traveling-wave component,
\begin{equation}
\label{eq:wavefunctionexp}
    \phi_{E_i}(\mC)
    \phi_{E_j}^{*}(\mC)
    \underset{E_i\approx E_j}{\approx}
    \frac{1}
    {2\pi e^{S_0}\bar D(\bar E)M}
    e^{iE_{ij}t_{\mC}}\,, \quad \text{with} \quad t_{\mC} \in(-\infty,\infty) \,. 
\end{equation}
Here, we have introduced a symmetric auxiliary coordinate $t_{\mC} \in(-\infty,\infty)$ on the full line in order to remove the imaginary contribution by hand. With this choice, the disconnected semiclassical contribution to the squared overlap contains a single phase and is therefore localized only around $t=t_{\mC}$ corresponding to the black-hole branch. The second peak at $t=-t_{\mC}$ corresponding to the white-hole branch is automatically absent in this simplified description. In other words, we instead use the absolute value $|t_{\mC}|$ to represent the physical complexity in the regime $\mC \in [0, \infty)$. The generating operator is then represented as
\begin{equation}
\begin{split}
    \langle E_i|
    \widehat{e^{-\alpha\mC }}
    |E_j\rangle
    &:=
    \int_{-\infty}^{\infty}d\mC\, \phi_{E_i}(\mC) \phi_{E_j}^{*}(\mC)
    e^{-\alpha|\mC|} \underset{E_i\approx E_j}{\approx}
    \frac{1}{\pi e^{S_0}\bar D(\bar E)}
    \frac{\talpha}{E_{ij}^2+\talpha^2}.
\end{split}
\label{eq:signedC_generating_matrix_element}
\end{equation}
The integration over the positive and negative auxiliary regions makes the imaginary part vanish and reproduces the same pair of poles as the real wavefunction \eqref{eq:wavefunction} restricted to the positive regime. 

In the following, we adopt this full-line representation with $t_{\mC} \in (-\infty, +\infty)$ to simplify the expressions, since it allows us to focus directly on the black hole evolution while retaining the same pole structure of the matrix elements.

\subsection{Probability distribution of fixed-$\mC$ states}
\label{sec:cod}

Having established the wavefunction representation of fixed-$\mC$ states, we can now apply it to formulate the codimension-zero holographic complexity directly in terms of overlaps of quantum states. For a single codimension-one observable, the squared overlap
\begin{equation}
    P^{(1)}(\mC,t):= \overline{ \langle \mathrm{TFD}(t)|\mC\rangle\langle \mC|\mathrm{TFD}(t)\rangle}
\end{equation}
provides a natural probability density for finding the time-evolved TFD state in a fixed-$\mC$ state. As discussed in the previous subsection, this interpretation is semiclassical: once nonperturbative wormhole corrections are included, the continuous family $|\mC\rangle$ becomes overcomplete and the resulting distribution need not remain normalizable.

For a codimension-zero observable, the decomposition \eqref{eq:Cdecomposition} involves two codimension-one quantities, $\CC_+$ and $\CC_-$. The corresponding generalization of the probability distribution $P^{(1)}(\mC,t)$ is therefore naturally defined by  
\begin{equation}
\begin{split}
    P(\CC_+,\CC_-,t) &:=
    \overline{\langle\mathrm{TFD}(t)|\CC_+\rangle \langle\CC_+|\CC_-\rangle \langle\CC_-|\mathrm{TFD}(t)\rangle }.
\end{split}
\label{eq:Pcodimzero}
\end{equation}
Unlike the codimension-one quantity, $P(\CC_+,\CC_-,t)$ is not defined as a squared overlap, owing to the additional overlap $\langle\CC_+|\CC_-\rangle$ between two fixed-complexity states. Instead, it takes the form of a Kirkwood--Dirac quasiprobability distribution,\footnote{For a quantum state $\rho$ and two bases $\{|a\rangle\}$ and $\{|b\rangle\}$, the Kirkwood--Dirac quasiprobability is defined as
\begin{equation*}
    P_{\rm KD}(a,b)= \langle b|a\rangle \langle a|\rho|b\rangle = \operatorname{Tr}\!\left(|b\rangle\langle b|\,|a\rangle\langle a|\, \rho \right),
\end{equation*}
up to the ordering convention for the two projectors. In contrast to a classical joint probability distribution, $P_{\rm KD}(a,b)$ can in general take complex values when the corresponding projectors do not commute. Nevertheless, it satisfies the normalization condition $\sum_{a,b}P_{\rm KD}(a,b)=1$ and is therefore referred to as a quasiprobability distribution. In our construction, the quantity $P(\CC_+,\CC_-,t)$ defined in \eqref{eq:Pcodimzero} does not obey this standard normalization before the Gram--Schmidt orthogonalization. We nevertheless refer to it as a quasiprobability distribution for simplicity, since it has the same characteristic overlap structure.} which provides a quantum generalization of a classical joint probability distribution. In the semiclassical limit, it is easy to find 
\begin{equation}
    \langle\CC_+|\CC_-\rangle_{\rm classical}
   = e^{S_0}\int dE \, D(E) \phi_{E}(\mC_-) \phi^\ast_{E}(\mC_+) \propto \delta(t_{\CC_+}-t_{\CC_-})\,. 
\end{equation}
because the fixed-$\mC$ states are mutually orthogonal. Accordingly, the quasiprobability distribution collapses onto the diagonal $t_{\CC_+}=t_{\CC_-}$ and reduces to the probability distribution for the codimension-one case, 
\begin{equation}
    P_{\rm classical}(\CC_+,\CC_-,t) \propto P^{(1)}_{\rm classical}(\CC_\pm,t)\,
    \delta(t_{\CC_+}- t_{\CC_-}) \,,
\end{equation}
in the semiclassical limit $ e^{S_0} \to \infty$. After including quantum corrections, however, the overlap between distinct fixed-complexity states becomes nontrivial due to nonperturbative Euclidean wormhole corrections. The resulting distribution therefore involves three independent energy variables and is governed by the same three-point spectral correlations that appeared in the codimension-zero generating function of section~\ref{sec:gen}.

Inserting energy eigenstates and using the microcanonical TFD state, we can decompose this generalized probability distribution in terms of 
\begin{equation}
\label{eq:Pdef}
\begin{split}
    P(\CC_+,\CC_-,t)
    ={}&\frac{e^{3S_0}}{Z} \int_{\mathcal I}dE_1\,dE_2\,dE_3\, e^{-iE_{12}t}\, \overline{\langle D(E_1)D(E_2)D(E_3)\rangle}
    \\
    &\times \phi_{E_1}(\CC_+)\phi_{E_3}^{*}(\CC_+)  \phi_{E_3}(\CC_-)\phi_{E_2}^{*}(\CC_-)\,.
\end{split}
\end{equation}
The dependence on the microscopic spectrum is entirely encoded in the ensemble-averaged three-point function. In particular, the disconnected part describes the semiclassical contribution, whereas its connected components incorporate the nonperturbative Euclidean wormhole corrections as discussed in section~\ref{sec:gen}. The decomposition of the three-point spectral correlator \eqref{eq:DDD} induces an analogous decomposition of the quasiprobability distribution,
\begin{equation}
\boxed{
    P_{\rm total} \equiv P(\CC_+,\CC_-,t)= P_{\rm classical} +P_{\rm bra-op}+P_{\rm op-ket} +P_{\rm bra-ket} +P_{\rm bra-op-ket} \,.
}
\label{eq:Ps}
\end{equation}
The five terms have the same connectivity interpretation as their counterparts in the generating function. We now examine their characteristic dependence on the three time variables $t_{\CC_+}$, $t_{\CC_-}$, and $t$.

Within the narrow-window approximation, we set the smooth density to its value at $E_0$ and neglect corrections of order $\Delta E/E_0$. The finite-$\Delta E$ expressions can be evaluated explicitly for the disconnected and pairwise connected sectors, but are unnecessarily lengthy. Since our purpose here is to expose the structure of the quasiprobability distribution and its relation to the generating function, we retain only the leading relevant forms.

The fully disconnected contribution is approximately given by the following form: 
\begin{equation}
\label{eq:Pclassical}
\begin{split}
    P_{\rm classical}
    (t_{\CC_+},t_{\CC_-},t)
    \approx{}&
    \frac{2}{\pi^2M_+M_-\Delta E}
    \frac{ \sin\left[   (t_{\CC_+}-t_{\CC_-})\frac{\Delta E}{2} \right]}{t_{\CC_+}-t_{\CC_-}}
    \\
    &\times
    \frac{
    \sin\left[ (t-t_{\CC_+})\frac{\Delta E}{2} \right]}{t-t_{\CC_+}}
    \times
    \frac{ \sin\left[ (t-t_{\CC_-})\frac{\Delta E}{2}\right]}{t-t_{\CC_-}}.
\end{split}
\end{equation}
It is sharply localized around
\begin{equation}
    t_{\CC_+}=t_{\CC_-}=t \,, 
\end{equation}
and therefore reproduces the leading semiclassical expectation
\begin{equation}
    \CC_\pm\simeq M_\pm t\,. 
\end{equation}
From the bulk perspective, this is precisely the linear growth of the two hypersurface contributions to the codimension-zero holographic complexity. Recall that eq.~\eqref{eq:Pclassical} has been obtained directly by using the approximate wavefunction \eqref{eq:wavefunctionexp}, for which we retain only the black-hole branch at positive time. If instead one uses the real wavefunction on the half-line \eqref{eq:wavefunction}, the probability distribution is obtained by including the corresponding sign-reflected contributions. In particular, an additional classical peak appears at $t_{\CC_+}=t_{\CC_-}=-t$, associated with the white-hole branch.


The pairwise connected sectors exhibit a different structure. It is convenient to display first the operator--ket contribution,
\begin{equation}
\label{eq:Popket}
\begin{split}
    P_{\rm op-ket}
    (t_{\CC_+},t_{\CC_-},t)
    \approx{}&
    \frac{1}{\pi M_+M_-\THH^2}
    \frac{\sin^2\left[  (t-t_{\CC_+})\frac{\Delta E}{2} \right]}
    {\Delta E \, (t-t_{\CC_+})^2}
    \\
    &\times
    \bigg\{
        \min\!\left[|t-t_{\CC_-}|,\THH\right]
        +
        \min\!\left[
            |t_{\CC_+}-t_{\CC_-}|,\THH
        \right]
    \bigg\}.
\end{split}
\end{equation}
The bra--operator contribution follows by exchanging the two codimension-one variables,
\begin{equation}
    P_{\rm bra-op}(t_{\CC_+},t_{\CC_-},t) = P_{\rm op-ket}(t_{\CC_-},t_{\CC_+},t) \,, 
\end{equation}
while the bra--ket contribution is obtained by the corresponding permutation of the three time arguments,
\begin{equation}
    P_{\rm bra-ket}(t_{\CC_+},t_{\CC_-},t) = P_{\rm op-ket}(t_{\CC_+},t,t_{\CC_-}) \,.
\label{eq:Ppairwise_relations}
\end{equation}
These relations make transparent the permutation structure inherited from the spectral three-point function. The fully connected bra--operator--ket sector is approximated by Fourier integrals, yielding
\begin{equation}
\label{eq:Pbraopket}
\begin{split}
    &P_{\rm bra-op-ket}
    (t_{\CC_+},t_{\CC_-},t)
    \approx
    -\frac{2}{M_+M_-\THH^3}
    \min\left[
        \frac{
        |t-t_{\CC_+}|
        +|t-t_{\CC_-}|
        +|t_{\CC_+}-t_{\CC_-}|
        }{2},
        \THH
    \right]
    \\
    &+
    \frac{1}{M_+M_-\THH^3}
    \bigg\{
        \min\!\left[|t-t_{\CC_+}|,\THH\right]
        +
        \min\!\left[|t-t_{\CC_-}|,\THH\right]
    +
        \min\!\left[
            |t_{\CC_+}-t_{\CC_-}|,\THH
        \right]
    \bigg\}.
\end{split}
\end{equation}
The dependence of the complete distribution is therefore organized by the three pairwise separations among $t$, $t_{\CC_+}$, and $t_{\CC_-}$. Since only two of these differences are independent, the structure of $P_{\rm total}$ can be characterized by fixing $t_{\CC_+}-t_{\CC_-}$ and studying its dependence on $t_{\CC_+}-t$.

\begin{figure}[t]
    \centering
    \includegraphics[width=5.0in]{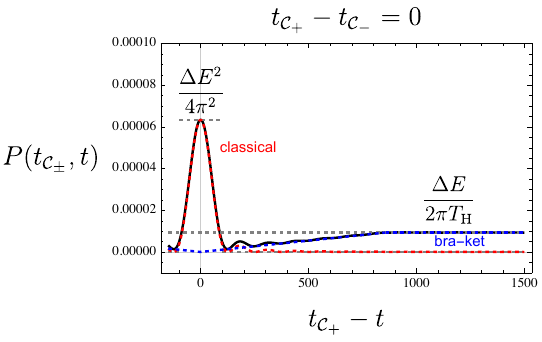}
    \caption{ Total probability distribution \eqref{eq:Ps} for $t_{\CC_+}-t_{\CC_-}=0$.
    We choose $S_0=3$, $\THH\simeq 856$, $\Delta E=0.05$, and $M_\pm=1$. The sharp peak at $t_{\CC_+}=t_{\CC_-}=t$ originates from $P_{\rm classical}$. The broad constant plateau is dominated by $P_{\rm bra-ket}$.}
    \label{fig:Ptotal0}
\end{figure}

We first consider the near-diagonal regime
\begin{equation}
    |t_{\CC_+}-t_{\CC_-}| \lesssim \frac{1}{\Delta E} \,. 
\label{eq:P_neardiagonal_regime}
\end{equation}
As illustrated in figure~\ref{fig:Ptotal0}, the distribution is dominated by the sharp classical peak at $t_{\CC_+}=t_{\CC_-}=t$ together with a broad contribution from the bra--ket sector. Far away from the classical peak, the latter approaches a constant of order $\THH^{-1}$. Although parametrically suppressed, this constant plateau persists over an infinitely long range. Consequently, $P_{\rm total}$ is not normalizable when the naive continuous family of fixed-$\CC$ states is integrated over the full range. This perspective provides a direct probability-space interpretation of the divergence encountered in section~\ref{sec:reg}. The source parameters $\alpha_\pm$ in the generating function suppress the large-$|\CC_\pm|$ region and thereby regulate the non-normalizable tail. The $1/\alpha_\pm$ singularities that appear when the regulators are removed in the limit $\alpha_\pm \to 0$ reflect the non-normalizability of the continuous fixed-$\CC$ spectrum. In section~\ref{sec:operator}, we will show that the same issue can instead be resolved at the Hilbert-space level by constructing a complete orthonormal basis from this overcomplete continuous family.

\begin{figure}[t]
    \centering
    \includegraphics[width=2.95in]{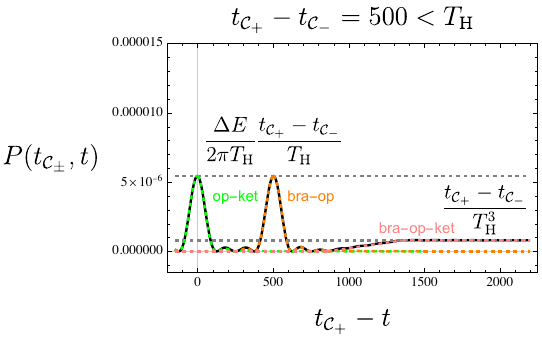}
    \includegraphics[width=2.95in]{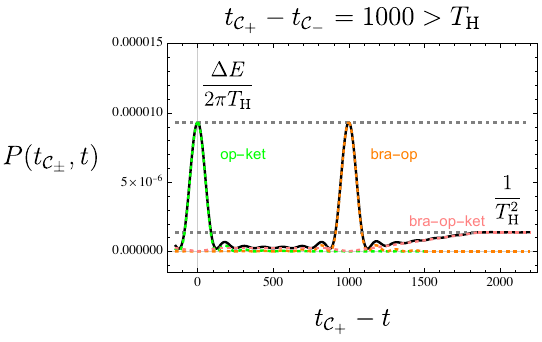}
    \caption{Total probability distribution away from the near-diagonal regime.
    Left: $1/\Delta E\lesssim  |t_{\CC_+}-t_{\CC_-}|<\THH$. Right: $|t_{\CC_+}-t_{\CC_-}|>\THH$. The remaining parameters are the same as in figure~\ref{fig:Ptotal0}. The operator--ket and bra--operator sectors generate peaks at $t_{\CC_+}=t$ and $t_{\CC_-}=t$, respectively. Their amplitudes grow with $|t_{\CC_+}-t_{\CC_-}|$ until this separation reaches the Heisenberg scale, after which they saturate. The fully connected bra--operator--ket contribution vanishes
    at both peak centers and approaches a constant sufficiently
    far from both peaks.}
    \label{fig:Ptotals}
\end{figure}


Away from the near-diagonal region,
\begin{equation}
    \frac{1}{\Delta E}
    \lesssim
    |t_{\CC_+}-t_{\CC_-}|,
\end{equation}
the oscillatory factor suppresses both
$P_{\rm classical}$ and $P_{\rm bra-ket}$.
The distribution is instead governed by the bra--operator,
operator--ket, and fully connected sectors, as shown in
figure~\ref{fig:Ptotals}. For
\begin{equation}
    \frac{1}{\Delta E}
    \lesssim
    |t_{\CC_+}-t_{\CC_-}|
    <
    \THH,
\end{equation}
the heights of the pairwise peaks grow approximately linearly with the separation $|t_{\CC_+}-t_{\CC_-}|$. Once this separation exceeds the Heisenberg time, their heights saturate. The fully connected sector provides an additional constant background of order $\THH^{-2}$ when $t_{\CC_+}-t$ is sufficiently far from the peaks.


Equipped with the quasiprobability distribution $P(\CC_+,\CC_-,t)$ associated with the codimension-zero complexity, one can thus evaluate the expectation values of operators associated with $\mC_{\mt{Any}}$. For example, the generating function of codimension-zero holographic complexity is then derived as 
\begin{equation}\label{eq:G0fromP}
    G^{(0)}(\alpha_\pm,t):=\int_{\mathcal{D}_\mC} d\CC_+ d\CC_- e^{-\alpha_+ |\CC_+|}e^{-\alpha_- |\CC_-| } P(\CC_+,\CC_-,t)\,.
\end{equation}
Equation~\eqref{eq:G0fromP} therefore provides the probability-space counterpart of the spectral representation \eqref{eq:exppm}. It is instructive to verify directly that the quasiprobability distribution reproduces the generating-function analysis of section~\ref{sec:gen}. We adopt the full-line convention introduced in section~\ref{sec:wavefunction_pole}, in which $\CC_\pm$ are auxiliary signed variables and $|\CC_\pm|$ denote the corresponding positive complexity measures. The leading contributions of the disconnected and bra--ket sectors are approximated by 
\begin{equation}
\begin{split}
    G_{\rm classical}
    &\approx 
    \int_{-\infty}^{\infty}
    d\CC_+\,d\CC_-\,
    e^{-\alpha_+|\CC_+|-\alpha_-|\CC_-|}
    P_{\rm classical}(\CC_+,\CC_-,t)\,,
    \\
    G_{\rm bra-ket}
    &\approx
    \int_{-\infty}^{\infty}
    d\CC_+\,d\CC_-\,
    e^{-\alpha_+|\CC_+|-\alpha_-|\CC_-|}
    P_{\rm bra-ket}(\CC_+,\CC_-,t)\,,
\end{split}
\label{eq:GfromP_leading}
\end{equation}
respectively.  Both distributions are strongly localized near $t_{\CC_+}=t_{\CC_-}$. In the leading order with $N \gg 1$, we can apply the following approximation
\begin{equation}
    \frac{
    \sin\left(N x \right)}{x} \sim  \pi \delta(x) \sim  \frac{
    \left(\sin\left(N x \right)\right)^2}{N x^2} \,. 
\end{equation}
Introducing the effective parameter 
\begin{equation}
    \talpha_{\rm eff} := \talpha_++\talpha_- \,, 
\end{equation}
it is easy to find that the above two integrals collapse onto a single coordinate, 
\begin{equation}
    G_{\rm classical}
    \approx
    \frac{2}{\pi\Delta E}
    \int_{-\infty}^{\infty}dt_{\CC}\,
    e^{-\talpha_{\rm eff}|t_{\CC}|}
    \frac{
    \sin^2\left[
        (t-t_{\CC})\frac{\Delta E}{2}
    \right]}
    {(t-t_{\CC})^2},
\label{eq:Gclassical_from_P}
\end{equation}
and
\begin{equation}
    G_{\rm bra-ket}
    \approx
    \frac{1}{\THH^2}
    \int_{-\infty}^{\infty}dt_{\CC}\,
    e^{-\talpha_{\rm eff}|t_{\CC}|}
    \min\!\left[|t-t_{\CC}|,\THH\right].
\label{eq:Gbraket_from_P}
\end{equation}
These are precisely the probability-space representations of the
codimension-one classical and quantum generating functions evaluated at
the effective pole parameter $\talpha_{\rm eff}$. Hence,
\begin{equation}
\begin{split}
    G_{\rm classical}
    &\approx
    G_{\rm classical}^{(1)}(\talpha_{\rm eff},t),
    \\
    G_{\rm bra-ket}
    &\approx
    G_{\rm quantum}^{(1)}(\talpha_{\rm eff},t),
\end{split}
\end{equation}
up to some negligible corrections already discussed in section~\ref{sec:gen}. In particular, $G_{\rm classical}$ can be explicitly evaluated as 
\begin{equation}
\begin{split}
    G_{\rm classical}\approx &-\frac{2}{\pi\Delta E}\frac{\sin^2{\frac{\Delta E}{2}t}}{t}+\frac{e^{-\tilde\alpha_{\rm eff} t}}{2\pi\Delta E}\bigl(2\tilde\alpha_{\rm eff}\mathrm{Ei}(\tilde\alpha_{\rm eff} t) \\
    & -(\tilde\alpha_{\rm eff}-i\Delta E)\mathrm{Ei}((\tilde\alpha_{\rm eff}-i\Delta E)t)-(\tilde\alpha_{\rm eff}+i\Delta E)\mathrm{Ei}((\tilde\alpha_{\rm eff}+i\Delta E)t)\bigr) \,, 
\end{split}
\end{equation}
which is dominated by 
\begin{equation}
    G_{\rm classical}  \approx e^{-\talpha_{\rm eff}t} \,, 
\end{equation}
in the slope regime with $\tilde\alpha_{\rm eff} t\sim \mathcal{O}(1)$. In addition, the bra--ket contribution generates the subsequent ramp and plateau and is derived as 
\begin{equation}
G_{\rm bra-ket} \approx  
\begin{cases}
\frac{1}{\THH^2}\left(\frac{2}{\tilde\alpha_{\rm eff}}t-\frac{e^{-\tilde\alpha_{\rm eff} (t+\THH)}+e^{\tilde\alpha_{\rm eff} (t-\THH)}-2e^{-\tilde\alpha_{\rm eff} t}}{\tilde{\alpha}_{\rm eff}^2}\right),\qquad t<\THH     \,, \\
\, \\ 
\frac{1}{\THH^2}\left(\frac{2\THH}{\tilde\alpha_{\rm eff}}-\frac{e^{-\tilde\alpha_{\rm eff} (t+\THH)}(e^{\tilde\alpha_{\rm eff} \THH}-1)^2}{\tilde{\alpha}_{\rm eff}^2}\right), \qquad t>\THH\,. 
\end{cases}
\end{equation}
The probability distribution therefore reproduces the leading slope-ramp-plateau structure of eq.~\eqref{eq:sloperampplateau_zero} and provides a complementary state-space interpretation of the same result.

The contributions from other sectors are parametrically subleading in the generating function with $\talpha_\pm\THH \gg 1$. This can be seen directly from
the operator--ket contribution. The sharp factor in eq.~\eqref{eq:Popket} localizes $t_{\CC_+}$ near $t$, so that
\begin{equation}
\begin{split}
    G_{\rm op-ket} &\approx  \int_{-\infty}^{\infty}
    d\CC_+\,d\CC_-\, e^{-\alpha_+|\CC_+|-\alpha_-|\CC_-|} P_{\rm op-ket}(\CC_+,\CC_-,t)
    \\
    &\approx
    \frac{e^{-\talpha_+t}}{\THH^2}
    \int_{-\infty}^{\infty}dt_{\CC_-}\,  e^{-\talpha_-|t_{\CC_-}|} \min\!\left[|t-t_{\CC_-}|,\THH\right].
\end{split}
\label{eq:Gopket_from_P}
\end{equation}
The remaining integral is precisely the quantum part of a codimension-one generating function, \ie 
\begin{equation}
    G_{\rm op-ket}
    \approx
    G_{\rm classical}^{(1)}(\talpha_+,t)\,
    G_{\rm quantum}^{(1)}(\talpha_-,t),
\end{equation}
which agrees with eq.~\eqref{eq:Gop-ket_result}. Exchanging $+\leftrightarrow-$ similarly gives
\begin{equation}
    G_{\rm bra-op}  \approx
    G_{\rm quantum}^{(1)}(\talpha_+,t)\,  G_{\rm classical}^{(1)}(\talpha_-,t),
\end{equation}
as found in eq.~\eqref{eq:Gbra-op_result}. The fully connected
probability is suppressed by $\mathcal{O}(\THH^{-2})$ and is likewise
subleading at finite $\talpha$.\footnote{We emphasize, however, that this hierarchy refers only to the generating function itself at fixed nonzero $\talpha$ with $\talpha_\pm\THH \gg 1$. As discussed in section~\ref{sec:reg}, all sectors of the derivative of generating function contribute at the same order to the regularized expectation value of the codimension-zero holographic complexity.}

In summary, we have shown that the fixed-$\mC$ probability distribution provides a state-space counterpart of the spectral generating-function analysis. Its semiclassical peak reproduces the linear growth of the black hole interior, while the connected spectral correlations generate broad nonperturbative tails. These tails are responsible for the non-normalizability of the continuous fixed-$\mC$ basis and hence for the divergence encountered in section~\ref{sec:reg}. More importantly, we now turn to the corresponding Hilbert space resolution of this problem by constructing proper holographic complexity operators from a complete orthonormal basis.

\subsection{Construction of holographic complexity operators}
\label{sec:operator}

The non-normalizability encountered in the previous subsection indicates
that the continuous family of fixed-$\CC$ states cannot be treated as a
complete set of independent states once nonperturbative corrections are
included. This is consistent with the finite-dimensional nature of the
microcanonical Hilbert space at finite $S_0$. While $\CC$ is treated as
a continuous variable in the semiclassical description, the number of
independent black hole microstates in the energy window $\mathcal I$ is
finite and given by
\begin{equation}
    Z
    \simeq
    e^{S_0}\bar D(E_0)\Delta E.
\end{equation}
The continuous fixed-$\CC$ description must therefore contain redundant
states at the nonperturbative level. This redundancy can be seen directly from the inner products between fixed-complexity states. Within the semiclassical continuum
approximation, distinct fixed-$\CC$ states are orthogonal. Euclidean wormhole corrections modify this relation by generating off-diagonal overlaps,
\begin{equation}
    \langle\CC'|\CC\rangle
    \approx 
    \delta(\CC-\CC')  +
    \,\raisebox{-0.29\height}{{
    \includegraphics[height=5em]{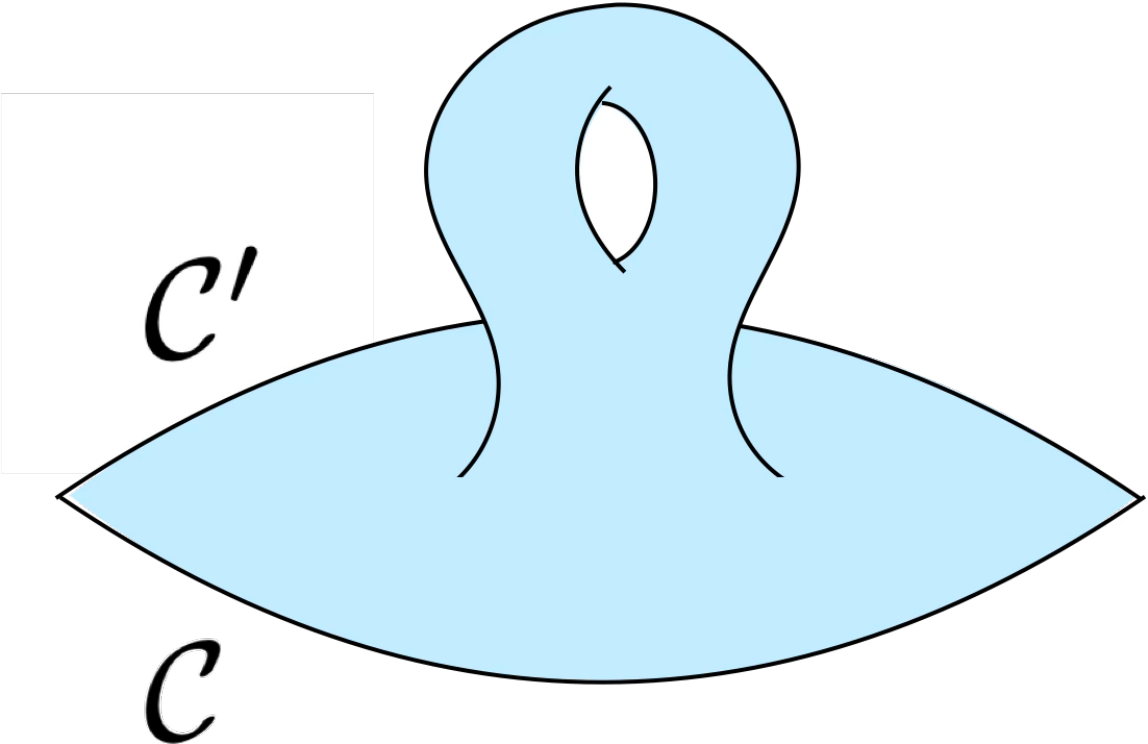}}}
    \,  +\cdots .
\label{eq:C_wormhole_overlap}
\end{equation}
Thus, the Gram matrix of the continuous fixed-$\CC$ family acquires nontrivial off-diagonal components \cite{Saad:2019pqd, Iliesiu:2021ari, Iliesiu:2024cnh}. Since this family contains infinitely many states while the underlying Hilbert space has only $Z$ independent directions, it becomes overcomplete. The same nonperturbative effects that produce the non-normalizable tails of $P_t(\CC)$ therefore also reveal the redundancy of fixed-complexity states.

A natural way to remove this redundancy is to extract a complete orthonormal basis from the infinite family of fixed-$\CC$ states \cite{Miyaji:2024ity, Balasubramanian:2024lqk, Miyaji:2025ucp}.  We denote the resulting discrete
orthonormal states by $|\CC_O\rangle$. To make the construction explicit, we first choose an ordered set of fixed-complexity states,
\begin{equation}
    |\CC_1\rangle,\,
    |\CC_2\rangle,\,
    |\CC_3\rangle,\ldots,
    \qquad
    \CC_1<\CC_2<\CC_3<\cdots,
\end{equation}
and apply the Gram--Schmidt procedure successively along this ordering. The first normalized state is simply given by 
\begin{equation}
    |\CC_{1,O}\rangle
    =
    \frac{|\CC_1\rangle}
    {\sqrt{\langle\CC_1|\CC_1\rangle}}\,.
\end{equation}
The $n$-th state is then obtained by subtracting from $|\CC_n\rangle$ its projections onto all previously constructed orthonormal states, namely 
\begin{equation}
    |\widetilde{\CC}_{n,O}\rangle
    :=
    |\CC_n\rangle
    -
    \sum_{m=1}^{n-1}
    |\CC_{m,O}\rangle
    \langle\CC_{m,O}|\CC_n\rangle,
\end{equation}
followed by normalization,
\begin{equation}
    |\CC_{n,O}\rangle
    :=
    \frac{|\widetilde{\CC}_{n,O}\rangle}
    {\sqrt{
        \langle\widetilde{\CC}_{n,O}
        |\widetilde{\CC}_{n,O}\rangle
    }}.
\label{eq:GS_complexity}
\end{equation}
By construction, 
\begin{equation}
    \langle\CC_{m,O}|\CC_{n,O}\rangle
    =
    \delta_{mn}.
\end{equation}
Because the nonperturbative microcanonical Hilbert space has only a finite dimension $Z$, this procedure can generate at most $Z$ linearly independent states. Beyond this point, any additional fixed-$\CC$ state lies in the span of the previously constructed
states, and its orthogonal component $|\widetilde{\CC}_{n,O}\rangle$ vanishes. In this sense, the Gram--Schmidt procedure extracts a complete finite orthonormal basis from the original over-complete and infinite fixed-$\CC$ family. 

The whole procedure is closely related to the Krylov state complexity \cite{Balasubramanian:2022tpr, Balasubramanian:2022dnj, Nandy:2024htc, Nandy:2024zcd, Balasubramanian:2024lqk, Rabinovici:2025otw}, when the $|\CC_{n,O}\rangle$ can be obtained by the action of the Hamiltonian on $|\CC_{m,O}\rangle$ with $m<n$. For the DSSYK model at $N\rightarrow\infty$, the chord number state indeed falls into this case \cite{Lin:2022rbf, Rabinovici:2023yex}. For non-perturbative chord number state in ETH matrix model for DSSYK, since these states are given by $q$-Hermite $n$-th order polynomial of the Hamiltonian, they also give the Krylov state complexity. On the other hand, in more generic operators this does not necessarily hold, and the basis is in general different from the Krylov basis. Furthermore, we also perform Gram-Schmidt orthogonalization not in increasing order in some cases. Examples are time-shift states \cite{Miyaji:2024ity, Miyaji:2025yvm} and chord number operator around finite chord number state \cite{Miyaji:2025ucp}. A related approach for angle variables was taken in \cite{Akers:2025ynh}. 

This construction has been carried out explicitly for fixed geodesic-length states in JT gravity and for fixed chord-number states in the double-scaled SYK model
\cite{Miyaji:2024ity,Miyaji:2025ucp}. The result depends on the ordering chosen for the original fixed-$\CC$ states. Our canonical prescription is to begin with the low-complexity states and orthogonalize successively toward larger $\CC$. Since the wormhole corrections are parametrically small in the semiclassical low-complexity regime, this choice preserves the classical interpretation of holographic complexity
eigenvalue as closely as possible.

\subsubsection{Codimension-one holographic complexity}
\label{sec:codim1_op}

We first consider the codimension-one holographic complexity measure. Given the
complete orthonormal basis $|\CC_O\rangle$, we can define the corresponding holographic complexity operator by
\begin{equation}
\label{eq:defineCO}
    \hat{\CC}_O := \sum_{\CC_O} \CC_O\,|\CC_O\rangle\langle\CC_O| \,. 
\end{equation}
This construction is closely analogous to Krylov complexity\footnote{For recent studies of Krylov complexity, see \eg \cite{Caputa:2021sib,Balasubramanian:2022tpr,Rabinovici:2022beu, Lin:2022rbf,Bhattacharjee:2022ave,Balasubramanian:2022dnj, Alishahiha:2022anw,Erdmenger:2023wjg,Hashimoto:2023swv, Caputa:2023vyr,Rabinovici:2023yex,Nandy:2024htc,
Camargo:2024deu,Nandy:2024zcd,Balasubramanian:2024lqk,
Ambrosini:2024sre,Baiguera:2025dkc,Rabinovici:2025otw,Craps:2025kub}.}, in the sense that the complexity operator is defined by assigning increasing complexity weight to an orthonormal basis. However, the orthonormal basis $|\CC_O\rangle$ constructed from fixed-complexity states can be more general and is not required to coincide with the Krylov basis. One crucial feature is that the Gram--Schmidt procedure is not unique, since it depends on the ordering chosen for the original overcomplete family. This ambiguity has a direct physical interpretation: different orthogonalization prescriptions generally define different quantum operators that could agree only in their common semiclassical regime, in close analogy with the choice of reconstruction map in non-isometric codes \cite{Akers:2021fut,Akers:2022qdl,Antonini:2024yif}. We adopt a canonical prescription in which the Gram--Schmidt procedure starts from the low-complexity states and proceeds successively toward larger $\CC$. Instead, the states $|\CC_O\rangle$ at larger complexity receive nonperturbative quantum modifications since they are adjusted to ensure orthogonality where Euclidean spacetime wormhole corrections modify quantum overlaps throughout the Gram--Schmidt procedure.

For the time-evolved TFD state, the probability of measuring the eigenvalue $\CC_O$ is
\begin{equation}\label{eq:P1O_definition}
   \begin{split}
      P^{(1)}(\CC_O,t)
    &:= \overline{ \left| \langle\CC_O|\mathrm{TFD}(t)\rangle \right|^2
    } \\
   &= \frac{e^{2S_0}}{Z} \int_{\mathcal I}dE_i\,dE_j\,
    e^{-iE_{ij}t}\,
    \overline{\langle D(E_i)D(E_j)\rangle}\,
    \phi_{E_i}^{*}(\CC_O)
    \phi_{E_j}(\CC_O).
   \end{split}
\end{equation}
Using this probability distribution, we can define the expectation value of the codimension-one holographic complexity operator with respect to the time-evolved TFD state as
\begin{equation}
\label{eq:C_O_expval}
    \langle\hat{\CC}_O\rangle
    :=  \langle\mathrm{TFD}(t)| \hat{\CC}_O |\mathrm{TFD}(t)\rangle = \sum_{\CC_O} \CC_O\,P^{(1)}(\CC_O,t)\,. 
\end{equation}
Since $|\CC_O\rangle$ is a complete orthonormal basis, the probability
distribution is now exactly normalized, namely 
\begin{equation}
   \boxed{ \sum_{\CC_O}P^{(1)}(\CC_O,t)=1 \,,} 
\label{eq:P1O_normalization}
\end{equation}
which is contrasted with the non-normalizable distribution defined with respect to the original continuous fixed-$\CC$ family.


In the semiclassical regime $t\ll\THH$, the spectral two-point function
as well as the normalized probability distribution $P^{(1)}(\CC_O,t)$ is dominated by its disconnected component. Moreover, for the low-complexity states relevant in this regime, corrections induced by the orthogonalization are non-perturbatively small. The dominating characteristic wavefunction structure \eqref{eq:wavefunctionexp}, and consequently the pole structure \eqref{eq:matrixelement}, therefore remain unchanged to leading semiclassical order. The discrete probability distribution is then well approximated by the continuum profile
\begin{equation}
\begin{split}
    P^{(1)}(\CC_O,t)
    \approx  \frac{2}{\pi M\Delta E}
    \frac{ \sin^2\left[ (t_{\CC_O}-t)\frac{\Delta E}{2}\right]}{(t_{\CC_O}-t)^2},
    \qquad   t_{\CC_O}:=\frac{\CC_O}{M}.
\end{split}
\label{eq:P1O_early}
\end{equation}
The distribution is sharply localized around $t_{\CC_O}=t$, showing that the expectation value of holographic complexity operator $\hat{\CC}_O$ reproduces the semiclassical linear growth, 
\begin{equation}
\label{eq:C_O_early}
    \langle\hat{\CC}_O\rangle  \big|_{t\ll\THH}  \approx   Mt \,, 
\end{equation}
at early times $t\ll\THH$.

As the evolution approaches the Heisenberg scale, the connected spectral correlations become comparable to the disconnected contribution. The finite dimension of the Hilbert space becomes essential in this regime. At sufficiently late times, the probability distribution can be organized as
\begin{equation}
\label{eq:P1O_late_general}
\begin{split}
    P^{(1)}(\CC_O,t)  \approx  \frac{1}{Z} +  \delta P^{(1)}(\CC_O,t) \,, 
\end{split}
\end{equation}
with 
\begin{equation}
 \delta P^{(1)}(\CC_O,t) :=  \frac{1}{Z}
    \int_{\mathcal I}dE_i\,dE_j\,  e^{-iE_{ij}t}\,  R_2(E_i,E_j)\,  \phi_{E_i}^{*}(\CC_O)
    \phi_{E_j}(\CC_O)\,.
\end{equation}
The first term originates from the diagonal part of the discrete energy sum. Its constant value $1/Z$ follows directly from the completeness of any complete orthonormal basis. The second term contains the off-diagonal spectral correlations.  Because chaotic spectra exhibit level repulsion,
\begin{equation}
    R_2(E_i,E_j)  \underset{E_i\to E_j}{\longrightarrow} 0,
\end{equation}
or more locally
\begin{equation}
    R_2(E_i,E_j) \sim  \mathcal{O}\!\left(|E_{ij}|^\beta\right),
\end{equation}
with $\beta=2$ for the GUE. Due to the rapidly oscillating phase $e^{-iE_{ij}t}$, the second term associated with a regular smooth kernel is always suppressed at late times. Schematically,
\begin{equation}
    P^{(1)}(\CC_O,t)  =  \frac{1}{Z}    +\mathcal{O}\left(\frac{1}{Z\, t}\right) ,  \qquad  t\gg\THH.
\label{eq:P1O_uniform}
\end{equation}
The probability distribution thus approaches the uniform distribution over the finite orthonormal basis. Although the probability at each fixed $\CC_O$ approaches its late-time value by dephasing, one may worry about the expectation value of the holographic complexity that requires an additional sum weighted by $\CC_O$. The corresponding off-diagonal contribution takes the form
\begin{equation}
\begin{split}
    \delta\langle\hat{\CC}_O\rangle  = \frac{1}{Z}
    \int_{\mathcal I}dE_i\,dE_j\, e^{-iE_{ij}t} R_2(E_i,E_j)
    \langle E_i|\hat{\CC}_O|E_j\rangle = \sum_{\CC_O} \CC_O\, \delta P^{(1)}(\CC_O,t) \,.
\end{split}
\end{equation}
In the semiclassical regime, the complexity matrix element inherits the near-diagonal double-pole structure
\begin{equation}
    \langle E_i|\hat{\CC}_O|E_j\rangle \sim  \frac{1}{E_{ij}^2},
\end{equation}
which is responsible for the classical linear growth. Consequently, the integrand cannot in general be regarded as a regular Fourier kernel near $E_{ij}=0$. The crucial role of the nonperturbative spectral correlations is to remove this singular contribution. As a result, we can find that the singular pole responsible for the semiclassical linear growth is thus canceled by level repulsion, \footnote{The role of level repulsion can also be understood directly from the probability distribution. To see this, consider first a hypothetical spectrum in which the connected spectral correlations are absent, so that the two-point function is replaced by its factorized contribution $R_2(E_i,E_j) \sim e^{2S_0}D(E_i)D(E_j)$. Using the semiclassical wavefunction \eqref{eq:wavefunction}, the corresponding off-diagonal probability takes the schematic form
\begin{equation*}
\delta P^{(1)}(\CC,t)  \approx  \frac{1}{\pi M}\int_{-\infty}^{+\infty} dE_{ij} \, e^{- i E_{ij} t}  \cos( t_{\mC} E_{ij})    \approx \frac{1}{M}\left( \delta (t- t_{\mC}) +\delta (t+t_{\mC}) \right)  \,, 
\end{equation*}
where we have taken the large-window limit for simplicity. For a finite microcanonical window, the delta functions are replaced by sharply localized peaks of width $\mathcal O(1/\Delta E)$.  The peak at $t_{\mC}=t$ therefore continues to propagate toward larger complexity and produces the linear growth $\langle\CC\rangle\sim Mt$. By contrast, using $R_2(E_i,E_j)\propto E_{ij}^2$, the near-diagonal contribution is then schematically transformed into
\begin{equation*}
    \delta P^{(1)}(\CC,t) \approx   -\frac{1}{M} \partial_t^2\left[ \delta(t-t_{\mC})+ \delta(t+t_{\mC}) \right]\,.
\end{equation*}
where the derivative automatically removes the contribution responsible for the linear growth.}
\begin{equation}
    R_2(E_i,E_j)
    \langle E_i|\hat{\CC}_O|E_j\rangle
    \underset{E_i\to E_j}{\sim} 
    \mathcal O(1).
\end{equation}

Once this singular contribution has been removed, the remaining off-diagonal terms are regular and are suppressed at late times. Consequently, the corresponding expectation value saturates,
\begin{equation}
\label{eq:C_O_late}
    \left.\langle\hat{\CC}_O\rangle\right|_{t\gg\THH} \approx \frac{1}{Z}\sum_{\CC_O}\CC_O \, .
\end{equation}
The existence of this late-time plateau is general, whereas its precise value is not universal. In particular, the plateau is determined by the spectrum of the regularized operator $\hat{\CC}_O$ and therefore depends on the prescription used to regularize the originally overcomplete fixed-complexity states \footnote{A precise example is provided by Krylov state complexity: for a finite Krylov basis of dimension $\mathcal{K}$ with eigenvalues $n=0,\ldots,\mathcal{K}-1$, an asymptotically uniform distribution gives
\begin{equation}
    \CC_{\rm K}^{\rm plateau} = \frac{1}{\mathcal{K}}
    \sum_{n=0}^{\mathcal{K}-1} n = \frac{\mathcal{K}-1}{2} \simeq \frac{\mathcal{K}}{2}.
\end{equation}}. It need not coincide with the plateau value obtained previously from a particular regularization of the holographic complexity generating function.

As a summary, we have shown that the proper holographic complexity operator $\hat{\CC}_O$ constructed from an orthonormal basis $\{ | \CC_O \rangle\}$ reproduces the universal behavior found in section~\ref{sec:gen}: its expectation value follows its semiclassical linear growth for exponentially long times, while the finite number of independent black hole microstates forces the evolution to saturate at late times. 

\subsubsection{Codimension-zero holographic complexity}
\label{sec:codim0_op}

Building on the previous construction of codimension-one holographic complexity operators $\hat{\mC}_O$, it is then straightforward to extend the construction to codimension-zero holographic complexity. As discussed in section~\ref{sec:gen}, a general codimension-zero gravitational observable decomposes into two boundary contributions associated with its future and past hypersurfaces $\Sigma_\pm$,
\begin{equation}
    \CC_{\rm Any}  =  \CC_+  + \CC_- \,. 
\end{equation}
The labels $+$ and $-$ refer to the future and past hypersurfaces $\Sigma_+$ and
$\Sigma_-$ as shown in figure~\ref{fig:CAnything}. At the operator level, we therefore introduce two independently orthogonalized codimension-one holographic complexity operators,
\begin{equation}
\label{eq:CpmO_operators}
    \hat{\CC}_{\pm,O}  := \sum_{\CC_{\pm,O}}  \CC_{\pm,O}\,  |\CC_{\pm,O}\rangle  \langle\CC_{\pm,O}|\,, 
\end{equation}
The two sets $\{|\CC_{+,O}\rangle\}$ and $\{|\CC_{-,O}\rangle\}$ are generally different complete orthonormal bases of the same nonperturbative microcanonical Hilbert space. They are therefore related by a unitary transformation, although the corresponding operators need not be simultaneously diagonalizable. 

The natural codimension-zero holographic complexity operator is thus given by their sum, 
\begin{equation}
\label{eq:CAnyO_operator}
    \hat{\CC}_{\rm Any,O} := \hat{\CC}_{+,O} + \hat{\CC}_{-,O} \,. 
\end{equation}
which immediately gives rise to 
\begin{equation}
\label{eq:CAnyOp}
\begin{split}
    \langle\hat{\CC}_{\rm Any,O}\rangle &= \langle\hat{\CC}_{+,O}\rangle  + \langle\hat{\CC}_{-,O}\rangle\,. 
\end{split}
\end{equation}
This relation is the operator counterpart of the reduction obtained from the regularized generating function in section~\ref{sec:reg}. The results about the codimension-one holographic complexity operator can therefore be applied to each term separately. Combining the two asymptotic regimes, we may summarize as
\begin{equation}
    \langle\hat{\CC}_{\rm Any,O}\rangle  \approx
    \begin{cases}
        (M_++M_-)t,
         & t\ll\THH,
        \\[4pt]
        \displaystyle
        \frac{1}{Z}\left(\sum_{\CC_{+,O}}\CC_{+,O}+\sum_{\CC_{-,O}}\CC_{-,O}\right),
        & t\gg\THH.
    \end{cases}
\label{eq:CAnyO_asymptotics}
\end{equation}
Thus, the orthonormal-basis construction provides a complementary operator interpretation of the same qualitative universal time evolution of holographic complexity obtained from the regularized generating functions, while clarifying how the late-time saturation arises within a finite-dimensional quantum Hilbert space.

The construction of the codimension-zero complexity operator $\hat{\CC}_{\rm Any,O}$ is particularly simple as a consequence of linearity. More importantly, its expectation value \eqref{eq:CAnyOp}, corresponding to the first moment of the complexity operator, depends only on two-point spectral correlations, even though the original codimension-zero generating function is governed by three-point spectral correlations. This reduction is special to the first moment and follows directly from the linearity of $\hat{\CC}_{\rm Any,O} =\hat{\CC}_{+,O}+\hat{\CC}_{-,O}$. Higher moments, by contrast, contain mixed products of $\hat{\CC}_{+,O}$ and $\hat{\CC}_{-,O}$ and therefore depend explicitly on the overlaps between the two orthonormal bases, $\langle \CC_{+,O}|\CC_{-,O}\rangle$.  For example, the mixed second moment is given by
\begin{equation}
\begin{split}
    \langle \hat{\CC}_{+,O}\hat{\CC}_{-,O} \rangle
    =
    \sum_{\CC_{\pm,O}} \CC_{+,O}\CC_{-,O}\,  P(\CC_{+,O},\CC_{-,O},t)\,,
\end{split}
\label{eq:Cpm_mixed_moment}
\end{equation}
where
\begin{equation}
\begin{split}
    P(\CC_{+,O},\CC_{-,O},t)
    &:=
    \overline{
    \langle\mathrm{TFD}(t)|\CC_{+,O}\rangle
    \langle\CC_{+,O}|\CC_{-,O}\rangle
    \langle\CC_{-,O}|\mathrm{TFD}(t)\rangle
    }.
\end{split}
\end{equation}
The overlap $\langle\CC_{+,O}|\CC_{-,O}\rangle$ couples the two orthonormal bases, so that this mixed moment involves the three-point spectral correlations discussed in section~\ref{sec:cod}. The same information reappears at quadratic order in the generating function $\langle e^{-\alpha_+ \CC_+ -\alpha_- \CC_-} \rangle$, through the ordered mixed contribution. Thus, the codimension-zero operator retains the additional spectral information carried by the three-point function, even though this information cancels from its first moment.


\section{Discussion}
\label{sec:discussion}

In this work, we have investigated the nonperturbative time evolution of codimension-zero holographic complexity within the complexity=anything framework. Starting from the linear decomposition $\CC_{\rm Any}=\CC_++\CC_-$, we formulated the corresponding generating function of holographic measures in the energy eigenbasis and showed that, unlike the codimension-one case, its spectral representation is governed by three-point correlations of the density of states. The decomposition of the three-point spectral correlator into disconnected, pairwise connected, and fully connected components provides a natural organization of the corresponding gravitational contributions in terms of Euclidean spacetime wormholes. The disconnected sector dominates the semiclassical linear growth, while the connected sectors encode nonperturbative quantum corrections that result in the late-time plateau after the Heisenberg time.

Combining the spectral correlations with the characteristic pole structure
of the matrix elements of complexity operator \eqref{eq:intro_pole_structure}, we derived the universal time evolution of codimension-zero holographic complexity from the regularized generating function 
\begin{equation}
G^{(0)}(\alpha_\pm;t)=
\langle e^{-\alpha_+ \CC_+ -\alpha_- \CC_-} \rangle
\end{equation}
with the expectation value of complexity obtained by taking the $\alpha_\pm\to0$ limit. As shown in section~\ref{sec:reg}, the semiclassical linear growth originates from the particular pole structure of the matrix elements \eqref{eq:intro_pole_structure} and persists up to times of order the
Heisenberg time scale. The connected spectral correlations become essential at this scale and modify the semiclassical result nonperturbatively. In particular, spectral level repulsion,
\begin{equation}
 R_2(E_i,E_j) \to 0 , \quad R_{3}(E_i,E_j,E_k)\to 0\,, \qquad \text{as} \quad   E_i\to E_j\,, 
\end{equation}
precisely cancels the near-diagonal contribution responsible for the linear growth of complexity, thereby producing the expected late-time plateau. 

We have also developed a complementary operator description in terms of fixed-complexity states $|\mC\rangle$. At the semiclassical level, these states form a continuously labeled family whose characteristic oscillatory wavefunctions \eqref{eq:wavefunction} reproduce the crucial pole structure of the generating function. Once nonperturbative Euclidean wormhole corrections are included, however, distinct fixed-complexity states acquire nontrivial overlaps, rendering the infinite family of fixed-complexity states $\{|\mC\rangle\}$ overcomplete. The key step is therefore to extract a complete orthonormal basis $\{|\mC_O\rangle\}$ through the Gram--Schmidt procedure, with respect to which well-defined holographic complexity operators $\hat{\mC}_O$ and $\hat{\mC}_{\rm Any,O}$ can be naturally constructed for the codimension-one and codimension-zero cases, respectively. This construction thus provides a direct quantum-mechanical interpretation of the late-time saturation in terms of the finite number of black-hole microstates.

The framework developed here raises several natural questions about the quantum structure of holographic complexity. In the following, we conclude by discussing a number of open questions and related directions that appear important or interesting from our perspective.

\subsection*{Switchback Effect}

In the main text, we have focused on the origin of the linear growth of holographic complexity and its nonperturbative quantum corrections at late times. Another characteristic property expected of holographic complexity is the switchback effect~\cite{Stanford:2014jda}, which provides an independent diagnostic of its dynamics. This behavior is also shared by the infinite family of codimension-one and codimension-zero holographic complexity measures in the complexity=anything proposal \cite{Belin:2021bga,Belin:2022xmt}. The switchback effect can be probed by considering perturbed TFD states of the form
\begin{equation}\label{eq:PertTFD}
\left|\text{TFD}\left(t_{\mathrm{L}}, t_{\mathrm{R}}\right)\right\rangle_{\rm pert} :=	e^{-i H_{\mathrm{L}} t_{\mathrm{L}}-i H_{\mathrm{R}} t_{\mathrm{R}}} \mathcal{O}\left(t_n\right) \ldots \mathcal{O}\left(t_1\right)\left|  \text{TFD}(0)\right\rangle \,,
\end{equation}
where $\mathcal{O}(t_i)$ denotes a simple boundary operator inserted at time $t_i$. When the perturbations are sufficiently separated, the semiclassical complexity follows the corresponding folded time contour, with a delay of order the scrambling time $t_\ast$ associated with each switchback. In particular, assuming $|t_{i+1}-t_i|\gg t_\ast$, the complexity of the perturbed TFD state takes the characteristic form
\begin{equation}
 \CC_{\rm pert}  \sim  \left|t_{\mt{R}}+t_1\right| +\left|t_2-t_1\right| +\cdots +\left|t_{\mt{L}}-t_n\right| -2n\,t_\ast \,.
\end{equation}
up to an overall normalization and time-independent contributions.

A natural generalization of the framework developed here is therefore to evaluate the generating function, and correspondingly the expectation value of the holographic complexity operator, in the perturbed TFD state:
\begin{equation}\label{eq:expPTFD}
    \left\langle e^{-\alpha\mC}\right\rangle_{\rm pert} :={}_{\rm pert}\!\left\langle
    \mathrm{TFD}(t_{\mt{L}},t_{\mt{R}})  \right| \widehat{e^{-\alpha\mC}} \left|
    \mathrm{TFD}(t_{\mt{L}},t_{\mt{R}})  \right\rangle_{\rm pert}.
\end{equation}
For a single perturbation, the key observation is that this quantity is naturally governed by a four-point function. More interestingly, the corresponding crossed amplitude provides the energy-basis kernel that generalizes the matrix elements $\langle E_i|\widehat{e^{-\alpha\mC}}|E_j\rangle$ considered in the main text, and exhibits the same characteristic pole structure,
\begin{equation}
    \boxed{
    \mathcal{A}_{\mathrm{cross}}^{\mathrm{pole}}
    \simeq
    \frac{\mathcal{R}(\alpha)}
    {\left(M_{\mt{L}}\alpha-i\omega_{\mt{L}}\right)
     \left(M_{\mt{R}}\alpha+i\omega_{\mt{R}}\right)}
    }\,,
\end{equation}
where $\omega_{\mt{L}}$ and $\omega_{\mt{R}}$ denote the relevant energy differences, analogous to $E_{ij}$ in the present analysis. In a related work for a different motivation, we find that these characteristic poles continue to determine the linear growth along the folded time contour, while their nontrivial residues encode the shift by the scrambling time and hence produce the expected switchback effect. A detailed derivation of this mechanism will be presented in a separate work in progress \cite{WIP2027}.

\subsection*{Wave Functions in Higher-Dimensional Gravity}

In the main text, we pointed out that the oscillatory wave function \eqref{eq:black_white_branches} and the associated pole structure \eqref{eq:pole_from_fixedC_wavefunction} are responsible for the linear growth of complexity measures. Such microscopic quantum descriptions are generally unavailable in full gravity except for a few low-dimensional examples discussed in Appendix~\ref{app:examples}. 
Nevertheless, we can make progress in higher dimensions by considering a symmetry-reduced sector. As a concrete example, consider the spherically symmetric sector of a four-dimensional asymptotically flat two-sided black hole. In \cite{Kuchar:1994zk}, the author showed that, after the reduction, the remaining phase space variables are the Schwarzschild mass $m$ and its conjugate boundary timeshift\footnote{This can be seen as an implementation of Birkhoff's theorem, which states that the Schwarzschild metric is the unique spherically symmetric vacuum solution. Interestingly, this phase space structure appears analogous to that of JT gravity \cite{Harlow:2018tqv}. In fact, similar reduction was known in general dilaton gravity \cite{Cavaglia:1998xj}.}. With future-directed
boundary times $t_\pm$ and compatible time origins, the reduced symplectic
form and Hamiltonian are
\begin{equation}
  \omega=dT\wedge dm,\qquad T:=t_++t_-,
  \qquad H=m.
\end{equation}

Consider the two-ended maximal slices anchored to the boundaries, which is studied in \cite{Beig:1997fp}, and define the volume $V(m,T)$ by subtracting a time-independent divergence. At late times, this volume grows linearly in the boundary timeshift as
\begin{equation}
  V(m,T)\to3\sqrt{3}\pi m^2T.
  \label{eq:higher_dimension_volume}
\end{equation}
The same behavior was obtained in \cite{Christodoulou:2014yia} for the interior volume anchored to the horizon. The anchoring differs, but the subtracted exterior volume approaches a finite limit and thus does not change the growth rate in this case.

On the expanding branch $T>0$, the leading relation $V\simeq 3\sqrt{3}\pi m^2T$ allows us to use the volume as a canonical coordinate. Requiring
$\omega=dV\wedge dP_V$, we can choose
\begin{equation}
  P_V\simeq-\frac{1}{3\sqrt{3}\pi m}\,,
\end{equation}
which leads to the following expression for the Hamiltonian at late times
\begin{equation}
    H_T(V,P_V)\simeq-\frac{1}{3\sqrt{3}\pi P_V}\,.
\end{equation}
The corresponding wave function at fixed $m$ is therefore
\begin{equation}
  \psi_m(V)\sim \exp\!\left(-\frac{iV}{3\sqrt{3}\pi m}\right)\,.
\end{equation}
The contribution of this asymptotic component to the matrix
element of $e^{-\alpha\hat V}$ takes the form
\begin{align}
\int_{-\infty}^{\infty}dV\,
    e^{-\alpha |V|}\psi^*_{m_1}(V)\psi_{m_2}(V)\sim
    \frac{\alpha}{\alpha^2+q_{12}^2},
  \qquad q_{12}:=\frac{1}{3\sqrt{3}\pi}\left(\frac1{m_1}-\frac1{m_2}\right),
\end{align}
where the integration region is taken to be $(-\infty,\infty)$ following the approach described in section~\ref{sec:wavefunction_pole}. With $\bar m=(m_1+m_2)/2$ and $m_{12}=m_1-m_2$, we can rewrite $q_{12}$ as
\begin{equation}
    q_{12}=-\frac{1}{3\sqrt3 \pi} \left(\frac{m_{12}}{\bar m^2-m_{12}^2/4}\right)\,, 
\end{equation}
which gives 
\begin{equation}
\frac{\alpha}{\alpha^2+q_{12}^2}=\frac{27\pi^2\alpha}{27\pi^2\alpha^2+\left(\frac{m_{12}}{\bar m^2-m_{12}^2/4}\right)^2}\,.
\end{equation}
In the semiclassical regime $\bar m\gg1$, this matrix element has poles
at $m_{12}=\pm i\alpha\,3\sqrt3\pi\bar m^2$. Combined with the
time-evolution phase $e^{im_{12}T}$, this pole yields
\begin{equation}
   \boxed{ \langle e^{-\alpha\hat V}\rangle \approx e^{-\alpha 3\sqrt3 \pi\bar m^2T}} 
\end{equation}
in agreement with the classical linear growth of the maximal volume in \eqref{eq:higher_dimension_volume}.

This demonstrates the emergence of the pole structure in symmetry-reduced higher-dimensional gravity. This type of reduction has also been developed for a negative cosmological constant, for example in \cite{Louko:1996dw}. It would be interesting to extend this analysis to asymptotically AdS black holes and investigate which specific geometric conditions are related to the pole structure of holographic complexity measures.

\subsection*{Quantum Corrections to Holographic Complexity}
Our analysis in this work has focused on nonperturbative quantum corrections associated with Euclidean spacetime wormholes. A broader question is how these effects fit together with other quantum corrections to holographic complexity. Classical holographic complexity proposals, including complexity=anything, are formulated in terms of geometric quantities in the bulk and scale parametrically as $1/\GN$. More generally, motivated by the structure of the generalized entropy prescription for holographic entanglement entropy, we expect the full quantum holographic complexity to receive contributions from several distinct sources. Schematically,
\begin{equation}
   \boxed{ \mC_{\rm holo}
    \sim
    \underbrace{ \frac{W_{\rm grav}}{\GN}}_{\text{gravitational sector}}
    +
    \underbrace{\mC_{\rm bulk}}_{\text{bulk quantum fields}}
    +
    \underbrace{\sum_{n\geq1} \,\mC_{\rm wh}^{(n)}}_{\text{Euclidean wormholes}}
    +\cdots ,}
\end{equation}
These three contributions have different physical origins and appear at parametrically distinct orders. Denoting $\mC_{\rm grav}\equiv W_{\rm grav}/\GN$, their schematic scaling is
\begin{equation}
    \mC_{\rm grav} \sim \mathcal O\left(\frac{1}{\GN}\right),
    \qquad
    \mC_{\rm bulk} \sim \mathcal O(\GN^0),
    \qquad 
    \mC_{\rm wh}^{(n)} \sim \mathcal O\left(e^{-nS_0}\right)\,. 
\end{equation}

The first term $\mC_{\rm grav}$ represents the leading gravitational contribution. In Einstein gravity, $W_{\rm grav}$ reduces to the classical geometric functional
specified by the complexity=anything proposal. In a more general higher-curvature theory, however, the gravitational functional itself can receive additional local geometric corrections, including terms constructed from intrinsic and extrinsic curvatures. Schematically,
\begin{equation}
   W_{\rm grav} = W_{\rm Any} +  \sum_k  \lambda_k\,W_{\rm{higher-curvature}}^{(k)}
    +\cdots \,, 
\end{equation}
where $\lambda_k$ denote the corresponding higher-derivative couplings. This structure is closely analogous to the replacement of the area functional by the Wald--Dong entropy functional \cite{Iyer:1994ys,Jacobson:1993vj,Dong:2013qoa} in higher-curvature theories of gravity.

The second term, $\mC_{\rm bulk}$, represents the quantum corrections from quantum fields in the bulk and plays a role analogous to the bulk entanglement entropy appearing in the generalized entropy and quantum extremal surface prescription. These first two ingredients have already arisen naturally in previous studies of holographic complexity. In particular, in the doubly holographic construction of Ref.~\cite{Hernandez:2020nem}, the holographic complexity prescription was generalized to higher-curvature gravity, leading to a generalized geometric functional determined by the effective gravitational action. The same analysis also suggested an additional contribution
$\mC_{\rm bulk}$ associated with the quantum state of bulk fields. While the gravitational contribution can be systematically generalized within the bulk effective theory, the precise quantum-information quantity captured by $\mC_{\rm bulk}$ remains less well understood.

The final term, $\mC_{\rm wh}$, represents a qualitatively different class of quantum corrections emphasized in the present work. The connected spectral correlations considered here correspond to Euclidean spacetime wormholes of different topologies in the gravitational path integral and generate nonperturbative corrections of order $e^{-nS_0}$. Unlike local higher-curvature corrections or perturbative quantum-field contributions in the bulk, these effects probe correlations between
different gravitational saddles and become essential at exponentially late times. In particular, they modify the semiclassical linear growth and give rise to the late-time saturation discussed in the main text.

It would therefore be interesting to develop a unified framework in which higher-curvature gravitational corrections, perturbative bulk quantum effects, and nonperturbative Euclidean-wormhole contributions are incorporated simultaneously. Such a formulation may provide a more complete quantum prescription for holographic complexity, analogous in spirit to the passage from classical extremal area to generalized entropy and quantum extremal surfaces.

\subsection*{Complexity of Mixed States and Subregions}

Another natural direction is to extend the present construction from pure states to quantum complexity of mixed states and holographic complexity of boundary subregions. Consider a boundary subregion $A$ described by a reduced density matrix $\hat{\rho}_A=\operatorname{Tr}_{A^c}\rho$. A natural mixed-state analogue of the generating function considered in this work is
\begin{equation}
    G_A(\alpha) :=\operatorname{Tr}_A \left(\hat{\rho}_A {e^{-\alpha \hat{\mC}_A}} \right)\,, 
\end{equation}
where $\hat{\mC}_A$ denotes a holographic complexity operator associated intrinsically with the subregion $A$. Its first moment is then obtained as
\begin{equation}
    \langle\hat{\mC}_A\rangle_{\hat{\rho}_A} = -\left. \partial_\alpha G_A(\alpha) \right|_{\alpha=0}.
\end{equation}
Holographically, one may expect the semiclassical expectation value of $\hat{\mC}_A$ to be related to an appropriate geometric complexity functional defined within the entanglement wedge of the boundary subregion $A$.

A different possibility is to induce a subregion generating operator directly from the reduction of the original holographic complexity. Formally, one may define
\begin{equation}
    \widehat{\mathcal G}_A(\alpha) := \mathcal E_A\left( \widehat{e^{-\alpha\mC}} \right),
\end{equation}
where $\mathcal E_A$ denotes an appropriate reduction map from operators on the full Hilbert space to operators acting on $\mathcal H_A$. The corresponding generating function is thus 
\begin{equation}
   \left\langle e^{-\alpha\mC}\right\rangle_{\hat{\rho}_A}  := \operatorname{Tr}_A \left[ \rho_A\, \widehat{\mathcal G}_A(\alpha) \right].
\end{equation}
For a finite-dimensional factorized Hilbert space, one possible and simple choice is the normalized partial trace
\begin{equation}
    \mathcal E_A(\mathcal{O})  = \frac{1}{\dim\mathcal H_{A^c}} \operatorname{Tr}_{A^c}\mathcal{O} \,. 
\end{equation}
Unlike ordinary local observables, however, complexity is intrinsically nonlocal, and there is no a priori reason for this reduction map to be unique. Determining whether a preferred prescription exists, and whether it admits a natural holographic interpretation in terms of subregion complexity, is therefore an interesting open question.

Another possibility is to formulate the problem directly in terms of the
entanglement spectrum, or equivalently the modular Hamiltonian
$K_A=-\log\rho_A$. In its eigenbasis, we can write the spectral decomposition as
\begin{equation}
    \rho_A = \sum_a p_a  | a \rangle \langle a | =  \sum_a e^{-k_a}|k_a\rangle\langle k_a| \,,
\end{equation}
where $k_a$ denote the eigenvalues of the modular Hamiltonian. Correspondingly, the matrix elements $\langle k_a|\widehat{e^{-\alpha{\mC}_A}}|k_b\rangle$ would play a role analogous to the energy-basis matrix elements studied in the present work. It would be interesting to determine whether the pole structure and spectral correlations developed here have a corresponding formulation in terms of the modular spectrum. Potentially, this framework may relate the time evolution of subregion complexity to correlations in the modular spectrum.

Alternatively, one may purify the reduced density matrix $\rho_A$ and apply the present construction in the enlarged Hilbert space $\mathcal H_A\otimes\mathcal H_{A'}$. For the canonical purification $|\sqrt{\rho_A}\rangle$, for example, we have 
\begin{equation}
    |\sqrt{\rho_A}\rangle = \sum_a e^{-k_a/2}\, |k_a\rangle_A |k_a\rangle_{A'} \,. 
\end{equation}
This has precisely the same spectral structure as TFD,
\begin{equation}
    |\mathrm{TFD}_\beta\rangle = \frac{1}{\sqrt{Z(\beta)}} \sum_i e^{-\beta E_i/2} |E_i\rangle_{\mt{L}} |E_i\rangle_{\mt{R}} \,. 
\end{equation}
Thus, in the purification construction, the modular spectrum of $\rho_A$ plays a role analogous to the energy spectrum in the TFD construction by $k_a \leftrightarrow \beta E_i$. In particular, for a thermal density matrix $\rho_\beta=e^{-\beta H}/Z(\beta)$, the canonical purification reduces precisely to the TFD state.

This analogy suggests that the spectral construction developed in this work may admit a natural generalization in which energy eigenstates and energy differences are replaced by modular eigenstates and modular energy differences. For example, one may consider
\begin{equation}
    G_A^{\rm pur}(\alpha)  =  \langle\sqrt{\rho_A}| e^{-\alpha {\mC}_{AA'}} |\sqrt{\rho_A}\rangle ,
\end{equation}
whose spectral representation involves matrix elements
\begin{equation}
    \langle k_a,k_a| \widehat{e^{-\alpha {\mC}_{AA'}}} |k_b,k_b\rangle \,. 
\end{equation}
It is then natural to ask whether these matrix elements exhibit a characteristic pole structure in the modular-energy difference $k_a-k_b$, and whether correlations of the modular spectrum play a role analogous to the energy-level correlations considered in the present work.

The purification construction also makes contact with the notion of purification complexity developed in the literature \cite{Agon:2018zso,Caceres:2019pgf,Ruan:2020vze}. Rather than assigning the complexity of a particular purification to the mixed state, one defines the purification complexity by minimizing over all possible purifications of $\rho_A$. The canonical purification $|\sqrt{\rho_A}\rangle$ introduced above therefore provides a natural starting point, but need not coincide with the optimal purification. It would be interesting to determine whether the spectral construction can be extended to this optimization problem, and whether the characteristic pole structure admits a formulation that is invariant under different choices of purification.

\subsection*{Beyond Spectral Ensemble Averaging}
Before closing the discussion, we would like to highlight an open question concerning the ensemble-averaging procedure adopted in our analysis. In our present analysis, the ensemble average is applied only to the spectral density correlations, while the matrix elements of the complexity generating operator are treated as universal smooth functions of the energies. For example, the codimension-one generating function is defined as 
\begin{equation} 
\begin{split}
    \langle e^{-\alpha\CC} \rangle&=\frac{e^{2S_0}}{Z}\int_{\mathcal{I}^2} dE_idE_j\,e^{-iE_{ij}t} \overline{\langle  D(E_i)D(E_j)\rangle} \times  M_\alpha (E_i, E_j)\,. 
\end{split}
\end{equation}
More precisely, this formulation assumes that, after coarse graining, the matrix element $M_\alpha(E_i,E_j)$ can be regarded as a universal smooth function that is statistically independent of the microscopic spectral fluctuations.

More generally, however, the spectrum and the complexity matrix elements may fluctuate jointly across the ensemble. A more complete ensemble average would then take the form
\begin{equation}
    \overline{G(\alpha,t)}  = \frac{e^{2S_0}}{Z} \int_{\mathcal I^2}dE_i\,dE_j\,
    e^{-iE_{ij}t}\,\overline{ \langle D(E_i)D(E_j) \, M_\alpha (E_i, E_j) \rangle} \,.
\end{equation}
The approximation adopted in the present work therefore amounts to neglecting connected correlations between the microscopic spectrum and the complexity matrix elements,
\begin{equation}
\begin{split}
\overline{\langle D_iD_jM_{ij} \rangle} = \overline{ \langle D_iD_j \rangle } \cdot\overline{M_{ij}} +\overline{ \langle D_iD_jM_{ij} \rangle}_{\rm conn}.
\end{split}
\end{equation}
Understanding the effects of the connected term is an interesting open question. In particular, if these mixed correlations modify the characteristic pole structure or its residue, they may generate additional quantum corrections to the time evolution of holographic complexity.

This distinction may also admit a gravitational interpretation. Averaging only the spectral density correlator isolates the Euclidean-wormhole contributions associated with spectral correlations, while the complexity operator insertion is treated through a fixed matrix element. Averaging the full product would instead capture possible correlations between the wormhole sector and the complexity insertion. In the gravitational path integral, these effects may be associated with more general connected spacetime geometries, in which the complexity operator insertion is incorporated directly into the Euclidean-wormhole configuration responsible for the spectral correlations.

A related but distinct subtlety concerns the distinction between annealed and quenched ensemble averaging. In general, the ensemble average of a normalized quantity need not coincide with the ratio of the corresponding ensemble averages, \ie
\begin{equation}
    \overline{\frac{\langle D_iD_jM_{ij}\rangle}{Z}}
    \qquad \text{vs.} \qquad
    \frac{\overline{\langle D_iD_jM_{ij}\rangle}}{\overline{Z}} \,.
\end{equation}
These two prescriptions generally differ because the numerator and normalization factor may fluctuate jointly across the ensemble. In our analysis, the microcanonical normalization $Z$ is treated as fixed under ensemble averaging, corresponding to a microcanonical sector with a fixed number of states. Under this assumption, the distinction does not arise. However, one may allow $Z$ to fluctuate, which could introduce additional corrections to the generating function and its late-time behavior. This may provide another interesting direction for future investigation.
\acknowledgments
We are grateful to Vijay Balasubramanian, Pawel Caputa, Bowen Chen, Michal Heller, Johanna Erdmenger, Jonathan Karl, Kazumi Okuyama, Robert Myers, Jacopo Papalini, Le-Chen Qu, Tim Schuhmann, Tadashi Takayanagi, Wayne Weng, and Zhuo-Yu Xian for valuable discussions and comments. MM is supported by JSPS KAKENHI Grant Numbers 24K17044 and 25K00997. SMR is supported by the National Natural Science Foundation of China (Grant No. 12675056) and Peking University under startup Grant No. 7101303985. KY and SS are supported by JST SPRING, grant number JPMJSP2125, ``THERS Make New Standards Program for the Next Generation Researchers''. 

\appendix 

\section{Fully Connected Three-Kernel Contribution}
\label{app:KKK}

For completeness, we present here the explicit result for the contribution involving the product of three kernels, $G_{K_{12}K_{23}K_{31}}$. To make the expression more compact, we introduce the dimensionless variables
\begin{equation}
    x:=\talpha_+\THH, \qquad y:=\talpha_-\THH, \qquad \tau:=\frac{t}{\THH}.
\end{equation}
For $t>\THH$, the expression takes the relatively simple form
\begin{equation}
\label{eq:GKKK_late}
    G_{K_{12}K_{23}K_{31}} =
    \frac{2e^{-(x+y)(\tau+1)}}{xy(x+y)} \left(e^x-1\right)\left(e^y-1\right) \left(e^{x+y}-1\right).
\end{equation}
For $t\leq\THH$, it is convenient to write
\begin{equation}
\label{eq:GKKK_early}
   G_{K_{12}K_{23}K_{31}}
    =\frac{2e^{-(x+y)(\tau+1)}}{x^2y^2(x^2-y^2)}\,
    \left( \mathcal A(x,y;\tau) -\mathcal A(y,x;\tau) \right),
\end{equation}
where
\begin{equation}
\label{eq:GKKK_A}
\begin{split}
    \mathcal A(x,y;\tau) ={}& xy(x+y)e^{x+2y\tau}
    \\[2pt]
    &+ x^2y
    \Big[-1-e^{2(x+y)\tau}+e^x+e^y +2e^{y(\tau+1)} \left(e^{2x\tau}-e^x\right)\Big]
    \\[2pt]
    &+
    2x^3e^{x(\tau+1)}
    \Big[e^{y\tau}+e^{2y\tau}-e^y+e^{y(\tau+1)}\left(-1+2y(1-\tau)\right)\Big].
\end{split}
\end{equation}

The two branches \eqref{eq:GKKK_late} and \eqref{eq:GKKK_early} match continuously at $t=\THH$. Although \eqref{eq:GKKK_early} contains an apparent singularity at $x=y$ (\ie $\talpha_+ = \talpha_-$), it is removable because the numerator is also antisymmetric under $x\leftrightarrow y$. The limit $\talpha_+=\talpha_-$ is therefore well defined. Moreover, \eqref{eq:GKKK_late} makes explicit that, for $\talpha_\pm>0$, the fully connected three-kernel contribution is exponentially suppressed at late times as $e^{-(\talpha_++\talpha_-)t}$.

\section{Some Explicit Examples}\label{app:examples}
There are some particular examples where we can explicitly compute the matrix element of the holographic complexity measures. In this appendix, we aim to show that these different matrix elements present the characteristic pole structure which determines the linear growth of complexity as discussed in the main text. 

\subsection{Geodesic length in JT gravity}\label{sec:JTlength}
In the two-dimensional two-sided eternal black hole, the renormalized geodesic length connecting left and right asymptotic boundaries can be computed as 
\begin{equation}
    \ell_{\rm classical}=2\log{\cosh{\frac{\pi}{\beta}t}}\,.
\end{equation}
The above expression shows that the geodesic length grows linearly at sufficiently late time $t\gg\beta$, \ie
\begin{equation}\label{eq:JTlinear}
    \ell_{\rm classical} \approx \frac{2\pi}{\beta}t\,,\qquad t\gg\beta\,.
\end{equation}
This linear growth implies that the geodesic length is a candidate of holographic complexity. This idea is verified in the context of DSSYK \cite{Rabinovici:2023yex}.

In this section, we consider JT gravity \cite{Jackiw:1984je,Teitelboim:1983ux}, where we have the quantum description of two-dimensional AdS space, and we show that the matrix element of geodesic length possesses the expected pole structure we discussed in the main text.

In JT gravity, the dynamics of geodesic length $\ell$ is governed by the Liouville Hamiltonian
\begin{equation}
    H=\frac{1}{2C}(-\partial_\ell^2+4e^{-\ell})\,,
\end{equation}
where $C$ is a Schwarzian coupling constant.
The corresponding wave function is 
\begin{equation}\label{eq:JT_wavefunction}
    \psi_E(\ell)=\sqrt{8e^{-S_0}}K_{2i\sqrt{2CE}}(4e^{-\frac{\ell}{2}})\,.
\end{equation}
Then the matrix element of the generating function of geodesic length can be expressed as follows.
\begin{equation}
\begin{split}
    \langle E_1| \widehat{e^{-\alpha\ell}} |E_2\rangle&=\int_{-\infty}^{\infty}d\ell\,e^{-\alpha\ell}\psi_{E_1}^*(\ell)\psi_{E_2}(\ell)\\
    &=\frac{e^{-S_0}}{2^{2\alpha-1}}\frac{|\Gamma(\alpha+ i\sqrt{2CE_1}+ i\sqrt{2CE_2})|^2|\Gamma(\alpha+ i\sqrt{2CE_1}- i\sqrt{2CE_2})|^2}{\Gamma(2\alpha)}\,.
\end{split}
\end{equation}
Since Gamma function has simple poles along the non-positive integers
\begin{equation}
    \Gamma(z)\sim\frac{(-1)^n}{n!}\frac{1}{z+n}\qquad \text{at} \qquad n=0,1,2,\ldots\,,
\end{equation}
there is an infinite sequence of simple poles on the complex plane
\begin{equation}
    \sqrt{E_1}+\sqrt{E_2}=\pm i (\alpha+n)\frac{1}{\sqrt{2C}},\qquad \sqrt{E_1}-\sqrt{E_2}=\pm i (\alpha+n)\frac{1}{\sqrt{2C}}, \qquad n=0,1,2,\ldots\,.
\end{equation}
At late times, the near-diagonal contribution $E_1\approx E_2$ becomes dominant. We thus can focus on the poles on the complex $E_{12}=E_1-E_2$ plane, which are approximately located at 
\begin{equation}\label{eq:JTpole}
    E_{12} \approx 2\sqrt{\bar E}(\sqrt{E_1}-\sqrt{E_2})=\pm i (\alpha+n)\sqrt{\frac{2\bar E}{C}}\,,
\end{equation}
where $\bar E=\frac{E_1+E_2}{2}$. 

As stated in the main text, this pole combined with the time evolution factor $e^{-iE_{12}t}$ results in the exponential decay $e^{-(\alpha+n)\sqrt{\frac{2E_0}{C}}t}$. This expression shows that the poles with $n\geq1$ are exponentially sub-leading contributions. Therefore, the leading contribution comes from the $n=0$ poles. Correspondingly, the matrix element can be approximated by the leading $n=0$ poles:
\begin{equation}\label{eq:JTpole1}
\begin{split}
    \langle E_1| \widehat{e^{-\alpha\ell}} |E_2\rangle
    &\approx\frac{e^{-S_0}}{2^{2\alpha-1}}|\Gamma(\alpha+ 2i\sqrt{2C\bar E})|^2\frac{\frac{4\bar E}{C}\alpha}{E_{12}^2+\frac{2\bar E}{C}\alpha^2}\,.
\end{split}
\end{equation}
In the semi-classical regime $C\bar E\gg1$, the gamma functions have the following asymptotic form which is proportional to the inverse of the density of states of JT gravity.
\begin{equation}
\begin{split}
    |\Gamma(\alpha+ 2i\sqrt{2C\bar E})|^2&\approx2\pi(2\sqrt{2C\bar E})^{2\alpha-1}e^{-2\pi\sqrt{2C\bar E}}\\
    &\propto1/D(\bar E)\,.
\end{split}
\end{equation}
Therefore, up to normalization, the matrix element takes a form
\begin{equation}\label{eq:JTpole2}
    \langle E_1| \widehat{e^{-\alpha\ell}} |E_2\rangle\sim \frac{\alpha}{\pi e^{S_0} D(\bar E)} \frac{1}{(\sqrt{\frac{2\bar E}{C}}\alpha+iE_{12})(\sqrt{\frac{2\bar E}{C}}\alpha-iE_{12})}
\end{equation}
which is consistent with the universal pole structure \eqref{eq:matrixelement}.
In particular, this pole structure gives rise to the classical linear growth of the geodesic length 
\begin{equation}\label{eq:JT length}
    \langle \hat\ell \rangle\approx \sqrt{\frac{2\bar E}{C}}t
\end{equation}
whose slope is determined by the ADM mass of the corresponding black hole.
Note that this is equivalent to the expectation from the geometric calculation \eqref{eq:JTlinear} under the thermodynamical relation 
\begin{equation}\label{eq:beta_E_relation}
    \beta=\sqrt{\frac{2\pi^2C}{E}}\,.
\end{equation}

In the main text, we mentioned that the pole structure of our interest originates from the oscillating form of the wave function \eqref{eq:wavefunction}.
In JT gravity, the pole structure \eqref{eq:JTpole} can be consistently derived from the asymptotic form of the wave function \eqref{eq:JT_wavefunction}.

In fact, using the asymptotic form of the Bessel function $K_{i\nu}(z)$ for $z\to0$,
\begin{equation}
    K_{\nu}(z)\overset{z\to0}{\to}\frac{1}{2}\Gamma(\nu)\left(\frac{z}{2}\right)^{-\nu}+\frac{1}{2}\Gamma(-\nu)\left(\frac{z}{2}\right)^{\nu}\,,
\end{equation}
\eqref{eq:JT_wavefunction}, at large $\ell$, becomes
\begin{equation}
    \psi_{E}(\ell)=\sqrt{8e^{-S_0}}K_{2ik}(4e^{-\frac{\ell}{2}})\to\sqrt{2e^{-S_0}}\left(\Gamma(2ik)\,2^{-2ik}e^{ik\ell}+\Gamma(-2ik)\,2^{2ik}e^{-ik\ell}\right)\,
\end{equation}
in which we introduced $k=\sqrt{2CE}$ for simplicity.
Thus, the product of two wave functions is expanded as
\begin{equation}
\begin{split}
    \psi_{E_1}^*(\ell)\psi_{E_2}(\ell)\to
    2e^{-S_0}&\left[A(k_1)A(k_2)^*e^{i(k_1-k_2)\ell}+A(k_1)^*A(k_2)e^{-i(k_1-k_2)\ell}\right.\\
    &\left.+A(k_1)A(k_2)e^{i(k_1+k_2)\ell}+A(k_1)^*A(k_2)^*e^{-i(k_1+k_2)\ell}\right]
\end{split}
\end{equation}
where
\begin{equation}
    A(k)=\Gamma(2ik)2^{-2ik}\,.
\end{equation}
The expected pole structure comes from the first two terms. Indeed, each term integrated with $e^{-\alpha\ell}$ results in a pole\footnote{The complete wave function \eqref{eq:JT_wavefunction} is doubly exponentially suppressed in negative large-$\ell$ region (See Appendix A in \cite{Miyaji:2025yvm}). Therefore, we may approximately drop the negative large-$\ell$ region from the integral. However, there is an ambiguity on where to start the integral region. For simplicity, we set the lower edge of the integral region to be zero since it only influences the results by an overall constant.}
\begin{equation}
    \int_{0}^{\infty}d\ell\,e^{-\alpha\ell}e^{\pm i(k_1-k_2)\ell}=\frac{1}{\alpha\mp i(k_1-k_2)}\,.
\end{equation}
Also, the coefficients evaluated on the poles become $|\Gamma(\alpha+2i\bar k)|^2$, with $\bar k=\frac{k_1+k_2}{2}$,
therefore, combining all the pieces together, we obtain
\begin{equation}
    \langle E_1| \widehat{e^{-\alpha\ell}} |E_2\rangle\approx\frac{e^{-S_0}}{2^{2\alpha-1}}|\Gamma(\alpha+ 2i\bar k)|^2\frac{2\alpha}{(k_1-k_2)^2+\alpha^2}\,.
\end{equation}
which is equivalent to \eqref{eq:JTpole1} in the $E=\frac{k^2}{2C}$ basis.

\subsection{CMC slices in JT gravity}
Another example in JT gravity is a constant-mean-curvature (CMC) slice. When the mean curvature is fixed to $K$, its classical renormalized length grows linearly as 
\begin{equation}\label{eq:CMClinear}
    \ell_{\rm classical}=\frac{2}{\sqrt{1+K^2}}\log{\cosh{\frac{\pi}{\beta}t}}\approx \frac{1}{\sqrt{1+K^2}}\frac{2\pi}{\beta}t, \qquad t\gg\beta\,.
\end{equation}

It was shown in \cite{Parrikar:2025xmz} that the length $\ell$ of CMC slices in JT gravity with a fixed mean curvature $K$ is governed by the following Hamiltonian.
\begin{equation}
    H=\frac{1}{2C}\left(-\frac{1}{1+K^2}\partial_\ell^2+4e^{-\ell\sqrt{1+K^2}}\right)\,.
\end{equation}
This expression is simply obtained by replacing $\ell$ by $\ell\sqrt{1+K^2}$ in the case of geodesic length, which is recovered as a special case, $K=0$, of the CMC slices. Therefore, the same argument as in Sec.~\ref{sec:JTlength} holds. We can obtain the final result by the following replacement
\begin{equation}
    \alpha\to\frac{\alpha}{\sqrt{1+K^2}}\,.
\end{equation}
Then, the matrix element of the length of the CMC slice possesses an associated pole structure
\begin{equation}
    \langle E_1| \widehat{e^{-\alpha\ell}} |E_2\rangle\sim \frac{\alpha}{\pi e^{S_0} D(E_0)} \frac{1}{(\sqrt{\frac{2E_0}{C(1+K^2)}}\alpha+iE_{12})(\sqrt{\frac{2E_0}{C(1+K^2)}}\alpha-iE_{12})}
\end{equation}
and this pole structure reproduces the classical linear growth of the CMC slice, \ie
\begin{equation}
    \langle \hat\ell \rangle\approx \sqrt{\frac{2E_0}{C(1+K^2)}}\,t\,,
\end{equation}
which is equivalent to \eqref{eq:CMClinear} under the relation \eqref{eq:beta_E_relation}.

\subsection{DSSYK and sine-dilaton gravity}\label{sec:DSSYK}
Another explicit example where we can compute the matrix element of holographic complexity is DSSYK or its holographic dual, sine-dilaton gravity \cite{Blommaert:2024whf,Blommaert:2024ymv}. In this duality, the chord number $n$ in DSSYK was identified with the geodesic length $L$ in sine-dilaton gravity through the relation
\begin{equation} \label{eq=n}
L=2|\log q| n,
\end{equation}
where $q$, with $0<q<1$, is a parameter of DSSYK.
It was also shown that the Krylov complexity associated with the TFD state is given by the expectation value of the chord number $n$. Together with the relation \eqref{eq=n}, this provides a concrete realization of the Complexity = Volume proposal in DSSYK \cite{Rabinovici:2023yex,Heller:2024ldz}.

The chord number in DSSYK, or equivalently the geodesic length in sine-dilaton gravity, is governed by the transfer matrix
\begin{equation}
T=-\frac{\cos P}{2|\log q|}+\frac{1}{4|\log q|}e^{iP} e^{-2|\log q|n}
\end{equation}
whose energy spectrum is parametrized by $\theta$ as
\begin{equation}
E(\theta)=-\frac{\cos\theta}{2|\log q|}\,.
\end{equation}
The corresponding left and right wave functions are expressed in terms of $q$-Hermite polynomials and $q$-Pochhammer symbols,
\begin{equation}
\psi^L_\theta(n)=\langle\theta|n\rangle=H_n(\cos\theta|q^2),\qquad
\psi_{\theta}^R(n)=\langle n|\theta\rangle=\frac{H_n(\cos\theta|q^2)}{(q^2;q^2)_n}\,.
\end{equation}
Then, the matrix element of the generating function of geodesic length in sine-dilaton gravity is given by
\begin{equation}
\begin{split}
    \langle\theta_1|e^{-\Delta L}|\theta_2\rangle&=\sum_{n=0}^{\infty} \psi^L_{\theta_1}(n) \psi^R_{\theta_2}(n) q^{2\Delta n}\\
    &=\frac{(q^{4\Delta};q^2)_{\infty}}{(q^{2\Delta }{e^{i(\pm\theta_1\pm\theta_2)}};q^2)_{\infty}}\,.
\end{split}
\end{equation}
Here, the denominator is understood as the product over all possible choices of signs.

The whole matrix element is meromorphic in $\Delta$, with its only singularities coming from the zeros of the four Pochhammers in the denominator, 
\begin{equation}
    (q^{2\Delta }{e^{i(\pm\theta_1\pm\theta_2)}};q^2)_{\infty}=\prod_{k=0}^{\infty}\left(1-q^{2\Delta+2k}e^{i(\pm\theta_1\pm\theta_2)}\right)=0\,.
\end{equation}
Equivalently, one finds the lattice of simple poles
\begin{equation}
    \Delta=-n+\frac{i(2\pi m-(\epsilon_1\theta_1+\epsilon_2\theta_2))}{2|\log q|},\qquad n=0,1,2,\ldots,m\in\mathbb{Z},\,\epsilon_i=\pm1\,.
\end{equation}
On the complex $\theta_{12}=\theta_1-\theta_2$ plane, the poles are located at
\begin{equation}\label{eq:thetapoles}
    \theta_{12}=2\pi m\pm 2i|\log q|\,(\Delta+n),\qquad m\in\mathbb{Z},\,n=0,1,2\ldots.
\end{equation}
Introducing $\bar\theta=(\theta_1+\theta_2)/2$, the energy difference $E_{12}=E(\theta_1)-E(\theta_2)$ is related to $\theta_{12}=\theta_1-\theta_2$ by
\begin{equation}
    E_{12}=\frac{\sin(\bar\theta)}{|\log q|}\sin\frac{\theta_{12}}{2}\,.
\end{equation}
Thus, the poles in the $E_{12}$ plane are located at
\begin{equation}
    E_{12}=\pm i\frac{\sin(\bar\theta)}{|\log q|}\sinh{\left(|\log q|(\Delta+n)\right)},\qquad n=0,1,2\ldots.
\end{equation}
When $|\log q|(\Delta+n)\ll1$, the pole locations reduce to
\begin{equation}
    E_{12}=\pm i\sin(\bar\theta)\,(\Delta+n),\qquad n=0,1,2\ldots.
\end{equation}
Therefore, the pole with $n=0$ is the leading contribution to the geodesic length and gives rise to the linear growth
\begin{equation}\label{eq:SDlength}
    \langle L\rangle\sim \sin(\bar\theta)\,t\,.
\end{equation}
This is consistent with the classical geodesic length in sine-dilaton gravity
\begin{equation}
    L_{\rm classical}=2\log\left(\cosh\left(\frac{\pi}{\beta}t\right)\right)-2\log\left(\frac{2\pi}{\beta}\right),\qquad \text{with} \qquad \beta=\frac{2\pi}{\sin\theta}\,.
\end{equation}

Also note that JT gravity is obtained by zooming in on the edge of the spectrum, $\theta\approx0$. In this limit, the JT energy $E$ is related to $\theta$ by
\begin{equation}
    \theta\approx\sin\theta\approx\sqrt{\frac{2E}{C}}\,.
\end{equation}
Thus, we recover the geodesic length in JT gravity \eqref{eq:JT length} from \eqref{eq:SDlength}.
Accordingly, the pole structure \eqref{eq:thetapoles} also reduces to the corresponding pole structure in JT gravity \eqref{eq:JTpole}.

\section{Dyson Classes}\label{app:Dyson}
In the main text, we specialize to the Gaussian unitary ensemble (GUE), or more generally to the unitary symmetry class with Dyson index $\beta=2$, because its spectral correlation functions admit a particularly simple determinantal representation. This makes it possible to evaluate the time dependence of the generating functions explicitly. The spectral mechanism responsible for the transition from semiclassical linear growth to a late-time plateau is, however, not restricted to the unitary class. After resolving all exact symmetries, chaotic Hamiltonians fall into one of the three Wigner--Dyson classes,
\begin{equation}
    \beta=1,\qquad \beta=2,\qquad \beta=4,
\end{equation}
corresponding, respectively, to the orthogonal, unitary, and symplectic symmetry classes. In the symplectic class, we count each Kramers pair once unless stated otherwise.

The dependence on the Dyson class enters the generating functions of complexity through the spectral correlation functions. In terms of the microscopic density of states $\rho(E)$, the codimension-one and codimension-zero generating functions are
defined as 
\begin{equation}
\begin{split}
G^{(1)}(\alpha;t)&=\frac{1}{Z} \int dE_i dE_j \, e^{-iE_{ij}t}
\overline{\left\langle \rho(E_i) \rho(E_j)\right\rangle} \,   \mathbf{M}_{\talpha}(E_i,E_j)
 \,, \\
G^{(0)}(\alpha_\pm;t) &= \frac{1}{Z}\int dE_1dE_2 dE_3 \, e^{-iE_{12}t} \overline{\langle \rho(E_1) \rho(E_2) \rho(E_3) \rangle} 
 \mathbf{M}_{\talpha_+} (E_1,E_3) \mathbf{M}_{\talpha_-} (E_3,E_2) \,.
\end{split}
\end{equation}
where
\begin{equation}
     \mathbf{M}_{\talpha}(E_i,E_j) :=\langle E_i|\widehat{e^{-\alpha\CC}}|E_j\rangle.
\end{equation}
All energy integrations are understood to lie within the microcanonical window as defined in the main text, and $Z=N$ denotes the corresponding number of levels.

The main structural distinction among the three classes can already be seen at the level of the factorial correlation functions. In the unitary class, the eigenvalues form a determinantal point process generated by a scalar Christoffel--Darboux kernel. In particular, the $n$-point factorial correlation function $R_n$ for GUE admits a particularly simple determinantal representation, 
\begin{equation}
    R_n^{(2)}(E_1,\ldots,E_n)
    =\det_{1\leq a,b\leq n}
    \left[K(E_a,E_b)\right],
    \qquad \beta=2.
    \label{eq:RnGUE}
\end{equation}
The orthogonal and symplectic classes, $\beta=1$ and $\beta=4$, instead define Pfaffian point processes. Their spectral correlation functions are generated by a $2\times2$ matrix-valued kernel, or equivalently by a self-dual quaternion kernel \cite{Tracy_1998,Widom:1998xm}. The correlation functions are Pfaffians, or equivalently quaternion determinants,
\begin{equation}
    R_n^{(\beta)}(E_1,\ldots,E_n) =
    \operatorname{Pf}
    \left[
        \mathbb{K}_{\beta}(E_i,E_j)
    \right]_{i,j=1}^{n},
    \qquad
    \beta=1,4.
\end{equation}
which are constructed from a $2\times2$ matrix-valued kernel $\mathbb{K}_{\beta}(E_i,E_j)$. The entries of the matrix-valued kernel  reduce in the universal bulk limit to the sine kernel and its derivative and integral.

In this appendix, we first derive the two-point spectral correlator in all three Dyson classes, including its unfolded bulk limit and short-distance level-repulsion behavior. We then evaluate the corresponding three-point Pfaffian and identify the connected structures that replace the simple GUE determinant expansion. The formulas below apply in the large-matrix-size bulk limit away from spectral edges and therefore extend beyond Gaussian potentials to invariant ensembles in the same universality class \cite{Tracy_1998,Eynard:2015aea}.

\subsection{Two-point spectral correlation functions in the three Dyson classes}
\label{app:dyson_two_point}
Consider the invariant $\beta$-ensemble
\begin{equation}
    \mathcal{P}_{\beta,N}(E_1,\ldots,E_N)
    =\frac{1}{\mathcal{Z}_{\beta,N}}
    \prod_{a=1}^{N}w(E_a)
    \prod_{a<b}|E_a-E_b|^{\beta},
    \qquad \beta=1,2,4.
    \label{eq:app_beta_ensemble}
\end{equation}
The $n$-level factorial correlation function is defined by
\begin{equation}
    R_n^{(\beta)}(E_1,\ldots,E_n)
    :=\frac{N!}{(N-n)!}
    \int dE_{n+1}\cdots dE_N\,
    \mathcal{P}_{\beta,N}(E_1,\ldots,E_N).
    \label{eq:app_Rn_definition}
\end{equation}
In particular,
\begin{equation}
    R_1^{(\beta)}(E)=\bar\rho(E),
    \qquad
    \int dE\,\bar\rho(E)=N.
    \label{eq:app_R1_density}
\end{equation}
The functions $R_n^{(\beta)}$ describe correlations among distinct eigenvalues. By contrast, correlation functions of the exact density operator
\begin{equation}
    \rho(E):=\sum_{a=1}^{N}\delta(E-E_a)
    \label{eq:app_exact_density}
\end{equation}
also contain contact terms.  For example, the two-point spectral correlation function is defined as 
\begin{equation}
   \boxed{ \overline{ \langle \rho(E_i)\rho(E_j) \rangle }
    =\bar\rho(E_i)\delta(E_i-E_j) +R_2^{(\beta)}(E_i,E_j)\,.}
    \label{eq:app_density_R2_relation}
\end{equation}
For $\beta=2$, the determinantal representation \eqref{eq:RnGUE} gives, in particular,
\begin{equation}
    R_2^{(2)}(E_i,E_j)
    =\bar\rho(E_i)\bar\rho(E_j)
    -K(E_i,E_j)K(E_j,E_i).
    \label{eq:app_R2_beta2}
\end{equation}
For $\beta=1,4$, the Vandermonde factor leads instead to a Pfaffian point process governed by a $2\times2$ matrix-valued kernel \cite{Tracy_1998,Widom:1998xm}. We now first review this construction below.

\subsubsection{Pfaffians and the matrix-valued correlation kernel}
\label{app:pfaffian_definition}
For an antisymmetric $2n\times2n$ matrix $A$, the Pfaffian is defined by
\begin{equation}
    \operatorname{Pf}A
    :=\frac{1}{2^n n!}
    \sum_{\sigma\in S_{2n}}
    \operatorname{sgn}(\sigma)
    \prod_{a=1}^{n}
    A_{\sigma(2a-1),\sigma(2a)},
    \qquad
    \bigl(\operatorname{Pf}A\bigr)^2=\det A.
    \label{eq:app_pfaffian_definition}
\end{equation}
The de Bruijn integration identities reduce the eigenvalue integrals for $\beta=1,4$ to Pfaffians. Algebraically, the relevant antisymmetric bilinear form pairs each basis function with an integral or derivative partner. The two components of this pair are conveniently packaged into a $2\times2$ kernel. This auxiliary matrix index is not a physical internal degree of freedom.

For $\beta=1,4$, introduce the self-dual $2\times2$ block kernel
\begin{equation}
    \mathbf{K}_{\beta}(E_i,E_j)
    :=
    \begin{pmatrix}
        S^{(\beta)}(E_i,E_j) & D^{(\beta)}(E_i,E_j)
        \\
        I^{(\beta)}(E_i,E_j) & S^{(\beta)}(E_j,E_i)
    \end{pmatrix} \,. 
    \label{eq:app_selfdual_kernel}
\end{equation}
Its entries obey
\begin{equation}
    D^{(\beta)}(E_i,E_j)=-D^{(\beta)}(E_j,E_i),
    \qquad
    I^{(\beta)}(E_i,E_j)=-I^{(\beta)}(E_j,E_i).
    \label{eq:app_DI_antisymmetry}
\end{equation}
It is convenient to define
\begin{equation}
    J:=
    \begin{pmatrix}
        0&1\\
        -1&0
    \end{pmatrix},
    \qquad
    \mathbb{K}_{\beta}(E_i,E_j)
    :=J\mathbf{K}_{\beta}(E_i,E_j) =
    \begin{pmatrix}
        I^{(\beta)}(E_i,E_j) & S^{(\beta)}(E_j,E_i)
        \\
        -S^{(\beta)}(E_i,E_j) & -D^{(\beta)}(E_i,E_j) 
         \end{pmatrix} \,.
    \label{eq:app_pfaffian_kernel}
\end{equation}
to obtain the antisymmetric block kernel which satisfies
\begin{equation}
    \mathbb{K}_{\beta}(E_i,E_j)
    =-\mathbb{K}_{\beta}(E_j,E_i)^{\mathsf T}.
\end{equation}
The factorial correlation functions are then 
\begin{equation}
    R_n^{(\beta)}(E_1,\ldots,E_n)
    =\operatorname{Pf}
    \left[
        \mathbb{K}_{\beta}(E_a,E_b)
    \right]_{a,b=1}^{n},
    \qquad \beta=1,4.
    \label{eq:app_Rn_pfaffian}
\end{equation}
Equivalently, one may express the same result as the quaternion determinant of the self-dual block matrix constructed from $\mathbf{K}_{\beta}$.

The one-point function follows immediately:
\begin{equation}
    R_1^{(\beta)}(E_i)
    =S^{(\beta)}(E_i,E_i)
    =\bar\rho(E_i).
    \label{eq:app_R1_matrix_kernel}
\end{equation}
For two energy levels, the Pfaffian of the corresponding $4\times4$ matrix gives
\begin{equation}
      R_2^{(\beta)}(E_i,E_j)=
    S^{(\beta)}(E_i,E_i)S^{(\beta)}(E_j,E_j)
    -S^{(\beta)}(E_i,E_j)S^{(\beta)}(E_j,E_i)
    +D^{(\beta)}(E_i,E_j)I^{(\beta)}(E_i,E_j).
    \label{eq:app_R2_matrix_kernel}
\end{equation}
It is therefore natural to define the dimensionful two-level cluster function
\begin{equation}
    \mathcal{Y}_2^{(\beta)}(E_i,E_j)
    :=S^{(\beta)}(E_i,E_j)S^{(\beta)}(E_j,E_i)-D^{(\beta)}(E_i,E_j)I^{(\beta)}(E_i,E_j),
    \label{eq:app_dimensionful_Y2}
\end{equation}
so that
\begin{equation}
    \boxed{R_2^{(\beta)}(E_i,E_j)
    =\bar\rho(E_i)\bar\rho(E_j)
    -\mathcal{Y}_{2}^{(\beta)}(E_i,E_j),}
    \label{eq:app_R2_dimensionful_cluster}
\end{equation}
Recasting the two-point spectral correlation function in a similar form as that in the main text, one can schematically write it as 
\begin{equation}
   \boxed{ \overline{\langle \rho(E_i)\rho(E_j) \rangle}
    =  \bar{\rho}(E_i)\bar{\rho}(E_j)
    +  \bar{\rho}(E_i)\delta(E_i-E_j) - \mathcal{Y}_2^{(\beta)}(E_i,E_j) \,.} 
\end{equation}
For $\beta=2$, the corresponding expression is recovered by setting
\begin{equation}
    \mathcal{Y}_2^{(2)}(E_i,E_j) :=K(E_i,E_j)K(E_j,E_i).
\end{equation}
with the sine-kernel 
\begin{equation}
K(E_i,E_j)
    \approx 
\frac{\sin\!\left[\pi\bar\rho(\bar E)(E_i-E_j)\right]}{\pi(E_i-E_j)} \,.
    \label{eq:app_dimensionful_sine_kernel}
\end{equation}
Thus, the derivative--integral term in \eqref{eq:app_dimensionful_Y2} is the essential distinction between the orthogonal or symplectic classes and the unitary class.

\subsubsection{Unfolded Spectrum}
In the following expressions, it is much more convenient to consider the unfolded spectrum. Introduce the unfolded coordinate
\begin{equation}
    x(E):=\int^{E}dE'\,\bar\rho(E'),
    \qquad
    \frac{dx}{dE}=\bar\rho(E),
    \label{eq:app_unfolding}
\end{equation}
and define
\begin{equation}
    x_i:=x(E_i),
    \qquad
    s_{ij}:=x_i-x_j
    =\int_{E_j}^{E_i}dE\,\bar\rho(E).
    \label{eq:app_exact_unfolded_separation}
\end{equation}
The unfolded $n$-point factorial correlation function is related to the original one by the Jacobian
\begin{equation}
 R_{n,\rm{unf}}^{(\beta)} (x_1, x_2\cdots x_n) = \frac{ R_{n}^{(\beta)} (E_1, E_2\cdots E_n) }{\prod_{a=1}^{n} \bar{\rho}(E_a)} \,. 
\end{equation}
A convenient representative of the unfolded Pfaffian kernel is obtained by
\begin{equation}
    \mathbb{K}_{\beta}^{\mathrm{unf}}(x_i,x_j)
    =
    \frac{
        \mathbb{K}_{\beta}(E_i,E_j)
    }{
        \sqrt{\bar\rho(E_i)\bar\rho(E_j)}
    } \,. 
    \label{eq:unfolded_matrix_kernel}
\end{equation}
In the large-$N$ bulk limit, the unfolded process is translation invariant. The two-point function therefore depends only on $s_{ij}$,
\begin{equation}
    R_{2,\mathrm{unf}}^{(\beta)}(s_{ij})
    =1-Y_2^{(\beta)}(s_{ij}),
    \label{eq:app_R2_Y2_unfolded}
\end{equation}
where $Y_2^{(\beta)}$ is dimensionless unfolded cluster function and is related to its dimensionful counterpart by 
\begin{equation}
    \mathcal{Y}_2^{(\beta)}(E_i,E_j)
    \simeq
    \bar\rho(E_i)\bar\rho(E_j)
    Y_2^{(\beta)}(s_{ij}).
    \label{eq:app_Y2_dimensional_unfolded}
\end{equation}
Consequently, the full two-point spectral correlator takes the universal local form
\begin{equation}\label{eq:app_rho_two_point_general_beta}
   \boxed{ \overline{\langle \rho(E_i)\rho(E_j) \rangle }
    \simeq 
    \bar\rho(E_i)\delta(E_i-E_j)  + \bar\rho(E_i)\bar\rho(E_j)  \left[ 1-  Y_2^{(\beta)}(s_{ij})  \right]} \,. 
\end{equation}
Furthermore, in a narrow spectral window centered at
\begin{equation}
    \bar{E}:=\frac{E_i+E_j}{2},
    \qquad E_{ij}:=E_i-E_j,
\end{equation}
one may further use the approximations 
\begin{align}
  s_{ij} = \bar\rho(\bar{E})E_{ij}+\mathcal{O}(E_{ij}^3)\,, \qquad s:=  \bar\rho(\bar E)(E_i-E_j)
    \label{eq:app_midpoint_approximation}
\end{align}
The resulting midpoint expression differs from \eqref{eq:app_rho_two_point_general_beta} only by subleading density gradient corrections.

\subsubsection{Universal matrix sine kernels}
\label{app:bulk_matrix_sine_kernels}

After unfolding the spectrum and taking the large-$N$ bulk scaling limit\footnote{By the large-$N$ microscopic bulk scaling limit, we mean that one fixes a reference energy $E_0$ in the interior (bulk) of the spectrum and considers $E_i \sim E_0$ while taking 
\begin{equation*}
  N \to \infty\,,  \quad  s_i = \bar\rho_N(E_0) (E_i -E_0)\,\, \text{fixed} \,.
\end{equation*}
However, at spectral edges, by contrast, the limiting correlations are described by Airy-type kernels at soft edges or Bessel-type kernels at hard edges, and are not translation invariant in this sense.}, the correlation functions become translation invariant. For the orthogonal and symplectic classes, the antisymmetric Pfaffian kernel associated with the unfolded factorial correlation functions can be represented as \footnote{Due to the equality $\operatorname{Pf}(B M B^{\mathsf T}) = \det(B)\operatorname{Pf}(M)$, the matrix kernel is not unique under local block transformations
\begin{equation*}
    \mathbb{K}_{\beta}(E_i,E_j)
    \longrightarrow
    A_i\mathbb{K}_{\beta}(E_i,E_j)A_j^{\mathsf T},
    \qquad \det A_i=1,
\end{equation*}
which leave all Pfaffian correlation functions invariant.}
\begin{equation}
    \mathbb{K}^{\mathrm{unf}}_{\beta}(s_{ij})
    =
 \begin{pmatrix}
        I^{(\beta)}(s_{ij}) & S^{(\beta)}(-s_{ij})
        \\
        -S^{(\beta)}(s_{ij}) & -D^{(\beta)}(s_{ij}) 
         \end{pmatrix}\,,
    \qquad \beta=1,4.
    \label{eq:app_bulk_matrix_kernel}
\end{equation}
where $S^{(\beta)}(s)$ is even, whereas $D^{(\beta)}(s)$ and $I^{(\beta)}(s)$ are odd. The unit one-point density of the unfolded spectrum fixes
\begin{equation}
     R_{1,\mathrm{unf}}^{(\beta)}(x) =\operatorname{Pf}\mathbb K_{\beta}^{\mathrm{unf}}(0)= S^{(\beta)}(0) = 1 \,. 
\end{equation}
while the oddness of $D^{(\beta)}$ and $I^{(\beta)}$ implies
\begin{equation}
    D^{(\beta)}(0)=I^{(\beta)}(0)=0.
\end{equation}
For two unfolded levels, the factorial correlation function is
\begin{equation}
    R_{2,\mathrm{unf}}^{(\beta)}(s)
    =
    \operatorname{Pf}
    \begin{pmatrix}
        \mathbb{K}^{\mathrm{unf}}_{\beta}(0)
        &
        \mathbb{K}^{\mathrm{unf}}_{\beta}(s)
        \\
        \mathbb{K}^{\mathrm{unf}}_{\beta}(-s)
        &
        \mathbb{K}^{\mathrm{unf}}_{\beta}(0)
    \end{pmatrix} 
    =
    1-S^{(\beta)}(s)S^{(\beta)}(-s)+ D^{(\beta)}(s)I^{(\beta)}(s)\,.
\end{equation}
Consequently, the two-level cluster function is given by 
\begin{equation}
  \boxed{  Y_2^{(\beta)}(s) =S^{(\beta)}(s)^2-D^{(\beta)}(s)I^{(\beta)}(s)\,.}
    \label{eq:app_unfolded_Y2_matrix_kernel}
\end{equation}

The scalar sine kernel,
\begin{equation}
    S(s):=\frac{\sin(\pi s)}{\pi s}\,,
    \label{eq:app_scalar_sine_function}
\end{equation}
remains the basic building block for all three Dyson classes, but in the orthogonal and symplectic classes it must be supplemented by its derivative and integral. 
For the orthogonal class,
\begin{align}
    S^{(1)}(s)&=S(s),
    \\
    D^{(1)}(s)&=\frac{d}{ds}S(s),
    \\
    I^{(1)}(s)&=\int_0^sdu\,S(u)-\frac{1}{2}\operatorname{sgn}(s).
    \label{eq:app_GOE_matrix_sine_kernel}
\end{align}
For the symplectic class, counting one representative from each Kramers pair and unfolding the resulting spectrum to unit mean spacing,
\begin{align}
    S^{(4)}(s)&=S(2s),
    \\
    D^{(4)}(s)&=\frac{d}{ds}S(2s),
    \\
    I^{(4)}(s)&=\int_0^sdu\,S(2u).
    \label{eq:app_GSE_matrix_sine_kernel}
\end{align}
If both members of each Kramers pair are retained, the unfolding convention and the factors of two in \eqref{eq:app_GSE_matrix_sine_kernel} must be modified accordingly.

Writing $r:=|s|$, one obtains
\begin{align}
    Y_2^{(1)}(s)
    &=S(r)^2
    +S'(r)\int_r^{\infty}du\,S(u),
    \label{eq:app_Y2_GOE}
    \\
    Y_2^{(2)}(s)
    &=S(s)^2,
    \label{eq:app_Y2_GUE}
    \\
    Y_2^{(4)}(s)
    &=S(2r)^2
    -S'(2r)\int_0^{2r}du\,S(u),
    \label{eq:app_Y2_GSE}
\end{align}
where $S'(u)$ denotes differentiation with respect to the argument. Equivalently,
\begin{align}
    Y_2^{(1)}(s)
    &=S(r)^2
    +S'(r)
    \left[
        \frac{1}{2}-\frac{1}{\pi}\operatorname{Si}(\pi r)
    \right],
    \\
    Y_2^{(4)}(s)
    &=S(2r)^2
    -\frac{1}{\pi}S'(2r)\operatorname{Si}(2\pi r),
\end{align}
with
\begin{equation}
    \operatorname{Si}(x):=\int_0^xdu\,\frac{\sin u}{u}.
\end{equation}
For $\beta=2$, the eigenvalue process is determinantal and the cluster function is the square of the scalar sine kernel. For $\beta=1$ and $\beta=4$, the correlation functions instead have a Pfaffian structure, and the additional derivative-integral terms arise from the corresponding matrix-valued sine kernels.

Substituting \eqref{eq:app_rho_two_point_general_beta} into the codimension-one generating function $G^{(1)}(\alpha ,t)$ gives the universal decomposition
\begin{equation}
   \boxed{ G^{(1)}_{\beta} (\alpha,t)
    \equiv G^{(1)}_{\rm classical}
    +G^{(1)}_{\rm delta}
    +G^{(1)}_{\mathcal{Y}_2^{(\beta)}} \,.}
    \label{eq:app_G1_beta_decomposition}
\end{equation}
where the first two terms have the same form in every Dyson class and the class dependence is carried by the last term: 
\begin{equation}
  G^{(1)}_{\mathcal{Y}_2^{(\beta)}} :=-\frac{1}{Z}\int dE_i\,dE_j\,
    e^{-iE_{ij}t}
    \mathcal{Y}_2^{(\beta)}(E_i,E_j)
    \mathbf{M}_{\widetilde{\alpha}}(E_i,E_j) \,. 
    \label{eq:app_G1_beta_components}
\end{equation}

\subsubsection{Level repulsion and late-time plateau}
The unfolded two-point correlation function is given by 
\begin{equation}
   \boxed{ R_{2,\mathrm{unf}}^{(\beta)}(s)
    =1-S^{(\beta)}(s)^2
    +D^{(\beta)}(s)I^{(\beta)}(s)
    =1-Y_2^{(\beta)}(s) \,.} 
    \label{eq:app_unfolded_R2_matrix_kernel}
\end{equation}
The distinction among the three Dyson classes is particularly transparent at small unfolded separation. Taking the $s \to 0$ limit and expanding the preceding formulas gives
\begin{align}
    R_{2,\mathrm{unf}}^{(1)}(s)
    &=\frac{\pi^2}{6}|s|+\mathcal{O}(s^2),
    \\
    R_{2,\mathrm{unf}}^{(2)}(s)
    &=\frac{\pi^2}{3}s^2+\mathcal{O}(s^4),
    \\
    R_{2,\mathrm{unf}}^{(4)}(s)
    &=\frac{16\pi^4}{135}s^4+\mathcal{O}(s^6).
    \label{eq:app_level_repulsion_expansions}
\end{align}
These expressions satisfy
\begin{equation}
    R_{2,\mathrm{unf}}^{(\beta)}(s) \propto |s|^{\beta},\qquad s\to0 \,. 
    \label{eq:app_Dyson_level_repulsion}
\end{equation}
In particular,
\begin{equation}
    R_{2,\mathrm{unf}}^{(\beta)}(0)=0,
    \qquad
    \beta=1,2,4.
    \label{eq:app_R2_zero}
\end{equation}

As shown in section~\ref{sec:reg}, the growth rate of the codimension-one holographic complexity at late-times is controlled by the factorial two-point correlation function $R_2$ evaluated at the relevant pole, $E_{ij}=-i\talpha$. In the limit $\alpha\to0$, this pole approaches the coincident-energy point. Consequently, the corresponding growth rate vanishes as a direct result of level repulsion, as expressed in eq.~\eqref{eq:app_R2_zero}. The classical linear growth is therefore precisely canceled by the quantum correction, leading to saturation at late times. This conclusion is not specific to the GUE. In all three Wigner--Dyson classes, corresponding to $\beta=1,2,4$, the distinct-level correlator vanishes when two energies coincide as a universal consequence of level repulsion. More generally, the same reasoning extends to the $\beta$-ensemble for any $\beta>0$, for which the local two-level correlation behaves as $R_2(E,E+s)\propto |s|^\beta$ at small energy separation. Although the detailed ramp profile and the approach to saturation depend on $\beta$, the cancellation of the classical linear growth, and hence the emergence of a late-time plateau \footnote{Here and throughout this paper, by a late-time plateau we mean the absence of persistent linear growth and an asymptotically vanishing growth rate. Subleading time-dependent contributions may nevertheless persist, producing residual growth or decay. For example, when $\beta=1$, a logarithmic contribution $\sim\log t$ may remain, while its derivative decays as $1/t$ at late times. More generally, additional subleading power-law corrections may also be present, provided that their time derivatives vanish asymptotically.}, follow universally from level repulsion, a generic feature of quantum-chaotic spectra.

\subsection{Three-point spectral correlation functions}
\label{app:dyson_three_point}
Finally, we move to the general discussion about the codimension-zero holographic complexity and the corresponding generating functions, which explicitly depend on the three-point spectral correlation function. For simplicity, the explicit time evolution is derived based on the sine kernel for GUE where the three-point factorial correlation function in the unitary class is the determinant of a $3\times3$ scalar-kernel matrix. For $\beta=1,4$, it is instead the Pfaffian of a $6\times6$ antisymmetric matrix,
\begin{equation}
    R_3^{(\beta)}(E_1,E_2,E_3)
    =\operatorname{Pf}
    \left[
        \mathbb{K}_{\beta}(E_a,E_b)
    \right]_{a,b=1}^{3},
    \qquad \beta=1,4.
    \label{eq:app_R3_pfaffian}
\end{equation}
To display the result compactly, define
\begin{equation}
    S_{ij}^{(\beta)}:=S^{(\beta)}(E_i,E_j),
    \qquad
    D_{ij}^{(\beta)}:=D^{(\beta)}(E_i,E_j),
    \qquad
    I_{ij}^{(\beta)}:=I^{(\beta)}(E_i,E_j).
\end{equation}
and
\begin{equation}
    \mathcal{Y}_{2,ij}^{(\beta)} 
    :=  \mathcal{Y}_{2}^{(\beta)}(E_i, E_j)=S_{ij}^{(\beta)}S_{ji}^{(\beta)}
    -D_{ij}^{(\beta)}I_{ij}^{(\beta)}.
\end{equation}
A direct evaluation of the Pfaffian gives
\begin{align}
  \boxed{  R_3^{(\beta)}(E_1,E_2,E_3)
    =\bar\rho_1\bar\rho_2\bar\rho_3
    -\bar\rho_1\mathcal{Y}_{2,23}^{(\beta)}
    -\bar\rho_2\mathcal{Y}_{2,31}^{(\beta)}
    -\bar\rho_3\mathcal{Y}_{2,12}^{(\beta)}
    +\mathcal{Y}_{3,123}^{(\beta)},}
    \label{eq:app_R3_cluster_decomposition}
\end{align}
where $\bar\rho_i:=\bar\rho(E_i)$ and
\begin{equation} \label{eq:app_dimensionful_Y3_general}
    \mathcal{Y}_{3,123}^{(\beta)}:=S_{12}^{(\beta)}S_{23}^{(\beta)}S_{31}^{(\beta)}
    +S_{13}^{(\beta)}S_{32}^{(\beta)}S_{21}^{(\beta)}  +\sum_{(i,j,k)\in\mathrm{cyc}}
    D_{ij}^{(\beta)}
    \left(
        I_{jk}^{(\beta)}S_{ki}^{(\beta)}
        -I_{ik}^{(\beta)}S_{kj}^{(\beta)}
    \right).   
\end{equation}
The cyclic sum runs over $(i,j,k)=(1,2,3),(2,3,1),(3,1,2)$. 

The full three-point density correlator also contains contact terms. With
\begin{equation}
    \delta_{ij}:=\delta(E_i-E_j),
\end{equation}
one has, in every Dyson class,
\begin{align}
    \overline{\langle \rho_1\rho_2\rho_3 \rangle }
    ={}&R_3^{(\beta)}(E_1,E_2,E_3)
    +\delta_{12}R_2^{(\beta)}(E_1,E_3)
    +\delta_{23}R_2^{(\beta)}(E_1,E_2)
    \nonumber\\
    &+\delta_{31}R_2^{(\beta)}(E_2,E_3)
    +\delta_{12}\delta_{13}\bar\rho_1.
    \label{eq:app_full_rho3_general_beta}
\end{align}
Equivalently, introducing connected cumulants,
\begin{equation}
   \boxed{ \overline{ \langle \rho_1\rho_2\rho_3 \rangle}
    =\bar\rho_1\bar\rho_2\bar\rho_3
    +\bar\rho_1{ \langle \rho_2\rho_3 \rangle}_{\mathrm c}
    +\bar\rho_2{ \langle \rho_3\rho_1 \rangle }_{\mathrm c}
    +\bar\rho_3{\langle \rho_1\rho_2 \rangle}_{\mathrm c}
    +{\langle \rho_1\rho_2\rho_3 \rangle }_{\mathrm c},}
    \label{eq:app_rho3_cumulant_decomposition}
\end{equation}
where
\begin{equation}
     {\langle \rho_i\rho_j \rangle }_{\mathrm c}
    =\bar\rho_i\delta_{ij}
    -\mathcal{Y}_{2,ij}^{(\beta)},
    \label{eq:app_rho2_connected_general_beta}
\end{equation}
and
\begin{equation}
  \boxed{ {\langle \rho_1\rho_2\rho_3 \rangle }_{\mathrm c}
    =\bar\rho_1\delta_{12}\delta_{13}
    +\mathcal{Y}_{3,123}^{(\beta)}
    -\delta_{12}\mathcal{Y}_{2,13}^{(\beta)}
    -\delta_{23}\mathcal{Y}_{2,12}^{(\beta)}
    -\delta_{31}\mathcal{Y}_{2,23}^{(\beta)} \,. }
    \label{eq:app_rho3_connected_general_beta}
\end{equation}
For $\beta=2$, eq.~\eqref{eq:app_rho3_cumulant_decomposition} reduces to the GUE expression as shown in eq.~\eqref{eq:DDD} after taking
\begin{equation}
    \mathcal{Y}_2^{(2)}(E_i,E_j)=K_{ij}K_{ji}\,, 
    \qquad
    \mathcal{Y}_3^{(2)}(E_1,E_2,E_3) =K_{12}K_{23}K_{31}+K_{13}K_{32}K_{21}\,.
\end{equation}

Substitution of \eqref{eq:app_rho3_cumulant_decomposition} into the codimension-zero generating function yields the same structural results as eq.~\eqref{eq:Gzerototal}
\begin{equation}
   \boxed{ G_\beta^{(0)}(\alpha_\pm;t) = G_{\rm classical}  +G_{\rm bra-op, \beta} +G_{\rm op-ket, \beta}  +G_{\rm bra-ket,\beta} +G_{\rm bra-op-ket, \beta}\,.}
     \label{eq:app_G0_beta_decomposition}
\end{equation}
The five terms have exactly the same connectivity interpretation as in the GUE analysis. The fully disconnected term $G_{\rm classical}$ is obtained from $\bar\rho_1\bar\rho_2\bar\rho_3$. The pairwise terms arise from $\bar\rho_2\langle {\rho_3\rho_1}\rangle_{\mathrm c}$, $\bar\rho_1\langle{\rho_2\rho_3}\rangle_{\mathrm c}$, and $\bar\rho_3\langle{\rho_1\rho_2}\rangle_{\mathrm c}$, corresponding, respectively, to the bra--operator, operator--ket, and bra--ket sectors. The final term is governed by \eqref{eq:app_rho3_connected_general_beta} and connects all three sectors.

The explicit Fourier transforms of the cluster functions, and hence the detailed ramp profiles and plateau coefficients, depend on $\beta$. Nevertheless, the short-distance suppression is common to all three classes. Whenever two arguments approach one another while the remaining energies are kept distinct,
\begin{equation}
    R_n^{(\beta)}(E_1,\ldots,E_a,\ldots,E_b,\ldots,E_n)
    \propto |E_a-E_b|^{\beta},
    \qquad E_a\to E_b.
    \label{eq:app_Rn_level_repulsion_general_beta}
\end{equation}
In particular,
\begin{equation}
    R_2^{(\beta)}(E,E)=0,
    \qquad
    R_3^{(\beta)}(E_1,E,E)=0.
    \label{eq:app_R2_R3_coincident_general_beta}
\end{equation}
These are the class-independent ingredients that underlie the cancellation mechanism discussed in the main text. The detailed codimension-one and codimension-zero time dependences need not coincide with the GUE polynomial, but the transition from classical linear growth to late-time saturation is expected to persist throughout the Wigner--Dyson classes.

\subsubsection{Unfolded three-level spectral correlations}
As in the two-point analysis, it is convenient to work with the unfolded bulk spectrum $x(E)$, for which the correlation functions become translation invariant in the large-$N$ microscopic bulk scaling limit. Writing $s_{ij}=x_i-x_j$, the scalar functions entering the Pfaffian kernels satisfy
\begin{equation}
   S_{ij}^{(\beta)}:=S^{(\beta)}(s_{ij})=S_{ji}^{(\beta)},
    \quad D_{ij}^{(\beta)}:=D^{(\beta)}(s_{ij})=-D_{ji}^{(\beta)},
    \quad  I_{ij}^{(\beta)}:=I^{(\beta)}(s_{ij})=-I_{ji}^{(\beta)}\,.
\end{equation} 
The unfolded three-level density can then be written as
\begin{equation}
   \boxed{ R_{3,\mathrm{unf}}^{(\beta)}(x_1,x_2,x_3)
    =1 +Y_3^{(\beta)}(x_1,x_2,x_3) - \sum_{(ij)=(12),(23),(31)}Y_2^{(\beta)}(s_{ij}) \,.}
    \label{eq:app_R3_unfolded_beta14}
\end{equation}
Here, the dimensionful and unfolded three-level cluster functions are related by
\begin{equation}
    \mathcal{Y}_{3,123}^{(\beta)}
    =
    \bar\rho(E_1)\bar\rho(E_2)\bar\rho(E_3)\,Y_3^{(\beta)}(x_1,x_2,x_3),
\label{eq:app_Y3_dimensional_unfolded}
\end{equation}
where
\begin{equation}   \label{eq:app_unfolded_Y3_beta14}
  \begin{split}
        &{}Y_3^{(\beta)}(x_1,x_2,x_3)
    =2S_{12}^{(\beta)}S_{23}^{(\beta)}S_{31}^{(\beta)}
    \\
    &+S_{12}^{(\beta)}\left(
        D_{23}^{(\beta)}I_{31}^{(\beta)}
        +I_{23}^{(\beta)}D_{31}^{(\beta)} \right)
    +S_{23}^{(\beta)}\left(
        D_{31}^{(\beta)}I_{12}^{(\beta)}
        +I_{31}^{(\beta)}D_{12}^{(\beta)} \right)
    +S_{31}^{(\beta)}\left(
        D_{12}^{(\beta)}I_{23}^{(\beta)}
        +I_{12}^{(\beta)}D_{23}^{(\beta)}\right),
  \end{split}
\end{equation}

For the unitary class, setting $D_{ij}^{(2)}=I_{ij}^{(2)}=0$ and $S_{ij}^{(2)}=S(s_{ij})$ reduces eq.~\eqref{eq:app_unfolded_Y3_beta14} to
\begin{equation}
    Y_3^{(2)}(s_1,s_2,s_3)   =   2S(s_{12})S(s_{23})S(s_{31})\,,
\label{eq:app_Y3_beta2}
\end{equation}
which is precisely the cyclic contribution obtained by expanding the GUE determinant. For the orthogonal and symplectic classes, fully explicit expressions
simply follow by substituting the corresponding universal functions $S^{(\beta)}$, $D^{(\beta)}$, and $I^{(\beta)}$ from eqs.~\eqref{eq:app_GOE_matrix_sine_kernel} and \eqref{eq:app_GSE_matrix_sine_kernel}. The resulting formulas for $Y_3^{(\beta)}$ and $R_{3,\mathrm{unf}}^{(\beta)}$ are straightforward but considerably more cumbersome, and do not provide additional structural insight. We therefore retain the compact kernel representation \eqref{eq:app_unfolded_Y3_beta14} here.



\bibliographystyle{jhep}
\bibliography{references}
\end{document}